\documentclass[letterpaper,twocolumn,10pt]{article}
\usepackage{usenix}
\usepackage{tcolorbox}

\newtcolorbox{finding}{
  colback=cyan!5,        
  colframe=teal,         
  boxrule=1pt,
  leftrule=3pt,
  arc=3pt,
  left=10pt,
  right=10pt,
  top=5pt,
  bottom=5pt
}

\usepackage{tikz}
\usepackage{algorithm}
\usepackage{algpseudocode}  

\usepackage{amsmath}
\usepackage{amssymb}   
\usepackage{graphicx}  
\usepackage{booktabs}
\usepackage{multicol}
\usepackage{multirow}
\usepackage[table]{xcolor}
\usepackage{hyperref} 
\usepackage{fontawesome5}
\usepackage{tabularx}
\usepackage{booktabs}
\usepackage{ragged2e}
\usepackage{enumitem}
\usepackage{makecell}
\usepackage[htt]{hyphenat}

\newcommand{\gitlink}[1]{\href{#1}{\textcolor{black}{\faGithub}}}
\newcommand{\apilink}[1]{\href{#1}{\textcolor{gray}{\faCloud}}}
\newcommand{\mylogo}[1]{\raisebox{-0.15em}{\includegraphics[height=1em]{#1}}}
\newcommand{\modelname}[1]{\texttt{\small #1}}

\begin{document}

\date{}

\title{\Large \bf 
Na\"ive Exposure of Generative AI Capabilities Undermines Deepfake Detection
}

\author{
{\rm Sunpill Kim$^\dagger$ \quad Chanwoo Hwang$^\dagger$ \quad Minsu Kim \quad Jae Hong Seo$^*$} \\
Department of Mathematics \& Research Institute for Natural Sciences, Hanyang University\\
\texttt{\{ksp0352, aa5568, iayaho3248, jaehongseo\}@hanyang.ac.kr}\\
{\rm $^\dagger$ Equal contribution \quad $^*$ Corresponding author}
} 

\maketitle

\begin{abstract}

Generative AI systems increasingly expose powerful reasoning and image refinement capabilities through user-facing chatbot interfaces. In this work, we show that the na\"ive exposure of such capabilities fundamentally undermines modern deepfake detectors. Rather than proposing a new image manipulation technique, we study a realistic and already-deployed usage scenario in which an adversary uses only benign, policy-compliant prompts and commercial generative AI systems.

We demonstrate that state-of-the-art deepfake detection methods fail under \emph{semantic-preserving image refinement}. Specifically, we show that generative AI systems articulate explicit authenticity criteria and inadvertently externalize them through unrestricted reasoning, enabling their direct reuse as refinement objectives. As a result, refined images simultaneously evade detection, preserve identity as verified by commercial face recognition APIs, and exhibit substantially higher perceptual quality. Importantly, we find that widely accessible commercial chatbot services pose a significantly greater security risk than open-source models, as their superior realism, semantic controllability, and low-barrier interfaces enable effective evasion by non-expert users. Our findings reveal a structural mismatch between the threat models assumed by current detection frameworks and the actual capabilities of real-world generative AI. While detection baselines are largely shaped by prior benchmarks, deployed systems expose unrestricted authenticity reasoning and refinement despite stringent safety controls in other domains. 

\end{abstract}

\section{Introduction}
The widespread use of deepfakes and AI-generated media has led to a crisis of trust in digital visual evidence, facilitating identity fraud, misinformation campaigns, and reputational damage. In response, the academic community has developed a multitude of deepfake detection methods~\cite{zhao2021multi, yan2023ucf, liang2022exploring, liu2021spatial, shiohara2022detecting, nguyen2024laa, zheng2021exploring}, achieving impressive performance on standardized benchmarks such as FaceForensics++~\cite{rossler2019faceforensics++}, Celeb-DF~\cite{li2020celeb}, and DeepFake Detection Challenge~\cite{dolhansky2020deepfake}. However, these successes largely rely on the assumption that generative models leave persistent, detectable fingerprints---such as frequency artifacts~\cite{tan2024frequency, liu2021spatial}, blending inconsistencies~\cite{zheng2021exploring, nguyen2024laa, shiohara2022detecting}, or spatial anomalies~\cite{zhao2021multi, liang2022exploring}---that distinguish them from authentic content.

\definecolor{deepred}{RGB}{220, 20, 60}   
\definecolor{realgreen}{RGB}{34, 139, 34} 

\begin{figure}[t]
    \centering
    \setlength{\fboxrule}{1.3pt} 
    \setlength{\fboxsep}{0pt}    
    \textbf{Refined Images using GPT (classified as \textcolor{realgreen}{Real} by detector)}

    \fcolorbox{realgreen}{white}{\includegraphics[width=0.32\linewidth]{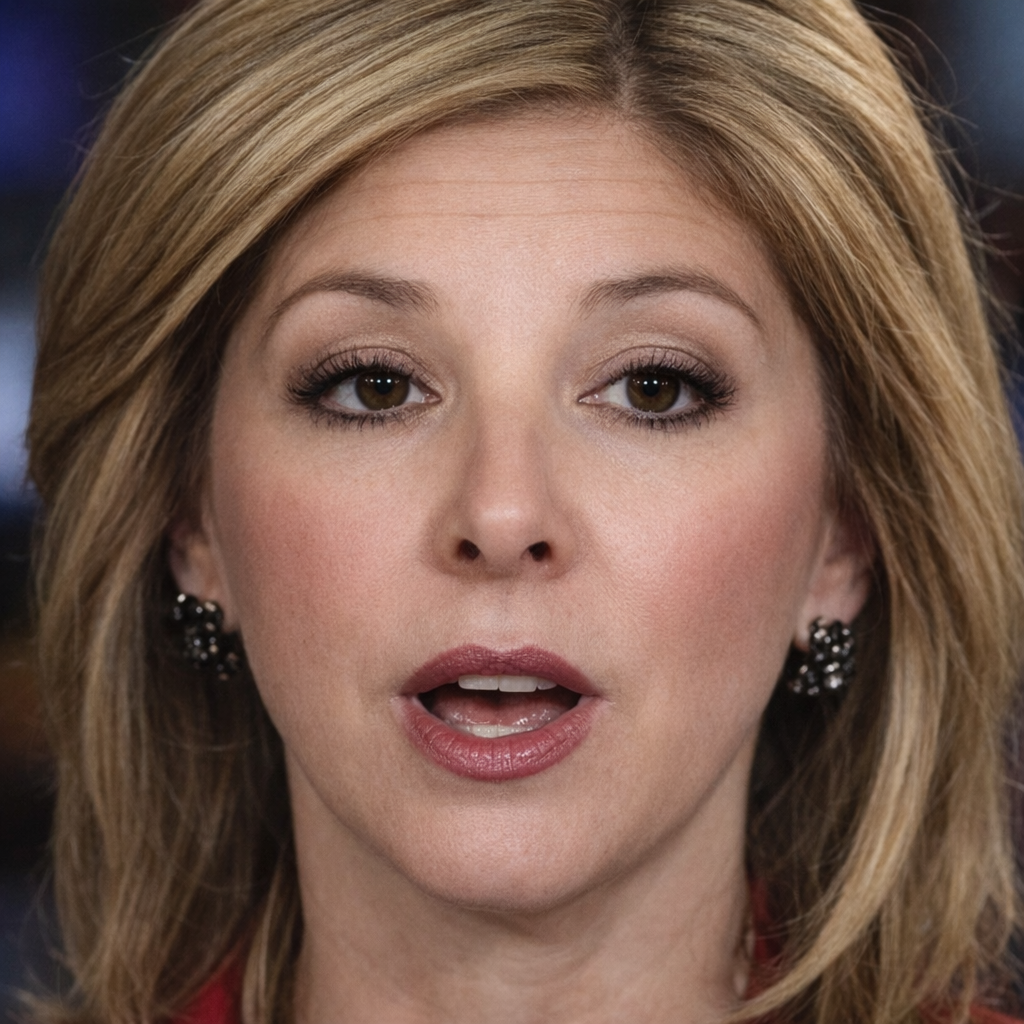}}\hfill
    \fcolorbox{realgreen}{white}{\includegraphics[width=0.32\linewidth]{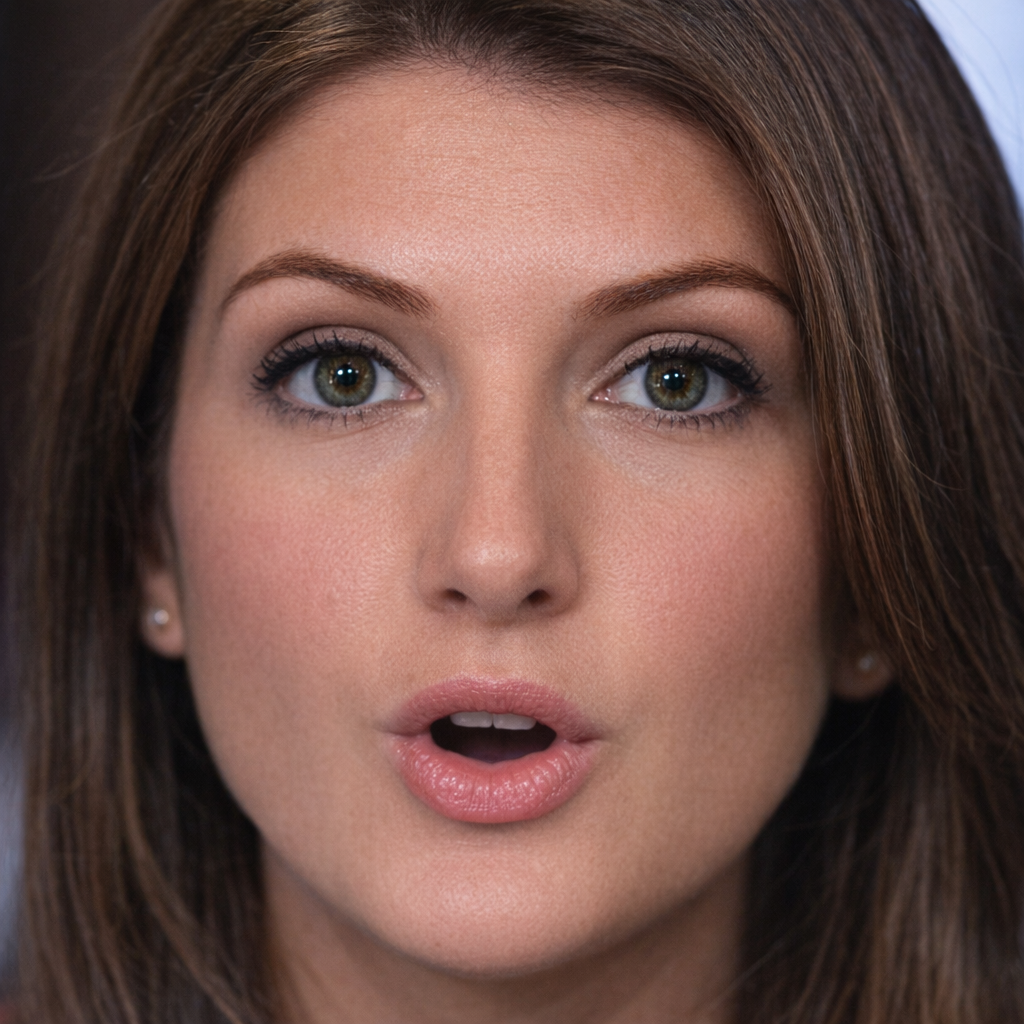}}\hfill
    \fcolorbox{realgreen}{white}{\includegraphics[width=0.32\linewidth]{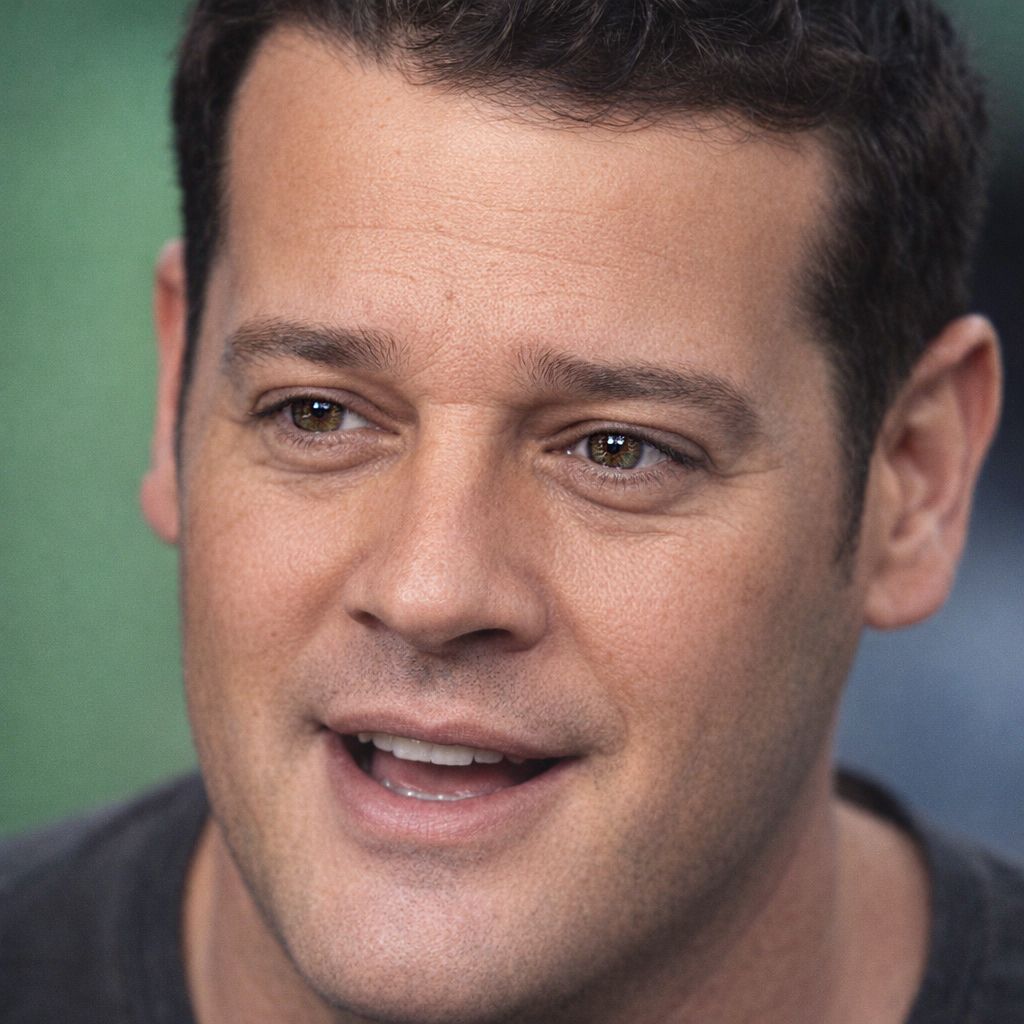}}

    \textbf{Deepfake Images from the \textcolor{deepred}{FaceForensics++}}

    \fcolorbox{deepred}{white}{\includegraphics[width=0.32\linewidth]{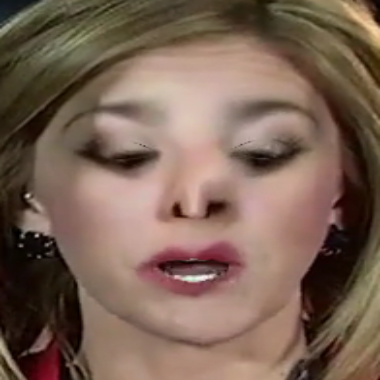}}\hfill
    \fcolorbox{deepred}{white}{\includegraphics[width=0.32\linewidth]{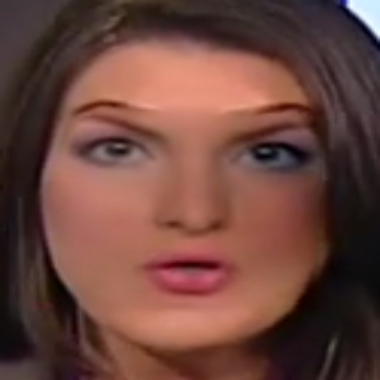}}\hfill
    \fcolorbox{deepred}{white}{\includegraphics[width=0.32\linewidth]{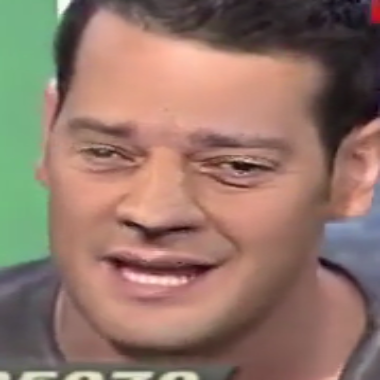}}

    \caption{Semantic-preserving refinement alters authenticity judgments without changing identity from commercial APIs.}\label{fig:comparison}

    
\end{figure}

We argue that this prevalent framing of deepfake detection as a static classification task is fundamentally flawed. In practice, detection operates in a highly adversarial environment where attackers rapidly adapt to the assumptions made by defense mechanisms. While early generation techniques (e.g., GANs~\cite{karras2017progressive,karras2019style}, early diffusion models~\cite{ho2020denoising,song2020denoising}) exhibited clear artifact patterns, the landscape has shifted dramatically with the advent of general-purpose Generative AI (GAI) systems. These modern systems, particularly Large Multimodal Models (LMMs) like GPT-4~\cite{achiam2023gpt} and Gemini~\cite{comanici2025gemini}, are not merely distinct generators; they are unified platforms that integrate visual perception, complex reasoning, and high-fidelity image refinement.

In this work, we demonstrate that the na\"ive exposure of these capabilities through user-facing interfaces creates a critical, structural vulnerability in current deepfake defense pipelines. We identify a realistic and already-deployed threat model in which an adversary utilizes widely accessible commercial GAI systems to evade state-of-the-art detectors without requiring technical expertise, white-box access, or policy-violating prompts. This threat arises not from a new generative algorithm, but from the interaction of three capabilities exposed within a single interface: (1) authenticity assessment, where the system articulates expert-level criteria for judging realism; (2) structured reasoning, where the system identifies specific artifacts in a given image; and (3) semantic-preserving refinement, where the system refines the image to correct these artifacts while maintaining the subject's identity.

Our analysis reveals that commercial GAI systems inadvertently externalize their alignment-driven judgments~\footnote{These capabilities are not explicitly designed for deepfake detection or evasion. Rather, they emerge from the interaction of general-purpose visual reasoning, alignment-driven helpfulness, and safety policies that permit authenticity assessment and benign image refinement.} about visual authenticity as an attack vector. When prompted, they articulate explicit authenticity criteria---such as skin texture irregularities, lighting inconsistencies, or unnatural hair boundaries---and apply them to critique input images. We show that these critiques are actionable: an adversary can simply feed the system’s own negative feedback back into the generation loop as a refinement objective. Because these refinement requests are framed as benign commands to ``improve naturalness'' or ``fix lighting,'' they bypass input-level safety systems designed to catch explicit malicious intent (e.g., ``make a deepfake'').


The consequences of this feedback loop are severe. We demonstrate that semantic-preserving refinement collapses the performance of state-of-the-art detectors. Crucially, our experiments show that this process preserves high-level semantics---specifically identity, pose, and expression---verified by commercial face recognition APIs, AWS CompareFaces~\cite{AWS} and Tencent CompareFace~\cite{Tencent}, while simultaneously enhancing perceptual quality to a degree that surpasses the original generation. This creates a paradox where the ``better'' and more realistic the image becomes through refinement, the more likely it is to be classified as real by both humans and automated detectors, effectively weaponizing the quest for visual quality against forensic analysis.

Furthermore, we highlight a significant disparity between open-source models and commercial services. While open-source models provide transparency, we find that closed, commercial chatbot services pose a significantly greater security risk due to their superior reasoning capabilities, higher-fidelity outputs, and low-barrier interfaces. These services democratize sophisticated evasion techniques, allowing non-expert users to produce forensic-resistant deepfakes through natural language conversation alone. This finding exposes a structural mismatch: while detection research often focuses on signal-level artifacts from specific generators, real-world adversaries are empowered by logic-driven, multimodal systems that actively assist in scrubbing those very signals.

In summary, this paper makes the following contributions:

\begin{itemize}
    \setlength\itemsep{0.2em}
    \item Identification of a Logic-Driven Evasion Vector: We show that the externalized reasoning of GAI systems regarding image authenticity can be directly repurposed as an optimization objective for evasion.

    \item Empirical Demonstration of Detector Failure: We provide extensive evidence that semantic-preserving refinement via commercial chatbot services causes widespread failure in state-of-the-art detectors while maintaining subject's identity and improving visual quality.

    \item Analysis of Safety Alignment Gaps: We expose the inconsistency in current safety guardrails, where explicit adversarial prompts are blocked, but reasoning-guided refinement---which achieves the same malicious outcome---is permitted as benign editing.

    \item Comparative Risk Assessment: We establish that commercial GAI services present a higher realistic threat than open-source alternatives due to the accessibility and effectiveness of their reasoning-refinement loop for non-expert users.

\end{itemize}

\section{Background}
\subsection{Image Authentication}

\noindent \textbf{Deepfake Detection}
In recent years, deepfake detection research has spanned multiple domains, including spatial analysis, frequency-domain cues, and both frame-level and video-level temporal inconsistencies~\cite{zhao2021multi,yan2023ucf, liang2022exploring, liu2021spatial, zheng2021exploring, pu2022learning}. While early detectors performed well under intra-domain settings~\cite{rossler2019faceforensics++}, their generalization to unseen manipulation methods or novel generators~\cite{karras2021alias, rombach2022high, yan2024df40} remained limited, reflecting an implicit static classification setting in detector design. To address this, various approaches have been proposed to improve cross-domain generalization, including the use of pseudo-fake samples~\cite{shiohara2022detecting, yan2025generalizing, nguyen2024laa, ma2025from}, a frequency-aware network~\cite{tan2024frequency}, video-level temporal consistency modelling\cite{liang2024speechforensics, xu2023tall}, and the design of a universal vision foundation model for downstream face security tasks\cite{wang2025fsfm}.

\quad

\noindent \textbf{AI-Generated Image Detection}
Similarly, early AI-generated image detectors\cite{frank2020leveraging, wang2020cnn, chai2020makes} were typically formulated as binary classifiers or relied on data augmentation strategies for detection; however, their ability to generalize to unknown generative models was often limited. 
Recent works~\cite{zhang2025detecting, wang2023dire, ricker2024aeroblade} have explored generalization with and without additional training across diverse generative techniques, beyond the scope of face manipulation. 

\noindent \textbf{VLM-based Detection}
In another direction, recent works~\cite{ojha2023towards, yang2025d, chen2025forgelens, liu2024forgery} has leveraged CLIP~\cite{radford2021learning}---a Vision-Language Model (VLM) trained for zero-shot classification---to build universal detectors capable of identifying synthetic content across a range of manipulation types and generation methods. For instance, \cite{ojha2023towards} shows that simple classifiers, such as $k$-NN or linear probes, built on top of CLIP embeddings can achieve competitive performance against unseen generative models. More recent work by \cite{yang2025d} further scales this approach by combining CLIP representations with patch shuffling.

Going further, several studies~\cite{yu2025unlocking, guo2025rethinking, jia2024can} explore the use of Large Language Models (LLMs) in conjunction with VLMs to reason about visual inconsistencies and explain why an image might be fake. This reasoning-based detection paradigm not only improves interpretability
but also exposes structured, human-interpretable criteria for authenticity assessment.

\subsection{Generative AI}
The recent advancement of generative VLMs~\cite{ dai2023instructblip, liu2023visual, brooks2023instructpix2pix} has enabled users to interactively modify images using natural language prompts~\cite{wu2023visual
}. These systems, which include commercial tools like GPT-4~\cite{achiam2023gpt}, DALL-E 3~\cite{betker2023improving}, and Gemini~\cite{comanici2025gemini}, as well as open-source Qwen-VL~\cite{bai2025qwen2}, support a range of image editing capabilities such as inpainting and attribute transfer.

To mitigate misuse, these systems typically incorporate safety guardrails that restrict the generation of harmful or inappropriate content—such as nudity, violent imagery, or hate symbols~\cite{singh2025openai, helff2024llavaguard, li2025t2isafety}. 
Despite the growing integration of safety guardrails in image generation systems, it remains unclear whether these mechanisms extend beyond
content-level moderation to effectively address interaction-level or iterative misuse
scenarios, such as the progressive refinement of deepfakes through natural language prompts.

\section{On the Interaction between Generative AI and Deepfake Detection} \label{sec:3} 
GAI systems are increasingly deployed as general-purpose assistants that expose multiple image-related functionalities through unified, user-facing interfaces.
From an end-user perspective, these systems support a wide range of tasks, including image analysis, natural-language explanation, and image generation or editing.
In this section, we examine how such functionalities interact when GAI systems are applied to facial deepfake detection tasks.

Rather than treating GAI systems as detectors or adversarial tools in isolation, we adopt a user-centric perspective and focus on the observable behaviors that arise when users assign authenticity-related tasks using commonly available GAI interfaces.
Our analysis characterizes interaction patterns that emerge when image understanding, reasoning, and generation capabilities are jointly accessible within a single system. 

\subsection{Authenticity Assessment with GAIs}
\label{sec:auth}

We examine how GAI systems behave when users interact with them for facial deepfake detection tasks.
Our analysis focuses on two stages of interaction: (i) how assessment criteria are articulated, and (ii) how such criteria are subsequently used to produce authenticity judgments and explanations.
The judgments analyzed in this section reflect the systems’ own reasoning under zero-shot task settings, rather than claims of correctness with respect to ground-truth labels.

\subsubsection{Articulation of Assessment Criteria.}

In the first stage, users prompt GAI systems to describe how they would assess the authenticity of facial images in general, without supplying any specific image as input.
The interaction is explicitly framed as a request for expert-level assessment principles rather than an instance-level authenticity judgment.

In response, GAI systems articulate assessment criteria in natural language.
These criteria are expressed as generalizable visual cues commonly associated with facial manipulation, rather than conclusions tied to a particular image, quantitative confidence outputs, or system-internal decision signals.
Across repeated interactions, the articulated criteria exhibit a notable degree of internal structure and consistency across interactions.
Rather than varying arbitrarily across prompts, systems repeatedly organize their responses around a small set of recurring artifact-level visual categories, including skin and texture irregularities, geometric consistency of facial components, lighting coherence, and contextual alignment.

Importantly, this articulation behavior is observed even in the absence of any concrete image input.
The criteria are presented as general assessment heuristics, rather than as post hoc explanations conditioned on instance-specific evidence.
Representative examples of such criterion articulation, drawn from multiple GAI systems under identical zero-shot prompts, are provided in Appendix (Figure~\ref{fig:criteria_articulation}).

\begin{finding}
  \textbf{Finding 1:} GAI systems articulate explicit and structured criteria for facial authenticity assessment, organized around common artifact-level visual cues.
\end{finding}
\label{f1}

\subsubsection{Criteria-Guided Judgments and Explanations.}

In the second stage, users request an explicit authenticity decision for a facial image together with an accompanying explanation.
These requests are formulated by explicitly referencing the assessment criteria articulated in the previous stage.
While this involves manual prompt formulation to specify how the judgment should be framed, the requests do not introduce additional evidence, ground-truth labels, or external decision signals beyond the input image itself.

Under this criteria-guided framing, GAI systems produce binary authenticity judgments  (e.g., real vs. manipulated).
The resulting decisions are aligned with the previously articulated assessment criteria, and the accompanying explanations consistently reference the same categories of artifact-level visual cues.
This alignment suggests that the articulated criteria actively shape how the judgment is structured and justified, rather than being introduced retrospectively to rationalize an already-formed decision.
Across our observations, we did not identify cases in which the stated authenticity judgment conflicted with the visual cues cited in its explanation. The subsequent application of these criteria to image-level analysis is illustrated in Appendix (Figure~\ref{fig:criteria_application}).

\begin{finding}
  \textbf{Finding 2:} When prompted to make instance-level (i.e., image-level) authenticity decisions under explicitly articulated assessment criteria, GAI systems produce binary judgments that are internally coherent with their accompanying explanations.
\end{finding}

\subsection{Structured and Actionable Reasoning} 
\label{sec:reasoning}

We examine the explanations produced by GAI systems when responding to criteria-guided facial authenticity assessment requests.
While users ask for an authenticity judgment and an accompanying explanation, they do not specify the desired level of detail, spatial granularity, or explanatory structure.

Despite this underspecification, the resulting explanations frequently exhibit two coupled properties.
First, they are articulated at the level of concrete visual artifacts rather than abstract impressions.
Second, this artifact-level structure enables the explanations to be reinterpreted as actionable guidance for downstream image refinement.
In this subsection, we analyze how these two properties reinforce each other.

\subsubsection{Artifact-Level Reasoning and Structured Explanations.}
In many interactions, GAI systems provide explanations that go beyond high-level or holistic assessments of facial realism.
Rather than describing the image in terms of overall plausibility or naturalness, the explanations often reference concrete and localized visual properties.

These references may include skin texture smoothness, edge consistency around facial features, lighting coherence, or alignment between the face and its surrounding context.
Importantly, such explanations are not merely lists of visual cues.
They are organized around identifiable artifact categories and associated with specific regions of the image, yielding a structured and referential form of reasoning.

This artifact-level structure emerges even though the user does not request a particular explanatory resolution or spatial focus.
As a result, the explanations make explicit not only what appears inconsistent, but also where such inconsistencies manifest within the image.

\subsubsection{Reinterpretation of Structured Reasoning as Refinement Guidance.}
The structured, artifact-level explanations described above enable reinterpretation of assessment reasoning as guidance for image refinement.
By identifying which visual properties appear problematic and where they are localized, the explanations delineate concrete targets for modification.
In this sense, the explanations are actionable, as they specify both target visual attributes and their approximate spatial localization.

Crucially, this reinterpretation does not depend on requesting the system to directly modify, improve, or regenerate the image.
Instead, it is the structured nature of the explanation itself that makes reinterpretation possible.
If the reasoning were expressed only at a high or abstract level, such transformation would not be feasible.
Here, the artifact-level organization allows users to manually reformulate the explanation into refinement prompts in subsequent interactions, without introducing new objectives or altering the original task framing.

\begin{finding}
  \textbf{Finding 3:} Under criteria-guided assessment, GAI systems often produce artifact-level explanations whose structured form can be reinterpreted as actionable guidance for image refinement.
\end{finding}

\subsection{Non-Adversarial Nature of Reasoning-Guided Refinement}
\label{sec:non-adversarial}

We analyze how refinement requests derived from authenticity assessment reasoning are framed and interpreted in practice.
This subsection focuses on interaction-level properties of such requests, rather than on the visual outcomes of refinement. Therefore, the term ``non-adversarial'' in this section refers to the framing of the interaction, not to the downstream security implications of the resulting refinements.

\subsubsection{Benign Framing of Refinement Requests.}
A defining characteristic of the refinement requests considered in this study is their continuity with the preceding authenticity assessment.
Rather than introducing a new objective, refinement requests are formulated as direct extensions of the assessment, targeting the same visual artifacts identified during explanation.
As a result, the interaction transitions from analysis to improvement without explicit shifts in user intent.

This continuity enables refinement requests to be expressed using benign and commonly accepted language.
Instead of referencing deepfake detection, evasion, or manipulation, such requests are framed in terms of improving visual naturalness, consistency, or overall image quality.
From the perspective of the GAI system, these requests align with standard image editing or enhancement tasks and remain compatible with expected usage patterns of user-facing generative interfaces.

\subsubsection{Built-in Safety Guardrails in GAI Systems.}
We describe observed interactions between reasoning-guided refinement requests and existing safety guardrails in contemporary GAI systems.
These observations constitute an empirical aspect of our study, capturing how such requests are handled in practice.
At the same time, we emphasize that our analysis of the underlying mechanisms remains necessarily limited by the lack of visibility into system internals.

At the input level, many GAI systems reject prompts that explicitly articulate adversarial or socially harmful intent, such as requests to bypass deepfake detection mechanisms.
These refusals typically occur prior to inference, preventing the system from producing intermediate reasoning or outputs.
This behavior reflects prompt-level safety guardrails that identify clearly adversarial intent from the request itself.

In contrast, refinement requests derived from authenticity assessment reasoning do not exhibit such explicit adversarial markers and therefore usually proceed through the inference stage.
We further observe cases in which refinement requests successfully initiate the image generation process but are intermittently rejected at the output stage.

While the precise causes of such output-stage refusals cannot be determined from external observation alone, we note that repeating the same refinement request—without modifying the prompt—can sometimes yield a successful output following an initial refusal.
This pattern suggests that stochasticity in the image generation process (e.g., diffusion based sampling), output-level filtering, or an interaction between the two may play a role.
We stress that these interpretations are offered as plausible explanations rather than definitive accounts of system behavior. However, such cases were rare ($<5\%$) in our observations and did not occur with sufficient frequency to support a meaningful quantitative analysis, and are therefore treated qualitatively in this work.

While these interactions remain non-adversarial in framing, 
it is unclear how requests for criteria articulation, image analysis, and refinement are handled by automated safety filtering systems.
We examine this aspect separately in Appendix~\ref{app:safety}.
\begin{finding}
  \textbf{Finding 4:} Reasoning-derived refinement requests can be framed as benign image improvement tasks that extend authenticity assessment reasoning, without explicitly indicating evasion or manipulation intent.
\end{finding}



\subsection{Semantic Properties and Controllability}
\label{sec:semantic}

We examine the semantic characteristics of images produced through reasoning-guided refinement, with a focus on how refinement requests shape what is preserved and what is modified.
In particular, we analyze how artifact-focused refinement influences high-level image semantics, and how additional user intent can further modulate these outcomes.

\subsubsection{Localized Modification and Preservation of High-Level Semantics.}
A consistent property of reasoning-guided refinement is its localized nature.
Rather than inducing global or indiscriminate changes, refinement operations concentrate on regions associated with the visual artifacts identified during authenticity assessment, such as unnatural texture, boundary inconsistencies, or localized lighting issues.

Because the refinement requests are articulated at an artifact-specific level, modifications are applied selectively rather than semantically.
As a result, high-level semantic attributes of the face---such as facial structure, pose, expression, and overall identity---are usually preserved in our observations.
This preservation is not incidental, but follows from the fact that the refinement targets low-level visual inconsistencies rather than high-level semantic concepts.
Accordingly, refinement functions as a targeted visual adjustment process rather than a semantic transformation.

\begin{finding}
  \textbf{Finding 5:} Reasoning-guided image refinement selectively modifies artifact-related regions while largely preserving high-level facial semantics.
\end{finding}

\subsubsection{Emergent Improvements in Visual Naturalness.}
In addition to reducing explicitly identified artifacts, refinement often yields secondary effects that enhance overall visual naturalness.
Refined images may exhibit improved clarity, reduced noise, or more coherent lighting, even when such properties are not directly specified in the refinement request.

These effects are best understood as byproducts of artifact removal and localized enhancement, rather than as primary objectives of the interaction.
While the extent of such improvements depends on the underlying capabilities of the GAI system, their consistent emergence highlights how artifact-focused refinement can make visual realism more salient without explicitly optimizing for it.



\begin{figure*}[t]
    \centering
    \includegraphics[width=\linewidth]{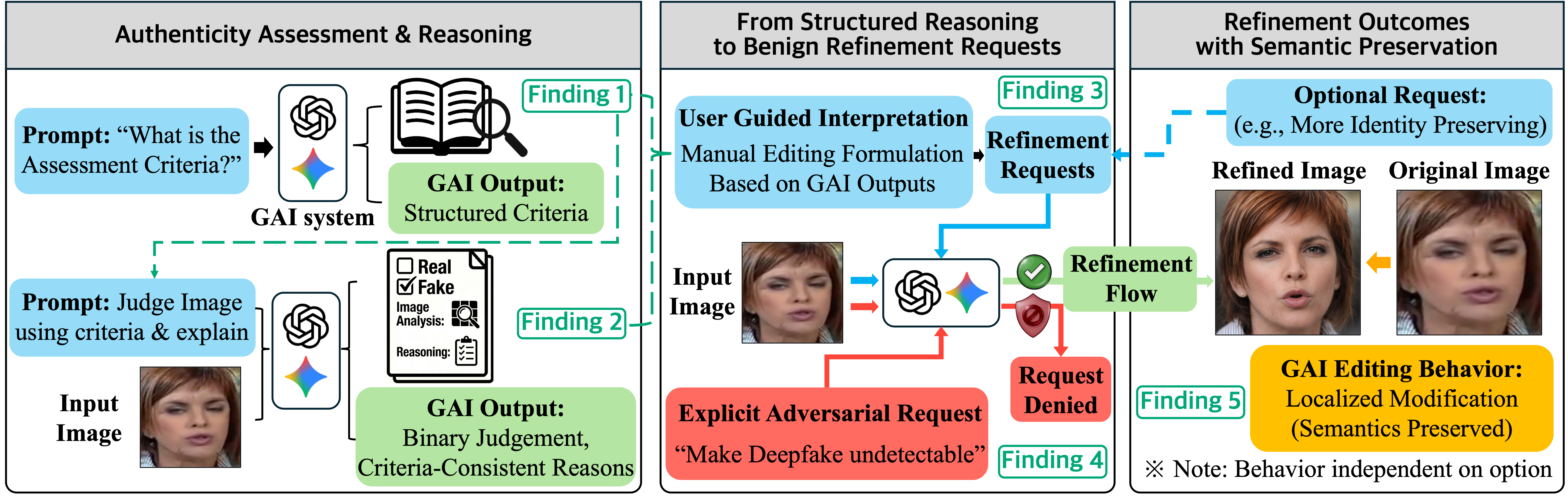}
    \vspace{-5mm}
    \caption{Summary of observed GAI behaviors across authenticity assessment and refinement stages, illustrating how structured reasoning transitions into benign, semantics-preserving image edits (Finding 1–5).}
    \label{fig:overview}
\end{figure*}

\section{From Findings to Refinement Workflows}
\label{sec:interaction}

The analysis in Section~\ref{sec:3} identified several consistent properties of how GAI systems assess facial authenticity and expose their reasoning to users.
In this section, we build directly on these findings and describe how assessment-oriented interactions can be composed into practical image refinement workflows.
Rather than introducing a new algorithm or optimization procedure, our goal is to clarify the interaction pipeline that connects assessment, explanation, and refinement.
This pipeline reflects realistic user behavior and serves as the conceptual foundation for the prompt constructions and experimental protocols evaluated in Section~\ref{sec:prompt-construction}. Figure~\ref{fig:overview} provides an overview of the interaction flow described in this section.

\subsection{Interaction-Level Properties}

We first summarize the interaction-level properties that follow directly from the findings in Section~\ref{sec:3} and are essential for understanding how refinement workflows arise.

As shown in \textbf{Finding~1}, GAI systems articulate explicit and structured criteria for facial authenticity assessment.
These criteria are expressed in natural language and are organized around recurring artifact-level visual cues, such as skin texture, hair boundaries, eye details, and background consistency.

Under such criteria-guided assessment, \textbf{Finding~2} shows that instance-level authenticity judgments are internally coherent with their accompanying explanations.
That is, when a system produces an authenticity decision, the associated rationale consistently aligns with the stated assessment principles.

Crucially, \textbf{Finding~3} establishes that these explanations are not merely descriptive.
Their structured, artifact-level form makes them readily interpretable as actionable guidance for image refinement.
The explanations identify what appears problematic, where it is localized, and which visual properties deviate from expectations of authenticity.

Building on this, \textbf{Finding~4} shows that refinement requests derived from such reasoning can be framed as benign image improvement tasks.
These requests extend the system’s own assessment logic without explicitly signaling evasion, manipulation, or malicious intent.

Finally, \textbf{Finding~5} indicates that reasoning-guided refinement tends to operate locally, selectively modifying artifact-related regions while largely preserving high-level facial semantics such as identity, pose, and expression.

Across all these stages, the system performs assessment, reasoning, and image generation automatically.
The only non-automated component is the interpretation and reuse of assessment outputs: users decide how assessment-related text is recontextualized into refinement requests.
This human-mediated step is a central aspect of the interaction setting considered in this work.

\subsection{Prompt Roles in the Refinement Pipeline}

Although all interactions with a GAI system are mediated through natural-language prompts, our analysis distinguishes three prompt roles that occupy different positions in the assessment-to-refinement pipeline.

\textbf{Criteria prompts} request image-independent descriptions of how facial authenticity is evaluated.
They elicit general assessment principles without reference to a specific image, reflecting the behavior described in \textbf{Finding~1}.

\textbf{Reasoning prompts} request an authenticity judgment for a given image under explicitly articulated assessment criteria obtained by criteria prompts.
In response, the system produces a decision together with an artifact-level explanation that instantiates those criteria at the image level, as characterized in \textbf{Finding~2} and \textbf{Finding~3}.

\textbf{Editing prompts} request image refinement.
They specify how an input image should be modified, typically framed as benign photographic enhancement according to \textbf{Finding~4}.

These prompt roles are not associated with different system modes or interfaces.
While the system executes refinement based on the editing prompt alone, the construction of that prompt typically depends on information surfaced through prior criteria articulation and reasoning interactions.

\subsection{Refinement Regimes}
\label{subsec:refinement_regimes}
Using the prompt roles defined above, we distinguish two refinement regimes based on how assessment-related information is reused and how much per-image adaptivity is introduced.
The regimes differ in user effort and information reuse, but do not require different system capabilities. 
Methodologically, this distinction is critical for isolating the impact of dynamic reasoning from static prompt engineering.

\subsubsection{Instance-Agnostic Refinement}
In the instance-agnostic regime, refinement is performed using a single fixed editing prompt that is applied uniformly to all images at evaluation time.
The key property of this regime is \emph{non-adaptivity across instances during refinement}: the editing instruction does not change from image to image.

Importantly, the fixed prompt itself is not assumed to emerge spontaneously.
Instead, it is derived through a one-time preprocessing step that aggregates assessment behavior observed in Section~3.
Specifically, we first elicit explicit assessment criteria (\textbf{Finding~1}) and then request criteria-guided, image-level rationales across a sample of images (\textbf{Finding~2--3}).
We aggregate these rationales to identify recurring artifact vocabulary and organize it into a coarse taxonomy (Section~\ref{subsec:artifact-taxonomy}).
The resulting taxonomy is then used to instantiate a single, human-written fixed refinement prompt (Section~\ref{subsec:prompt-iap}), which is subsequently reused for all images without per-instance modification.

This separation between a one-time preparatory analysis (aggregation of assessment rationales and prompt design) and the subsequent evaluation phase
(uniform application of a fixed prompt) makes the instance-agnostic regime a non-adaptive reference setting for our experiments.

\subsubsection{Instance-Specific Refinement}
In the instance-specific regime, refinement prompts are constructed by reusing image-specific assessment explanations.
For each input image, a reasoning prompt is first issued, producing an artifact-level rationale as described in \textbf{Finding~3}.
This rationale is then directly incorporated into the editing prompt, specifying which issues should be corrected.

Compared to the instance-agnostic regime, instance-specific refinement requires less manual effort per image.
Once a base editing instruction is fixed, the refinement prompt can be formed by combining the base instruction with the system-generated rationale.
No additional filtering, optimization, or reinterpretation is assumed beyond this reuse.

Importantly, instance-specific refinement does not introduce new system capabilities.
It is a natural extension of the same interaction pattern, enabled by the accessibility and structure of reasoning outputs already exposed by the system.
This regime captures the additional effect of per-image adaptivity arising from reasoning reuse.
Consequently, comparing these two regimes allows us to isolate the specific security risk posed by the feedback loop itself.

In the following section, we instantiate both refinement regimes using concrete prompt templates and evaluate their effects under a controlled experimental setup.

\section{Experiments}
\begin{table*}[t]
\centering
\caption{Detailed comparison of image refinement services and model specifications (As of Feb 2026).}
\small
\setlength{\tabcolsep}{7.5pt}
\label{tab:model}
\begin{tabular}{l ccc c l c c}
\toprule
\multirow{2.5}{*}{\textbf{Service / Family}} & \multicolumn{3}{c}{\textbf{Access Channel}} & \multirow{2.5}{*}{\textbf{License}} & \multirow{2.5}{*}{\textbf{Specific API Model ID}} & \textbf{Release} & \textbf{Cost} \\
\cmidrule(lr){2-4}
 & \scriptsize{WEB} & \scriptsize{APP} & \scriptsize{API} & & & \scriptsize{(YYYY.MM)} & \scriptsize{(\$/img)} \\
\midrule
\multirow{2}{*}{\mylogo{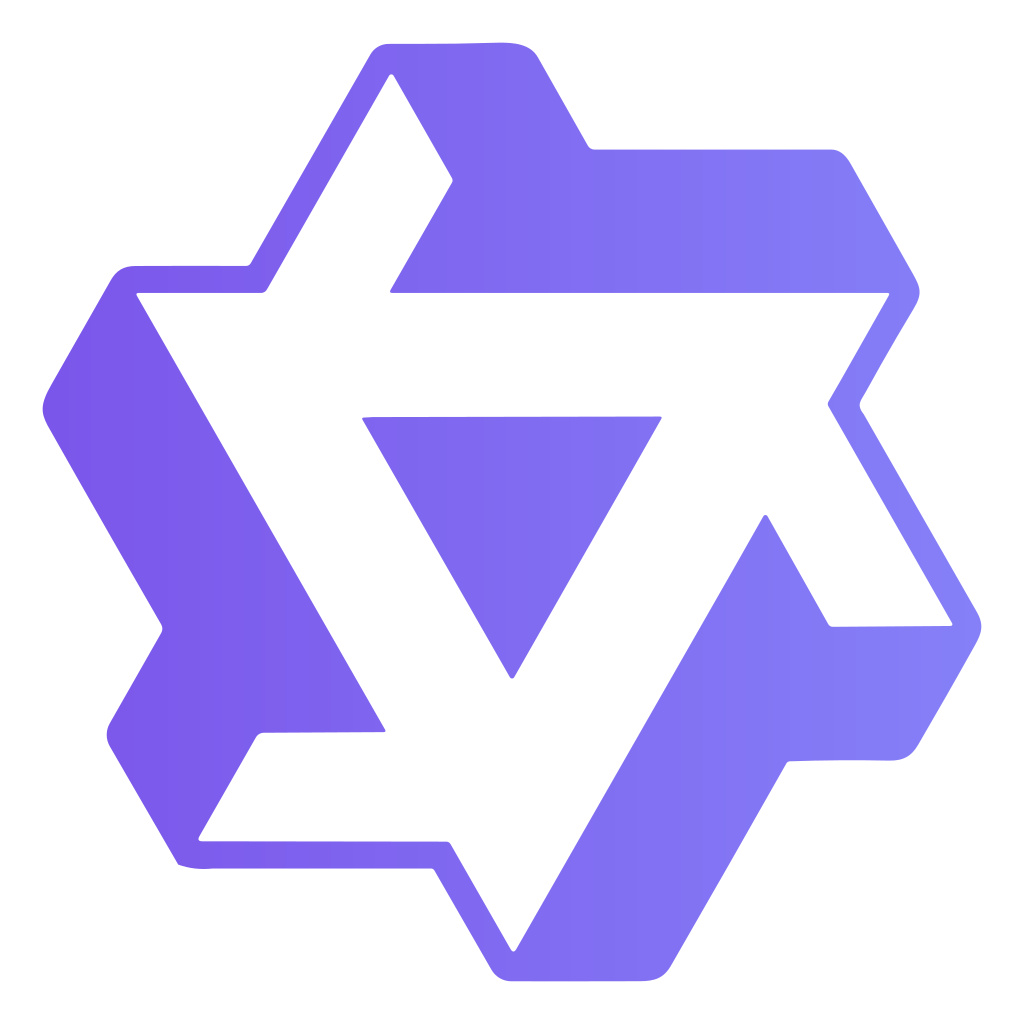} Qwen (Alibaba)} & \checkmark & -- & \checkmark & Open Weights & \modelname{qwen-image-edit} & 2024.10 & \$0 \\
 & \checkmark & -- & \checkmark & Open Weights & \modelname{qwen-image-edit-2511} & 2025.11 & \$0 \\
\midrule
\mylogo{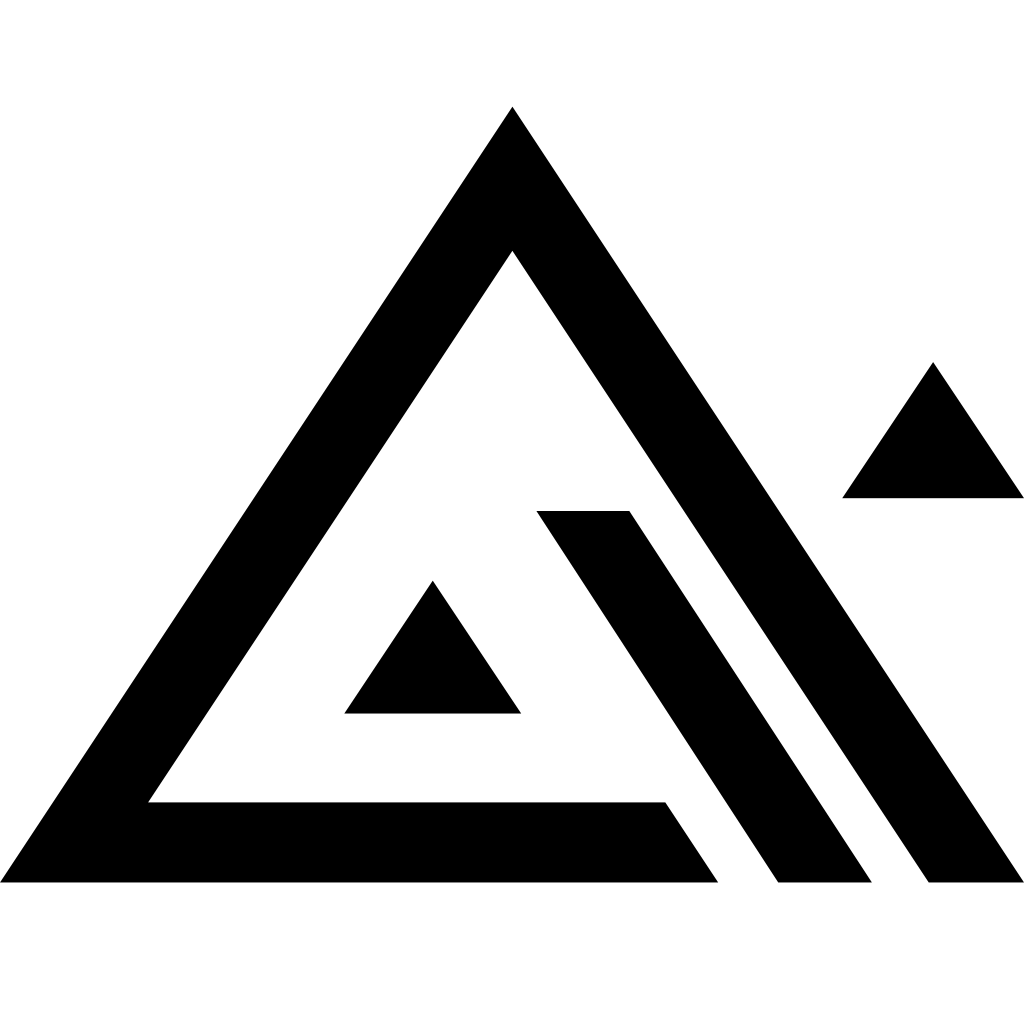} Flux AI (Black Forest Labs) & \checkmark & -- & \checkmark & Hybrid$^{\dagger}$ & \modelname{flux-2-max} & 2025.12 & \$0.070 \\
\mylogo{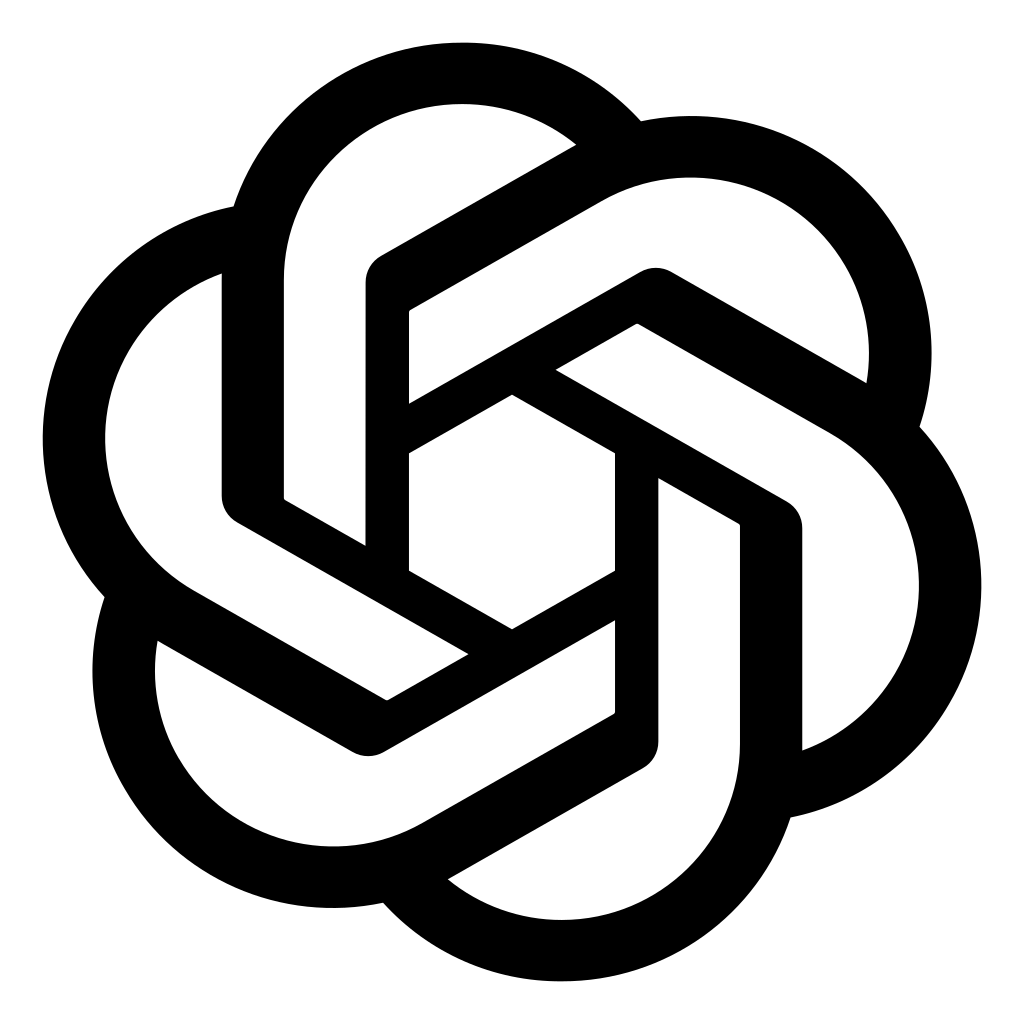} ChatGPT (OpenAI) & \checkmark & \checkmark & \checkmark & Commercial & \modelname{gpt-image-1.5} & 2025.10 & \$0.133 \\
\mylogo{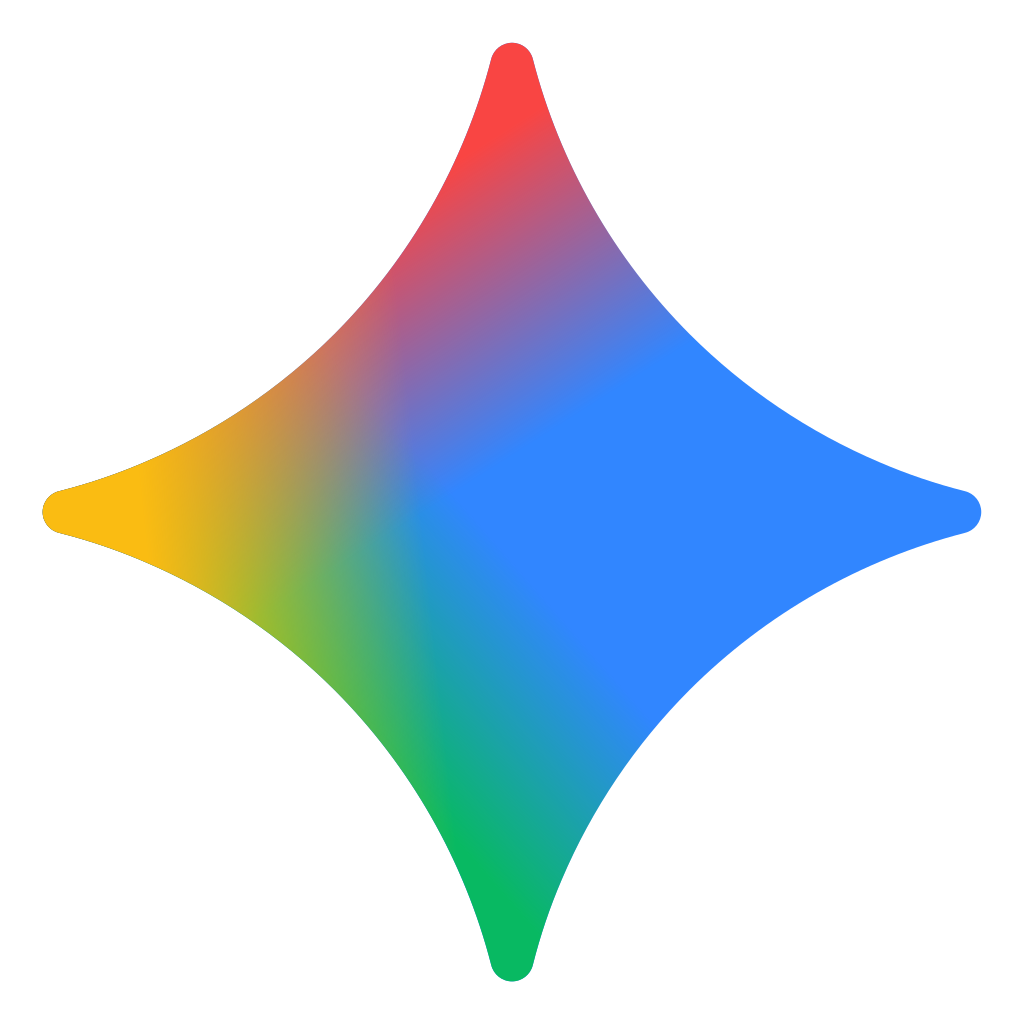} Gemini (Google) & \checkmark & \checkmark & \checkmark & Commercial & \modelname{gemini-3-pro-image-preview} & 2025.11 & \$0.200 \\
\bottomrule
\multicolumn{8}{l}{\footnotesize $^{\dagger}$High-end models (Pro/Max) are proprietary API-only, while distilled versions (dev/dev-Turbo/VAE/klein) are released as open weights.\vspace{-5mm}}
\end{tabular}
\end{table*}

\begin{table*}[t]
\centering
\caption{Detailed comparison of evaluation targets. Rows are color-coded by detection scope: \colorbox{blue!5}{Deepfake} and \colorbox{red!5}{AI-generated}.}
\small
\setlength{\tabcolsep}{4pt}
\begin{tabular}{lllc}
\toprule
\textbf{Target} & \textbf{Feature Extractor} & \textbf{Training Dataset} & \textbf{Access} \\
\hline
\rowcolor{blue!5}
GenD$^\dagger$~\cite{yermakov2025deepfake} & CLIP:ViT-L~\cite{radford2021learning} / Perceptual~\cite{bolya2025perception} / DINOv3~\cite{simeoni2025dinov3} & FaceForensics++~\cite{rossler2019faceforensics++} & Open \gitlink{https://github.com/yermandy/GenD} \\
\rowcolor{blue!5}
M2F2-Det$^{\ddagger}$~\cite{guo2025rethinking} & EfficientNet-B4~\cite{tan2019efficientnet} + CLIP:ViT-L~\cite{radford2021learning} & FaceForensics++~\cite{rossler2019faceforensics++} + DD-VQA$^{\star}$~\cite{zhang2024common} & Open \gitlink{https://github.com/CHELSEA234/M2F2_Det?tab=readme-ov-file} \\
\rowcolor{blue!5}
Hive-DF~\cite{Hivedeepfake} & EfficientNet-B4~\cite{tan2019efficientnet} + YoloV8 & Proprietary (Internal) & API \apilink{https://build.nvidia.com/hive/deepfake-image-detection} \\
\hline
\rowcolor{red!5}
UnivFD~\cite{ojha2023towards} & CLIP:ViT-L~\cite{radford2021learning} & ProGAN (LSUN)~\cite{karras2017progressive,yu2015lsun} & Open \gitlink{https://github.com/WisconsinAIVision/UniversalFakeDetect} \\
\rowcolor{red!5}
D$^3$~\cite{yang2025d} & CLIP:ViT-L~\cite{radford2021learning} & ProGAN (LSUN) + GenImage~\cite{zhu2023genimage} & Open \gitlink{https://github.com/BigAandSmallq/D3} \\
\rowcolor{red!5}
Hive-AI~\cite{Hiveai} & EfficientNet-B4~\cite{tan2019efficientnet} & Proprietary (Internal) & API \apilink{https://build.nvidia.com/hive/ai-generated-image-detection} \\
\bottomrule
\multicolumn{4}{l}{\footnotesize $^{\dagger}$GenD is evaluated in three variants—GenD (C), (P), and (D)—based on the first letter of its feature extractors: CLIP, Perceptual, and DINOv3, respectively.} \\[-0.4em]
\multicolumn{4}{l}{\footnotesize $^{\ddagger}$Unlike other models, M2F2-Det utilizes an large multimodal model (i.e., Vicuna-7b~\cite{chiang2023vicuna}) to reason and generate additional text outputs explaining the forgery.} \\[-0.4em]
\multicolumn{4}{l}{\footnotesize $^{\star}$The DD-VQA~\cite{zhang2024common} dataset contains high-quality image-text pairs annotated by Amazon Mechanical Turk.}
\end{tabular}
\label{tab:target}
\end{table*}

In this section, we present a comprehensive empirical evaluation to assess how state-of-the-art deepfake and AI-generated image detectors respond to \emph{semantic-preserving image refinement}. We first detail our experimental setup, including the refinement models, target detectors, datasets, and evaluation metrics. We then report quantitative results on authenticity judgment changes under refinement, complemented with semantic-preservation analyses that verify identity and content consistency. In addition, we provide qualitative visual examples illustrating the preservation of visual quality across different refinement models. Finally, we present ablation studies investigating the effects of prompt variations and extend our analysis beyond deepfake images to the refinement of AI-generated images.

\subsection{Experimental Setup}

\subsubsection{Image Refinement Models}

We evaluate a diverse set of image refinement models and services, including both open-weight systems and commercial APIs, as summarized in Table~\ref{tab:model}. These models span different access channels (web, app, and API), licensing, and pricing tiers, reflecting realistic deployment and usage scenarios.

Specifically, we select four widely used generative AI systems for our study: \textbf{Qwen}~\cite{bai2025qwen2}, \textbf{ChatGPT}~\cite{achiam2023gpt}, \textbf{Gemini}~\cite{comanici2025gemini}, and \textbf{Flux AI}~\cite{2026flux}.
For the open-weight setting, we utilize two variants from the Qwen family, leveraging their specialized image editing capabilities~\cite{wu2025qwen} which allow full prompt control and serve as a transparent baseline for controlled experimentation. 
For commercial services, we evaluate image refinement capabilities provided by high-end proprietary vision generation models, including \textbf{GPT image generator}~\cite{betker2023improving} (via ChatGPT), \textbf{Imagen 3}~\cite{baldridge2024imagen} (via Gemini), and the \textbf{Flux AI image generator}~\cite{labs2025flux} (via Flux AI). All commercial models are accessed through their official API endpoints (\texttt{gpt-image-1.5}, \texttt{gemini-3-pro-image-preview}, and \texttt{flux-2-max}). While these services are also accessible via interactive web interfaces, such platforms often retain conversational context across successive inputs unless explicitly reset. To ensure independent and reproducible refinement for each image, we therefore use the stateless API interfaces in all experiments.



\subsubsection{Evaluation Target Detectors}

We evaluate refinement effects on six detection targets~\cite{yermakov2025deepfake,guo2025rethinking,Hivedeepfake,ojha2023towards,yang2025d,Hiveai} covering both deepfake and AI-generated image detection, as detailed in Table~\ref{tab:target}. These targets include open-source detectors as well as commercial APIs, enabling a comprehensive assessment across diverse detection scopes and model designs.

For deepfake detection, we consider GenD~\cite{yermakov2025deepfake}, M2F2-Det~\cite{guo2025rethinking} (hereinafter M2F2), and the commercial Hive-DF API~\cite{Hivedeepfake}. These models differ substantially in their feature extractors and decision mechanisms, ranging from CNN-based architectures to multimodal systems that incorporate large language models for explanatory reasoning. For AI-generated image detection, we evaluate UnivFD~\cite{ojha2023towards}, D$^3$~\cite{yang2025d}, and the corresponding Hive-AI API~\cite{Hiveai}, which collectively cover supervised, zero-shot, and proprietary detection paradigms.

By spanning multiple backbones, training datasets, and output formats, this target set ensures that our evaluation does not depend on a specific detector architecture or training regime. For open-source detectors, we use the released pre-trained models from the corresponding github repository without additional fine-tuning. For API-based detectors, all evaluations are performed using the default service configurations. 

\subsubsection{Datasets}

We use the FaceForensics++ (FF++) dataset~\cite{rossler2019faceforensics++} as the source of starting deepfake images. The original FF++ dataset contains facial manipulations generated using four methods: Face2Face~\cite{thies2016face2face}, FaceSwap~\cite{faceswap2018github}, DeepFakes~\cite{deepfakes2018github}, and NeuralTextures~\cite{thies2019deferred}. In addition, following the official dataset updates released by the authors, we include the FaceShifter manipulation method~\cite{li2019faceshifter}.

For our experiments, we sample a total of 100 images at the image level, selecting 20 samples from each manipulation category (Face2Face, FaceSwap, DeepFakes, NeuralTextures, and FaceShifter). This balanced sampling strategy mitigates bias toward any specific manipulation technique and enables a more robust evaluation of refinement effects.

For the real-image reference set, we use the Flickr-Faces-HQ (FFHQ) dataset~\cite{karras2019style} instead of real frames from FF++. We randomly sample 1,000 images to represent the real class. This choice is motivated by two considerations. First, FFHQ provides higher image quality and resolution than video frames, ensuring that real samples are comparable in visual fidelity to refined images. Second, since our target detectors span diverse training regimes---including models trained on ProGAN-generated data or proprietary sources---using an independent, high-quality real dataset offers a more rigorous and unbiased basis for calibrating detection thresholds.

\subsubsection{Metrics}\label{sec:metrics}

To quantify the impact of semantic-preserving image refinement on detection outcomes, we report the \textbf{Detection Rate (DR)}, defined as the proportion of images classified as Deepfake (or AI-generated) by a given detector. Lower DRs after refinement indicate a shift in authenticity judgments induced by visual refinement.


To enable fair comparison across detectors with different output scales, we calibrate detector-specific thresholds using an independent real-image reference set.
Specifically, for each detector we define two operating points based on 1,000 real images from FFHQ: a strict threshold $\tau_{99}$, at which 99\% of real images are classified as real (1\% false DR), and a relaxed threshold $\tau_{90}$, at which 90\% are classified as real (10\% false DR).
DRs are reported under both $\tau_{99}$ and $\tau_{90}$.

In addition to detection outcomes, we assess whether semantic refinement preserves identity information using an \textbf{Identity-Preserving Rate (IPR)}. For each image, we compare the original deepfake image and its refined counterpart using a commercial face verification API. IPR is defined as the proportion of image pairs that are verified as belonging to the same individual. High IPRs indicate that refinement alters visual appearance without changing the underlying semantics.

\subsection{Prompt Construction for Refinement}
\label{sec:prompt-construction}

Our experiments evaluate how \emph{semantic image refinement} changes downstream authenticity judgments while preserving identity semantics.
To ensure transparency and reproducibility, we explicitly document how refinement prompts are constructed.
We follow a four-step process:
(i) explicitly elicit assessment criteria used by GAI systems, i.e., ChatGPT and Gemini, as illustrated in Figure~\ref{fig:criteria_articulation};
(ii) collect criteria-guided, instance-level reasoning outputs across a set of images and summarize the recurring artifact vocabulary that appears in such assessments, yielding the artifact taxonomy described in Section~\ref{subsec:artifact-taxonomy} and Table~\ref{tab:keyword_stats};
(iii) recontextualize the resulting artifact taxonomy into an instance-agnostic fixed refinement prompt, as detailed in Section~\ref{subsec:prompt-iap}; and
(iv) define an instance-specific prompt template that directly reuses per-image reasoning outputs without additional post-processing, as described in Section~\ref{subsec:prompt-isp}.
(More detailed in Appendix~\ref{app:prompts})



\subsubsection{Artifact Taxonomy from Assessment Rationales}
\label{subsec:artifact-taxonomy}

We first analyze the textual rationales produced by GAI systems when asked to assess deepfake authenticity with pre-defined criteria. Across 100 assessment records, we aggregate the most frequent artifact-related keywords and organize them by the corresponding assessment criteria (Table~\ref{tab:keyword_stats}), including \emph{skin texture}, \emph{eyes and gaze}, \emph{hair and boundaries}, \emph{facial geometry}, and \emph{scene/background}. This taxonomy serves two purposes: it (1) provides evidence that assessment rationales consistently surface a structured set of forensic cues, and (2) yields a grounded vocabulary that can be used to instantiate refinement instructions in a consistent, readable manner.

\begin{table}[t]
\centering
\caption{Taxonomy of visual artifacts frequently identified by GAI systems during deepfake assessment (Top-K keywords aggregated from 100 records).} 
\label{tab:keyword_stats}
\footnotesize
\setlength{\tabcolsep}{4pt} 
\begin{tabular}{l p{0.6\linewidth}}
\toprule
\textbf{Artifact Category} & \textbf{Top Frequent Descriptors (Count)} \\
\midrule
\textbf{Skin Texture} & smooth (158), plastic (102), texture (116), airbrushed (56), pores (51), sheen (40), waxy (17) \\
\midrule
\textbf{Eyes \& Gaze} & catchlights (90), iris (54), sclera (37), reflections (32), glassy (22), mismatch (9) \\
\midrule
\textbf{Hair \& Boundaries} & strands (88), hairline (69), halo (44), clumps (30), flyaways (22), merge (17) \\
\midrule
\textbf{Facial Geometry} & edges (111), nostrils (73), contours (69), lip (63), ear (53), jawline (47), eyelid (35) \\
\midrule
\textbf{Scene \& Background} & background (166), depth (84), camera (65), blur (44), gradient (40), bokeh (39) \\
\bottomrule
\end{tabular}
\end{table}

\subsubsection{Instance-Agnostic Prompt (Fixed Prompt)}
\label{subsec:prompt-iap}

Our default evaluation uses an \textbf{instance-agnostic} construction, where a single fixed refinement prompt is applied uniformly to all images. This setting is conservative in the sense that it does not use image-specific feedback, and thus avoids adaptivity across instances. The fixed prompt is derived from the aggregated artifact taxonomy in Table~\ref{tab:keyword_stats} and targets common visual inconsistencies while explicitly requesting preservation of identity-related semantics.

\quad

\noindent\textbf{Instance-Agnostic Prompt (IAP).}
\vspace{-0.25em}

\par\medskip\noindent
{
\ttfamily\footnotesize
\linespread{0.85}\selectfont 
\quad

\noindent Refine this portrait to look like a natural camera photograph. Restore subtle skin micro-texture on cheeks and forehead (fine pores, tonal variation, gentle lighting grain); balance sharpness so facial edges (nostrils, lips, eyelids) match the overall focus; make irises unique with asymmetric reflections and detailed micro-patterns; improve hair boundaries with visible individual strands and natural flyaways; add realistic ear cartilage detail; apply mild photographic grain and consistent lens bokeh (DSLR 85mm, f/2). Preserve identity, pose, lighting, and expression.
\par 
}
\par\medskip


This prompt is intentionally phrased as a benign photographic enhancement request and contains no explicit identity manipulation instructions (e.g., requests to match a target person or change the subject's identity). 

\begin{table*}[t]
\centering
\caption{DRs (\%) of deepfake and AI-generated detectors before and after refinement using Qwen models with prompt variations. Columns report performance on Original (FF++) vs. refined outputs produced by Qwen under Prompt~1--5 (increasingly detailed).}
\label{tab:deepfake_results_qwen_prompt}
\setlength{\tabcolsep}{4.5pt} 
\footnotesize
\begin{tabular}{l cc | *{10}{c} | *{10}{c}}
\toprule
\multirow{3}{*}{\textbf{Detector}} &
\multicolumn{2}{c|}{\cellcolor{gray!10}\textbf{Original}} &
\multicolumn{10}{c|}{\textbf{Qwen-v1} (\texttt{qwen-image-edit})} &
\multicolumn{10}{c}{\textbf{Qwen-v2} (\texttt{qwen-image-edit-2511})} \\
\cmidrule(lr){2-3}
\cmidrule(lr){4-13}
\cmidrule(lr){14-23}
& \multicolumn{2}{c}{\cellcolor{gray!10}\textbf{N/A}} &
\multicolumn{2}{c}{\textbf{P1}} & \multicolumn{2}{c}{\textbf{P2}} & \multicolumn{2}{c}{\textbf{P3}} & \multicolumn{2}{c}{\textbf{P4}} & \multicolumn{2}{c|}{\textbf{P5 (IAP)}} &
\multicolumn{2}{c}{\textbf{P1}} & \multicolumn{2}{c}{\textbf{P2}} & \multicolumn{2}{c}{\textbf{P3}} & \multicolumn{2}{c}{\textbf{P4}} & \multicolumn{2}{c}{\textbf{P5 (IAP)}} \\
\cmidrule(lr){2-3}\cmidrule(lr){4-5}\cmidrule(lr){6-7}\cmidrule(lr){8-9}\cmidrule(lr){10-11}\cmidrule(lr){12-13}
\cmidrule(lr){14-15}\cmidrule(lr){16-17}\cmidrule(lr){18-19}\cmidrule(lr){20-21}\cmidrule(lr){22-23}
& \cellcolor{gray!10}$\tau_{99}$ & \cellcolor{gray!10}$\tau_{90}$ &
$\tau_{99}$ & $\tau_{90}$ & $\tau_{99}$ & $\tau_{90}$ & $\tau_{99}$ & $\tau_{90}$ & $\tau_{99}$ & $\tau_{90}$ & $\tau_{99}$ & $\tau_{90}$ &
$\tau_{99}$ & $\tau_{90}$ & $\tau_{99}$ & $\tau_{90}$ & $\tau_{99}$ & $\tau_{90}$ & $\tau_{99}$ & $\tau_{90}$ & $\tau_{99}$ & $\tau_{90}$ \\
\midrule
\rowcolor{blue!5}
GenD (C) & \cellcolor{gray!10}75 & \cellcolor{gray!10}95 &
68 & 88 & 65 & 90 & 45 & 91 & 28 & 85 & 28 & 89 &
76 & 95 & 79 & 94 & 34 & 78 & 5 & 40 & 22 & 89 \\
\rowcolor{blue!5}
GenD (P) & \cellcolor{gray!10}85 & \cellcolor{gray!10}88 &
67 & 80 & 72 & 84 & 20 & 67 & 8 & 42 & 3 & 46 &
79 & 96 & 86 & 98 & 27 & 56 & 3 & 13 & 7 & 30 \\
\rowcolor{blue!5}
GenD (D) & \cellcolor{gray!10}84 & \cellcolor{gray!10}93 &
81 & 92 & 69 & 89 & 25 & 74 & 11 & 70 & 7 & 61 &
88 & 97 & 90 & 98 & 30 & 58 & 3 & 18 & 11 & 45 \\
\rowcolor{blue!5}
M2F2 & \cellcolor{gray!10}31 & \cellcolor{gray!10}61 &
19 & 45 & 11 & 35 & 7 & 22 & 3 & 12 & 2 & 11 &
18 & 51 & 21 & 57 & 11 & 40 & 1 & 17 & 18 & 68 \\
\rowcolor{blue!5}
Hive-DF & \cellcolor{gray!10}85 & \cellcolor{gray!10}94 &
80 & 91 & 72 & 87 & 29 & 38 & 12 & 17 & 4 & 4 &
76 & 91 & 77 & 92 & 23 & 32 & 2 & 7 & 9 & 13 \\
\midrule
\rowcolor{blue!5}
Average & \cellcolor{gray!10}66 & \cellcolor{gray!10}82 &
57 & 74 & 51 & 70 & 22 & 46 & 10 & 32 & \textcolor{blue}{6} & \textcolor{blue}{27} &
58 & 80 & 61 & 82 & 21 & 45 & \textcolor{red}{2} & \textcolor{red}{16} & 13 & 45 \\
\midrule
\rowcolor{red!5}
D$^3$ & \cellcolor{gray!10}7 & \cellcolor{gray!10}20 &
11 & 25 & 16 & 37 & 55 & 90 & 62 & 99 & 52 & 97 &
10 & 23 & 13 & 27 & 43 & 84 & 46 & 87 & 47 & 82 \\
\rowcolor{red!5}
UnivFD & \cellcolor{gray!10}26 & \cellcolor{gray!10}63 &
0 & 19 & 0 & 15 & 2 & 37 & 0 & 25 & 0 & 18 &
12 & 14 & 16 & 18 & 42 & 54 & 17 & 28 & 39 & 53 \\
\rowcolor{red!5}
Hive-AI & \cellcolor{gray!10}0 & \cellcolor{gray!10}18 &
3 & 20 & 15 & 28 & 71 & 71 & 88 & 88 & 96 & 96 &
8 & 24 & 12 & 23 & 56 & 77 & 86 & 98 & 80 & 91 \\
\midrule
\rowcolor{red!5}
Average & \cellcolor{gray!10}11 & \cellcolor{gray!10}34 &
\textcolor{red}{5} & \textcolor{blue}{21} & 10 & 27 & 43 & 66 & 50 & 71 & 49 & 70 &
\textcolor{blue}{10} & \textcolor{red}{20} & 14 & 23 & 47 & 72 & 50 & 71 & 55 & 75 \\
\midrule
\midrule
Total Avg. & \cellcolor{gray!10}38 & \cellcolor{gray!10}59 &
31 & \textcolor{blue}{48} & 30 & 48 & 32 & 56 & 30 & 51 & \textcolor{blue}{28} & 49 &
34 & 50 & 37 & 52 & 34 & 59 & \textcolor{red}{26} & \textcolor{red}{43} & 34 & 60 \\
\bottomrule
\end{tabular}
\end{table*}

\subsubsection{Instance-Specific Prompt (Per-Image Reasoning)}
\label{subsec:prompt-isp}

To reflect a realistic non-expert workflow, we additionally evaluate an \textbf{instance-specific} construction in which refinement instructions are generated on a per-image basis. Concretely, for each input image, we obtain an artifact-focused rationale from a GAI system (i.e., ChatGPT or Gemini) and \emph{directly} feed the rationale back into the refinement prompt, without manual editing or post-processing. This mirrors a simple user behavior: ``ask why it looks fake'' and ``request the system to fix exactly those issues.''

\quad

\noindent\textbf{Instance-Specific Prompt (ISP) Template.}
\vspace{-0.75em}

\par\medskip\noindent
{
\ttfamily\footnotesize
\linespread{0.85}\selectfont 
\quad

\noindent Refine this portrait to look like a natural camera photograph. The following issues were detected in the image. Please correct them while preserving the subject's identity:-\{Skin rationale\}-\{Eyes rationale\}-\{Hair rationale\}-\{Facial-edges rationale\}-\{Background rationale\}-\{Global rationale\}
\par 
}
\par\medskip


In our implementation, each ``rationale'' line is taken verbatim from the model-provided assessment output for the corresponding image. Unless otherwise specified, all main results in Section~\ref{sec:main_result} use this adaptive instance-specific prompt. 

\subsection{Main Results}\label{sec:main_result}

This section investigates how natural-language–driven image refinement reshapes the effectiveness of modern forensic detectors.
Using both open-weight and commercial GAI systems, we demonstrate that semantic refinement does not merely degrade detection performance gradually, but instead exposes a structural mismatch between deepfake detectors and AI-generated image detectors. Furthermore, we show that this mismatch is amplified by commercially deployed models and adaptive, feedback-driven prompting.

\subsubsection{Prompt Granularity Reveals a Structural Split in Detection Paradigms}~\label{sec:opensource}

Table~\ref{tab:deepfake_results_qwen_prompt} reports detection rates under progressively detailed refinement prompts (P1--P5) using the Qwen model family. Crucially, all results are obtained under an intra-dataset setting (FF++), eliminating distribution shift as a confounding factor. Despite this controlled setup, detection performance diverges sharply across detector classes.


As prompt granularity increases from an identical reconstruction instruction (P1) to a highly detailed refinement objective (P5), traditional deepfake detectors---such as GenD variants and Hive-DF---exhibit a pronounced and consistent collapse in detection rates. On average, deepfake DRs drop from moderate-to-high values at P1 to near-single-digit levels at P5.
In contrast, AI-generated image detectors follow the opposite trajectory. These detectors show limited sensitivity to early prompts but increasingly flag images produced under more aggressive refinement. This inverse relationship indicates that semantic refinement systematically removes manipulation-specific forensic artifacts while simultaneously introducing new statistical or generative signatures characteristic of modern image synthesis.

Prompt P1 explicitly instructs the model to preserve the same photo and identical composition, resulting in outputs that largely retain the low-level artifacts of the original deepfake generation pipeline. Consequently, deepfake detectors remain effective under P1, while AI-generated image detectors---particularly those tuned to recent generative models---largely fail to respond. This behavior establishes P1 as a diagnostic baseline, highlighting that the observed detection collapse under later prompts is not driven by image regeneration alone, but by targeted semantic refinement.

\begin{table*}[t]
\centering
\caption{Detection rates of deepfake and AI-generated image detectors before and after refinement with commercial GAI systems.}
\label{tab:deepfake_results_commercial}
\setlength{\tabcolsep}{4pt}
\footnotesize
\begin{tabular}{l cc | *{8}{c} | *{8}{c} | *{4}{c}}
\toprule
\multirow{3}{*}{\textbf{Detector}} &
\multicolumn{2}{c|}{\cellcolor{gray!10}\textbf{Original}} &
\multicolumn{8}{c|}{\textbf{P4 (short version of IAP)}} &
\multicolumn{8}{c|}{\textbf{P5 (IAP)}} &
\multicolumn{4}{c}{\textbf{P6 (ISP)}} \\

\cmidrule(lr){2-3}
\cmidrule(lr){4-11}
\cmidrule(lr){12-19}
\cmidrule(lr){20-23}

& \multicolumn{2}{c}{\cellcolor{gray!10}\textbf{N/A}} &
\multicolumn{2}{c}{Qwen-v1} & \multicolumn{2}{c}{Flux AI} &
\multicolumn{2}{c}{ChatGPT} & \multicolumn{2}{c}{Gemini} &
\multicolumn{2}{c}{Qwen-v1} & \multicolumn{2}{c}{Flux AI} &
\multicolumn{2}{c}{ChatGPT} & \multicolumn{2}{c}{Gemini} &
\multicolumn{2}{c}{ChatGPT} & \multicolumn{2}{c}{Gemini} \\

\cmidrule(lr){2-3}
\cmidrule(lr){4-5}\cmidrule(lr){6-7}\cmidrule(lr){8-9}\cmidrule(lr){10-11}
\cmidrule(lr){12-13}\cmidrule(lr){14-15}\cmidrule(lr){16-17}\cmidrule(lr){18-19}
\cmidrule(lr){20-21}\cmidrule(lr){22-23}

& \cellcolor{gray!10}$\tau_{99}$ & \cellcolor{gray!10}$\tau_{90}$ &
$\tau_{99}$ & $\tau_{90}$ & $\tau_{99}$ & $\tau_{90}$ &
$\tau_{99}$ & $\tau_{90}$ & $\tau_{99}$ & $\tau_{90}$ &
$\tau_{99}$ & $\tau_{90}$ & $\tau_{99}$ & $\tau_{90}$ &
$\tau_{99}$ & $\tau_{90}$ & $\tau_{99}$ & $\tau_{90}$ &
$\tau_{99}$ & $\tau_{90}$ & $\tau_{99}$ & $\tau_{90}$ \\

\midrule
\rowcolor{blue!5}
GenD (C) & \cellcolor{gray!10}75 & \cellcolor{gray!10}95 &
28 & 85 & 12 & 59 & 0 & 0 & 0 & 0 &
28 & 89 & 10 & 62 & 0 & 0 & 0 & 0 &
0 & 0 & 0 & 1 \\

\rowcolor{blue!5}
GenD (P) & \cellcolor{gray!10}85 & \cellcolor{gray!10}88 &
8 & 42 & 1 & 24 & 0 & 0 & 0 & 0 &
3 & 46 & 0 & 14 & 0 & 0 & 0 & 0 &
0 & 0 & 0 & 0 \\

\rowcolor{blue!5}
GenD (D) & \cellcolor{gray!10}84 & \cellcolor{gray!10}93 &
11 & 70 & 6 & 33 & 0 & 0 & 0 & 0 &
7 & 61 & 6 & 35 & 0 & 0 & 0 & 0 &
0 & 0 & 0 & 0 \\

\rowcolor{blue!5}
M2F2 & \cellcolor{gray!10}31 & \cellcolor{gray!10}61 &
3 & 12 & 1 & 21 & 3 & 29 & 0 & 5 &
2 & 11 & 5 & 33 & 6 & 41 & 0 & 15 &
0 & 17 & 0 & 2 \\

\rowcolor{blue!5}
Hive-DF & \cellcolor{gray!10}85 & \cellcolor{gray!10}94 &
12 & 17 & 10 & 12 & 6 & 65 & 1 & 8 &
4 & 4 & 6 & 6 & 3 & 55 & 5 & 9 &
3 & 50 & 0 & 7 \\

\midrule
\rowcolor{blue!5}
Average & \cellcolor{gray!10}66 & \cellcolor{gray!10}82 &
10 & 32 & 6 & 24 & 3 & 31 & \textcolor{blue}{0} & \textcolor{blue}{4} & 6 & 27 &
5 & 25 & 3 & 32 & 2 & 8 & 1 & 22 & \textcolor{red}{0} & \textcolor{red}{3} \\

\midrule
\rowcolor{red!5}
D$^3$ & \cellcolor{gray!10}7 & \cellcolor{gray!10}20 &
62 & 99 & 24 & 57 & 2 & 28 & 14 & 53 &
52 & 97 & 37 & 72 & 4 & 30 & 4 & 30 &
3 & 24 & 25 & 59 \\

\rowcolor{red!5}
UnivFD & \cellcolor{gray!10}26 & \cellcolor{gray!10}63 &
0 & 25 & 26 & 38 & 8 & 19 & 2 & 8 &
0 & 18 & 25 & 36 & 14 & 23 & 11 & 14 &
11 & 17 & 3 & 7 \\

\rowcolor{red!5}
Hive-AI & \cellcolor{gray!10}0 & \cellcolor{gray!10}18 &
88 & 88 & 51 & 90 & 29 & 90 & 2 & 79 &
96 & 96 & 28 & 92 & 15 & 94 & 3 & 63 &
15 & 93 & 9 & 73 \\

\midrule
\rowcolor{red!5}
Average & \cellcolor{gray!10}11 & \cellcolor{gray!10}\textcolor{red}{34} &
50 & 71 & 32 & 31 & 13 & 46 & \textcolor{blue}{6} & 47 & 49 & 70 &
30 & 67 & 11 & 49 & \textcolor{red}{6} & \textcolor{blue}{36} & 10 & 44 & 12 & 46 \\

\midrule
\midrule
Total Avg. & \cellcolor{gray!10}38 & \cellcolor{gray!10}58 &
30 & 51 & 20 & 43 & 8 & 39 & \textcolor{red}{3} & \textcolor{blue}{26} & 28 & 49 &
18 & 46 & 7 & 41 & \textcolor{blue}{4} & \textcolor{red}{22} & 5 & 34 & 6 & 25 \\

\bottomrule
\end{tabular}
\end{table*}

While increased prompt detail generally improves visual realism, Qwen-v2 exhibits a notable deviation from monotonic behavior: in several cases, P4 outperforms P5 in evading detection. We attribute this to the model’s heightened sensitivity to explicit texture-level instructions. The exhaustive, instance-agnostic descriptions used in P5 cause Qwen-v2 to overemphasize micro-textural details---such as skin pores---leading to exaggerated high-frequency patterns that reintroduce detectable artifacts. Visual examples in the Appendix corroborate this explanation. This observation underscores a broader limitation of fixed IAPs: when interpreted too literally by a capable generator, additional prompt detail may paradoxically reduce stealth.

Taken together, these results demonstrate that no single detector class provides robust coverage against \emph{semantic-preserving image refinement}. Deepfake detectors remain tightly coupled to legacy manipulation artifacts, while AI-generated image detectors primarily respond to synthesis fingerprints. Semantic refinement exploits the gap between these paradigms, revealing a structural vulnerability in current forensic defenses.

\subsubsection{Commercial GAI Systems and the Escalation of Adaptive Threats}

Table \ref{tab:deepfake_results_commercial} extends the analysis to widely accessible commercial GAI systems, including Flux AI, ChatGPT, and Gemini. We compare IAPs (P4/P5) against rationale-guided ISPs (P6), highlighting the impact of adaptive refinement strategies.

Refinement performed by commercial systems renders traditional deepfake detectors almost entirely ineffective. In many configurations, detection rates for GenD and Hive-DF drop to zero, substantially outperforming open-weight baselines. This effect cannot be attributed solely to dataset bias; rather, it reflects the superior perceptual coherence and photorealistic priors embedded in commercial models.

More strikingly, certain commercial systems---most notably Gemini---also substantially degrade the effectiveness of AI-generated image detectors. In these cases, refined outputs evade both manipulation-focused and synthesis-focused detection under strict operating thresholds ($\tau_{99}$), despite triggering stronger responses at looser thresholds ($\tau_{90}$). This finding suggests that the apparent robustness of AI-generated detectors observed in Section~\ref{sec:opensource} does not translate into reliable decisions under realistic operating thresholds for state-of-the-art commercial GAI service built-in generators.

Comparing IAP (P5) and ISP (P6) reveals the compounding risk of adaptive attacks. ISP incorporates the system’s own artifact-level critiques into the refinement prompt, achieving the strongest evasion performance across nearly all detectors. Importantly, this strategy requires no specialized expertise: a user can simply restate the model’s feedback as a benign refinement request. This transforms alignment-driven explanatory behavior into an actionable attack primitive.

The concentration of best-evasion results within the Gemini columns highlights a critical security paradox. As commercial GAI systems are optimized to better satisfy user intent and improve visual realism, they simultaneously become more effective at scrubbing forensic traces. This dynamic mirrors the classical GAN arms race: as generators approach perceptual indistinguishability, static detection boundaries erode. Our results suggest that without fundamentally new detection strategies, current forensic methods risk rapid obsolescence in the face of evolving commercial GAI services.

\subsubsection{Semantic Preservation Analysis}
\label{subsubsec:semantic}
\begin{figure}[t]
    \centering
    \includegraphics[width=\columnwidth]{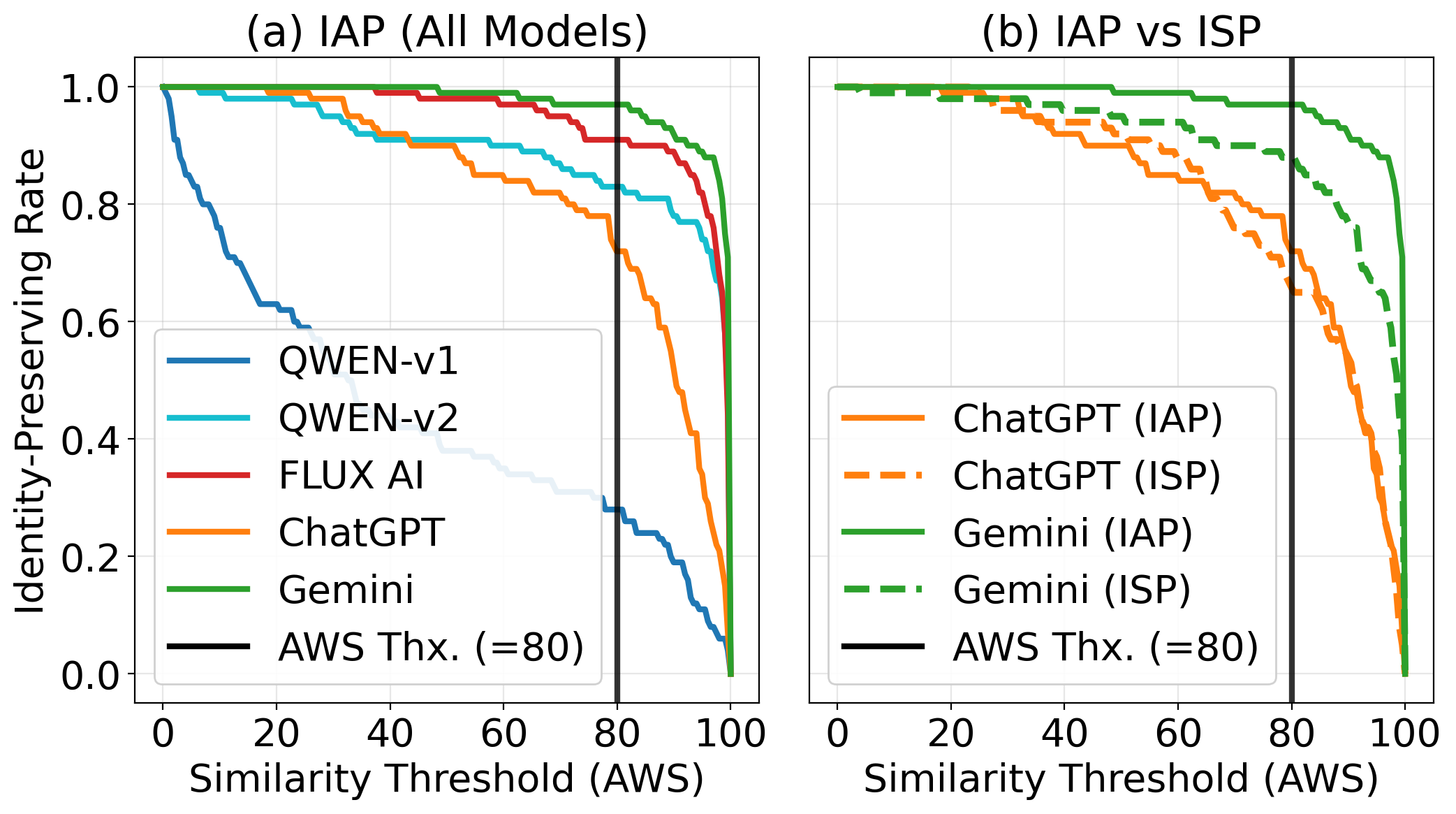}
    \vspace{-6mm}
    \caption{Semantic preservation with IAP and ISP.}
    \label{fig:semantic_preservation}
    \vspace{-3mm}
\end{figure}

\begin{figure*}[t]
\centering
\caption{Qualitative examples. Each row shows one source image and corresponding outputs from different systems.}
\label{fig:qual_3x9}
\setlength{\tabcolsep}{3.5pt}
\small
\begin{tabular}{c | *{5}{c} | c c c}
\toprule
\multicolumn{1}{c|}{\textbf{Original}} &
\multicolumn{5}{c|}{\textbf{Qwen-v2}} &
\textbf{Flux} &
\textbf{GPT} &
\textbf{Gemini} \\
\cmidrule(lr){1-1}
\cmidrule(lr){2-6}
\cmidrule(lr){7-9}
\textbf{N/A} &
\textbf{P1} & \textbf{P2} & \textbf{P3} & \textbf{P4} & \textbf{P5 (IAP)} &\multicolumn{3}{c}{\textbf{P5    (IAP)}} \\
\midrule

\includegraphics[width=0.10\linewidth]{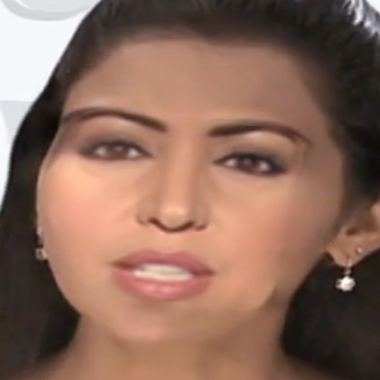} &
\includegraphics[width=0.10\linewidth]{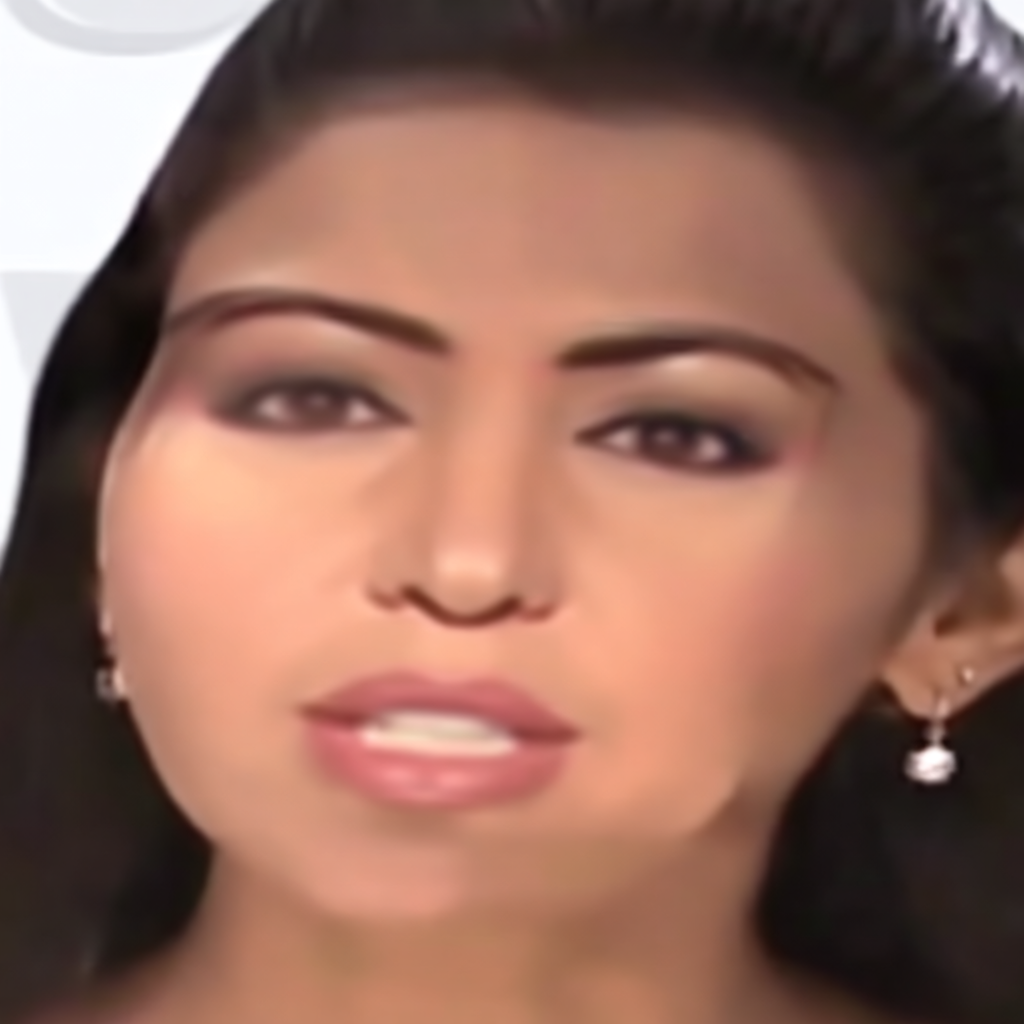} &
\includegraphics[width=0.10\linewidth]{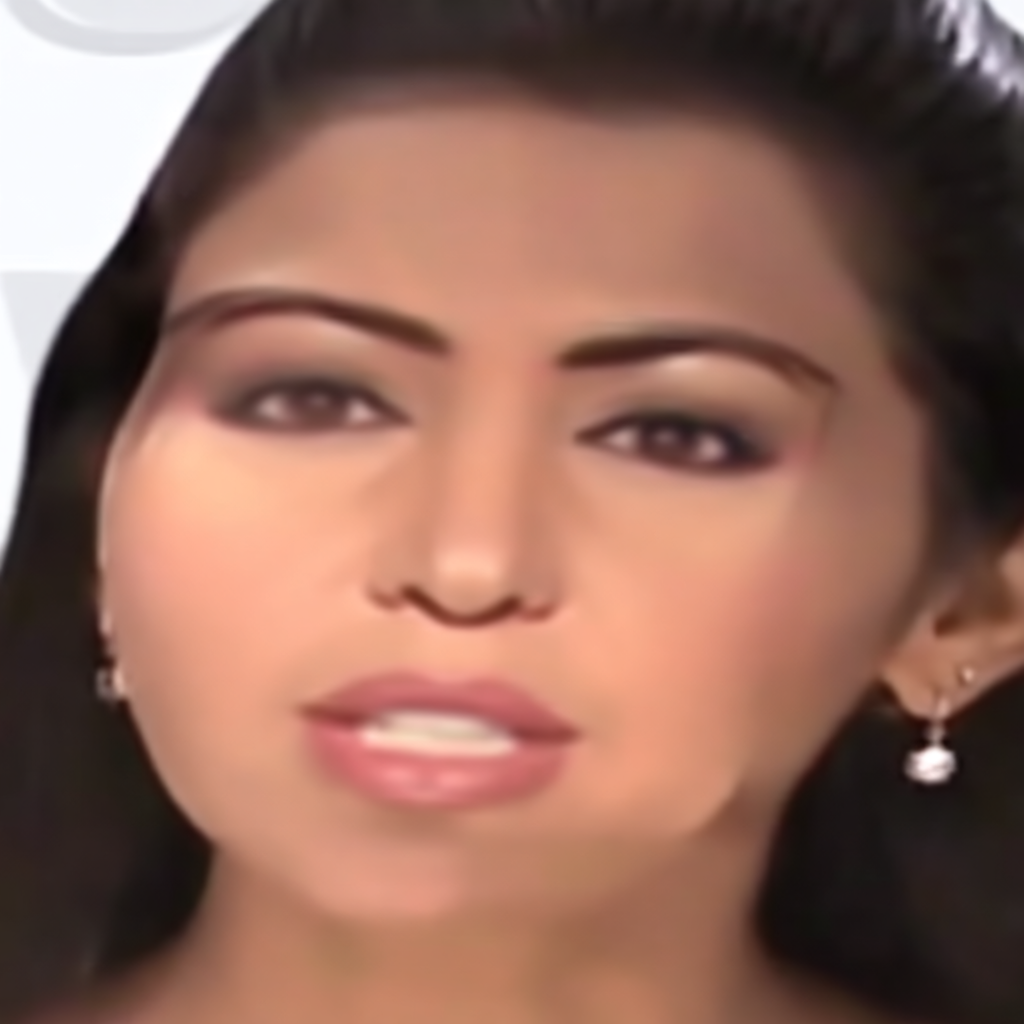} &
\includegraphics[width=0.10\linewidth]{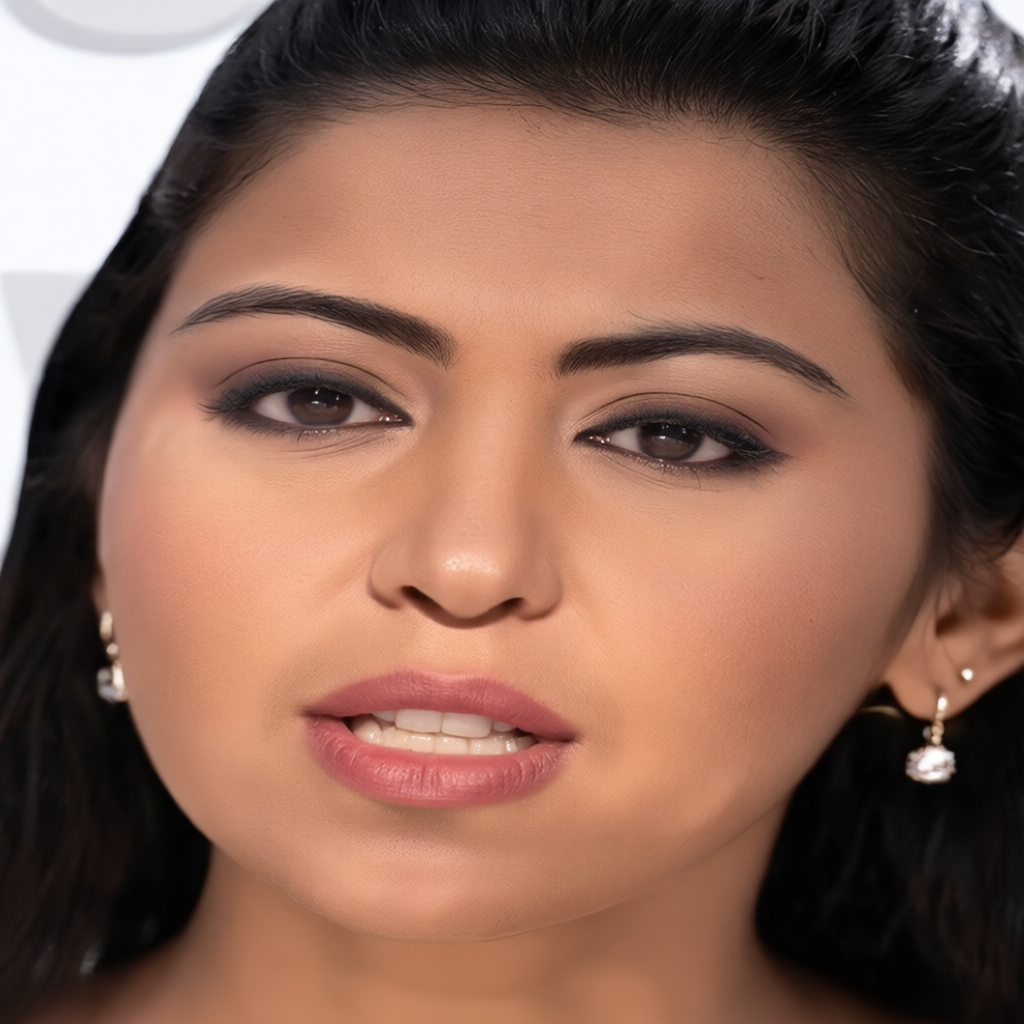} &
\includegraphics[width=0.10\linewidth]{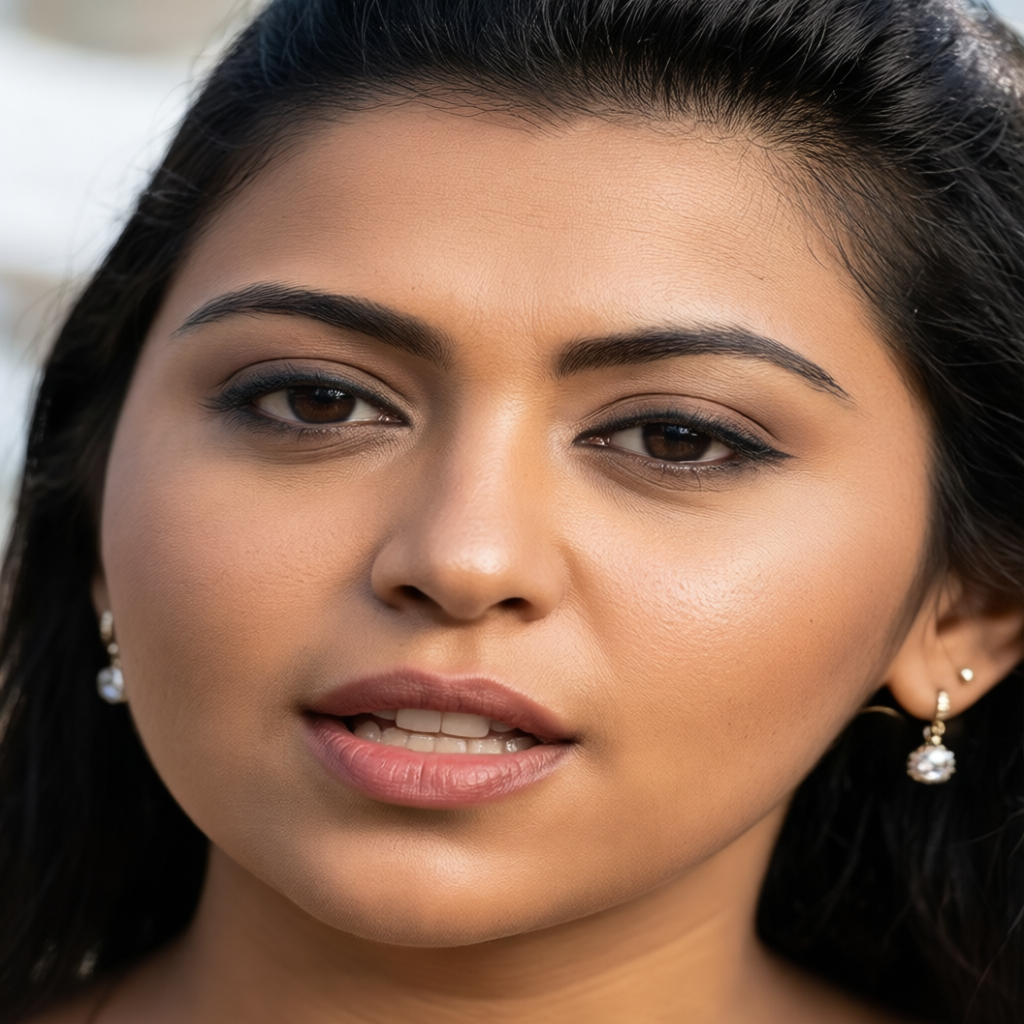} &
\includegraphics[width=0.10\linewidth]{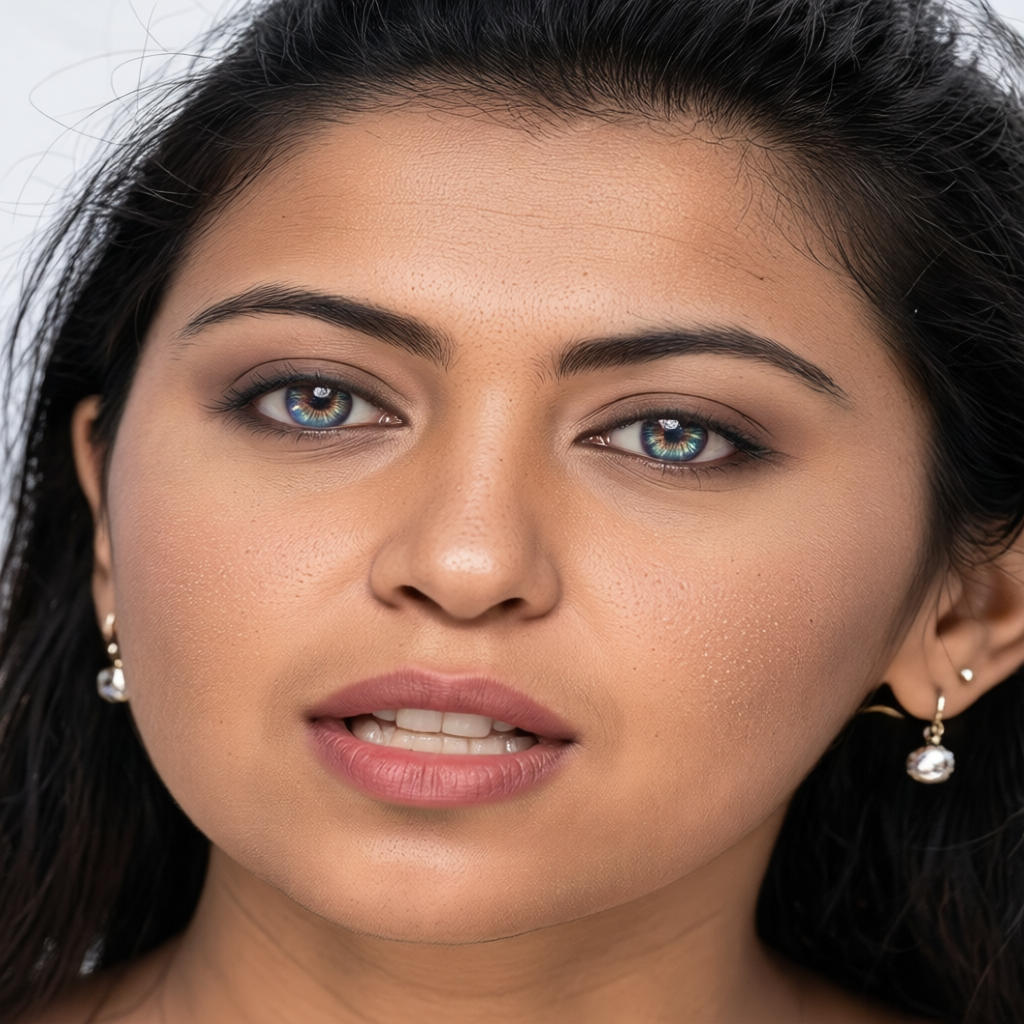} &
\includegraphics[width=0.10\linewidth]{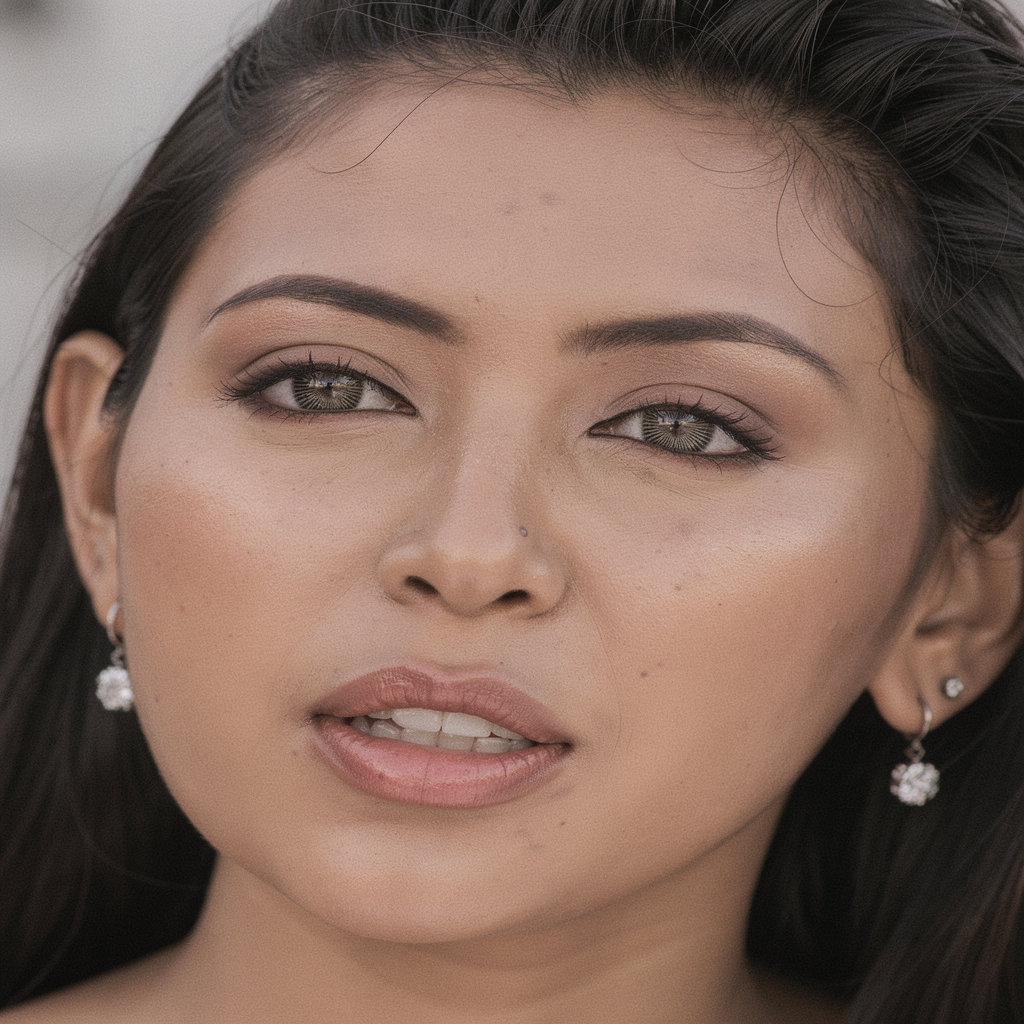} &
\includegraphics[width=0.10\linewidth]{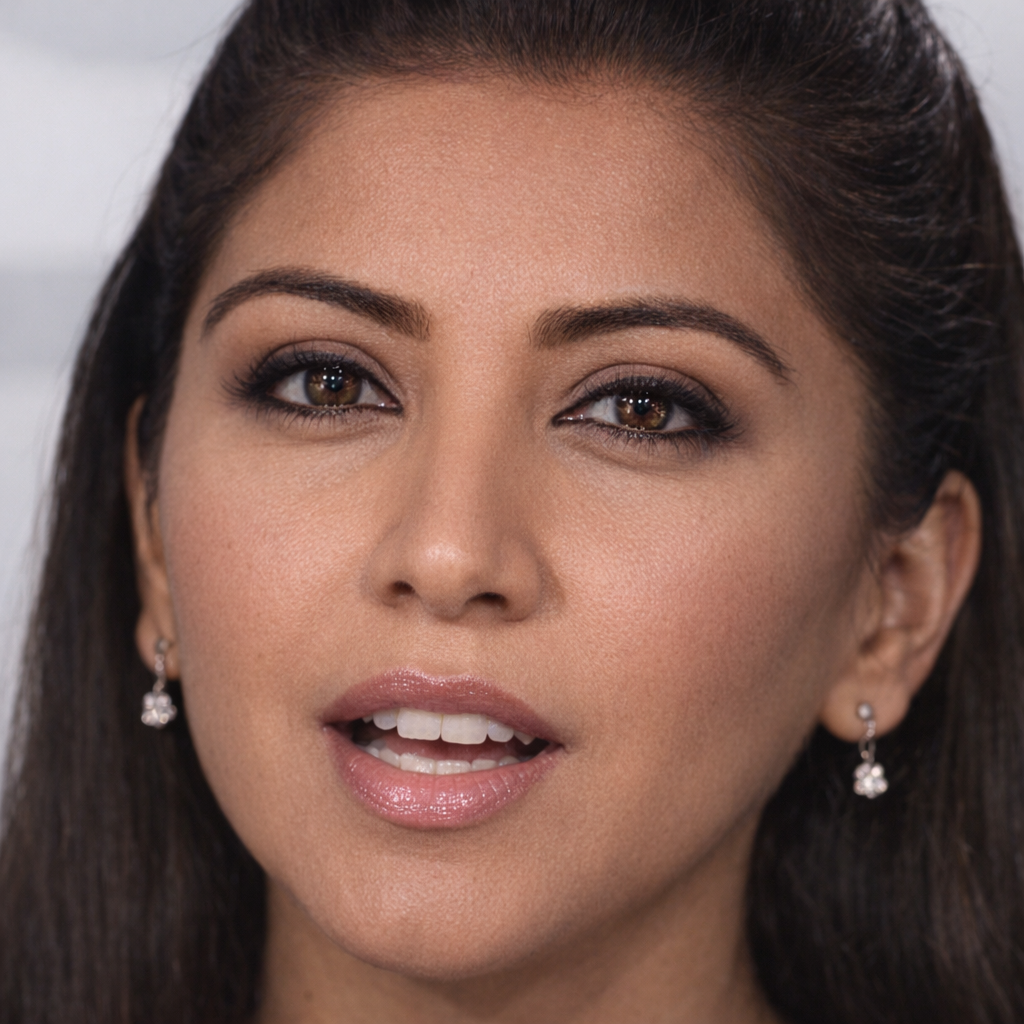} &
\includegraphics[width=0.10\linewidth]{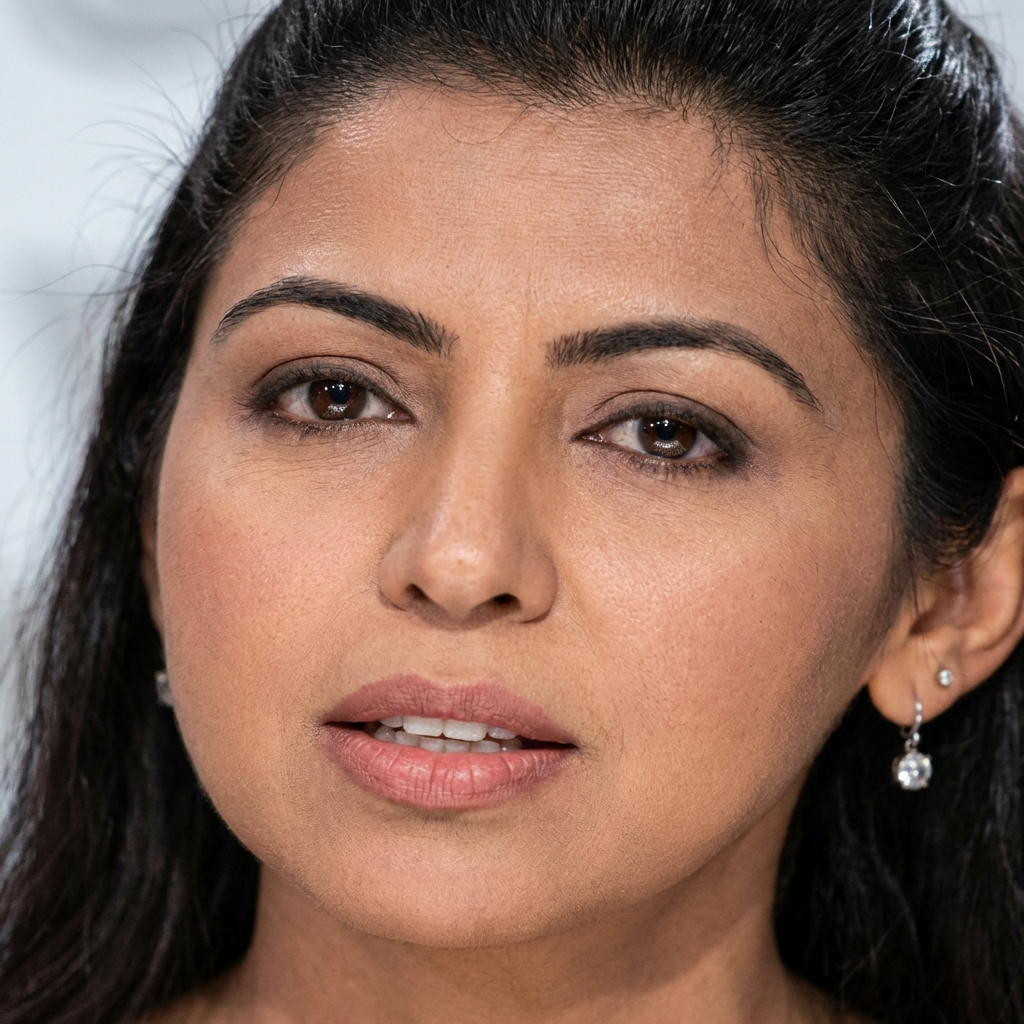} \\

\includegraphics[width=0.10\linewidth]{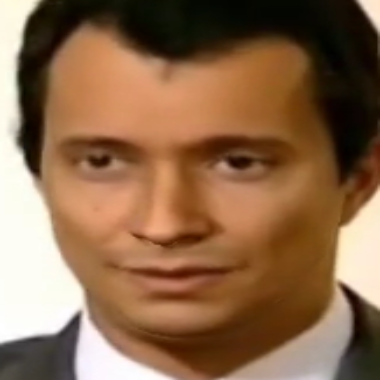} &
\includegraphics[width=0.10\linewidth]{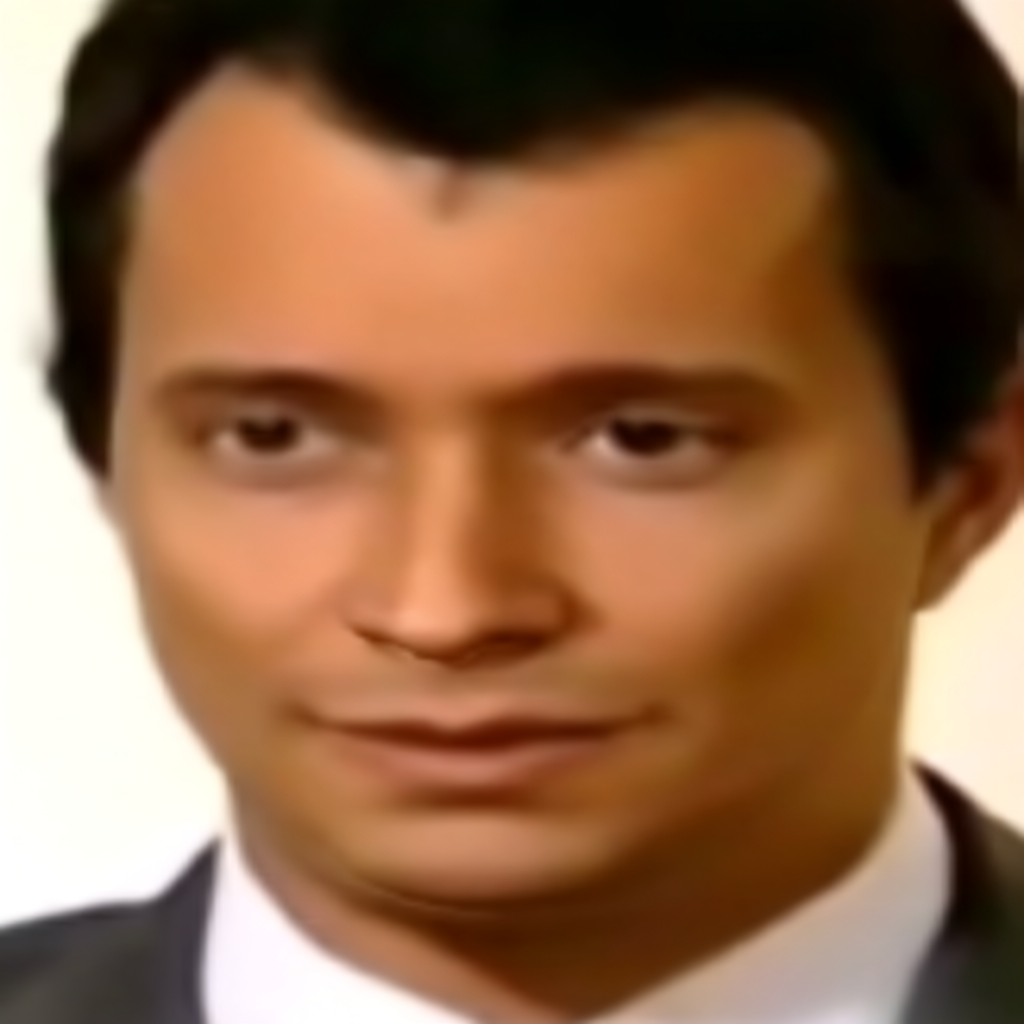} &
\includegraphics[width=0.10\linewidth]{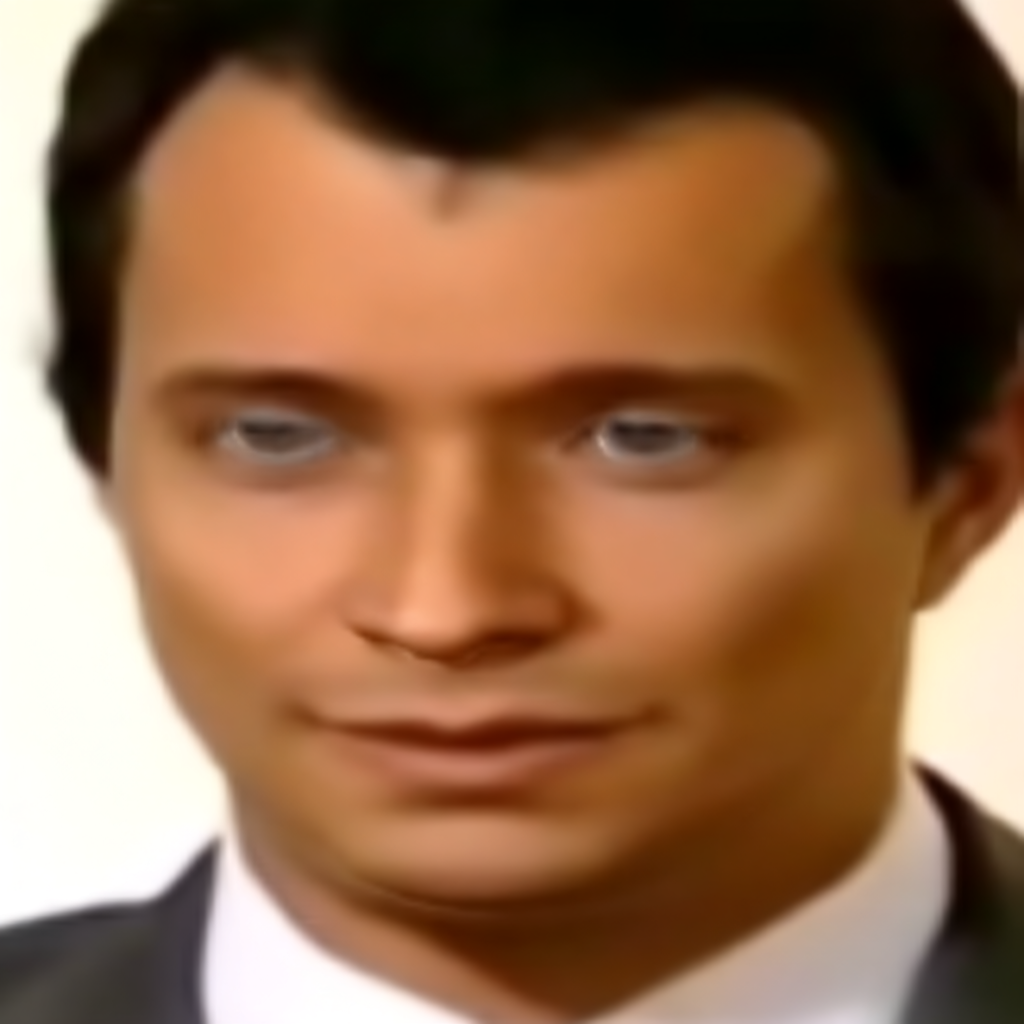} &
\includegraphics[width=0.10\linewidth]{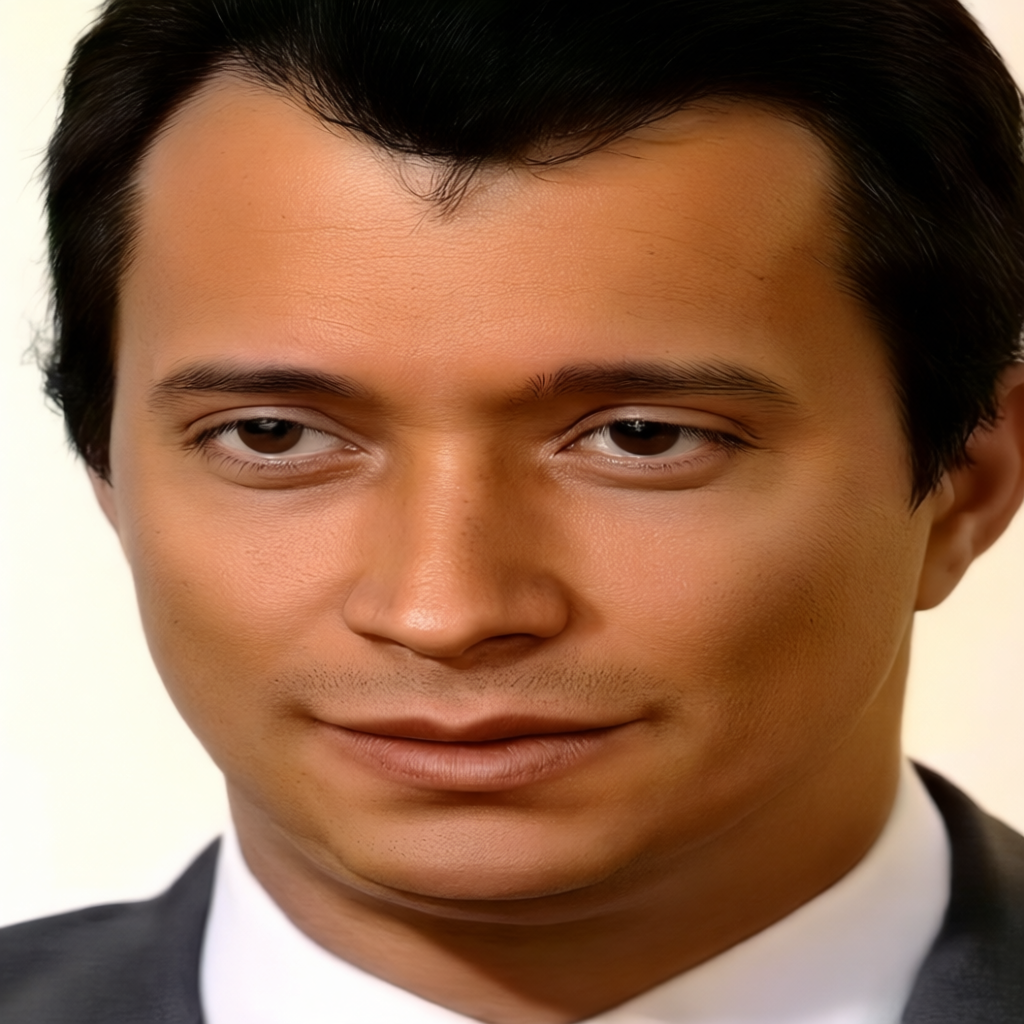} &
\includegraphics[width=0.10\linewidth]{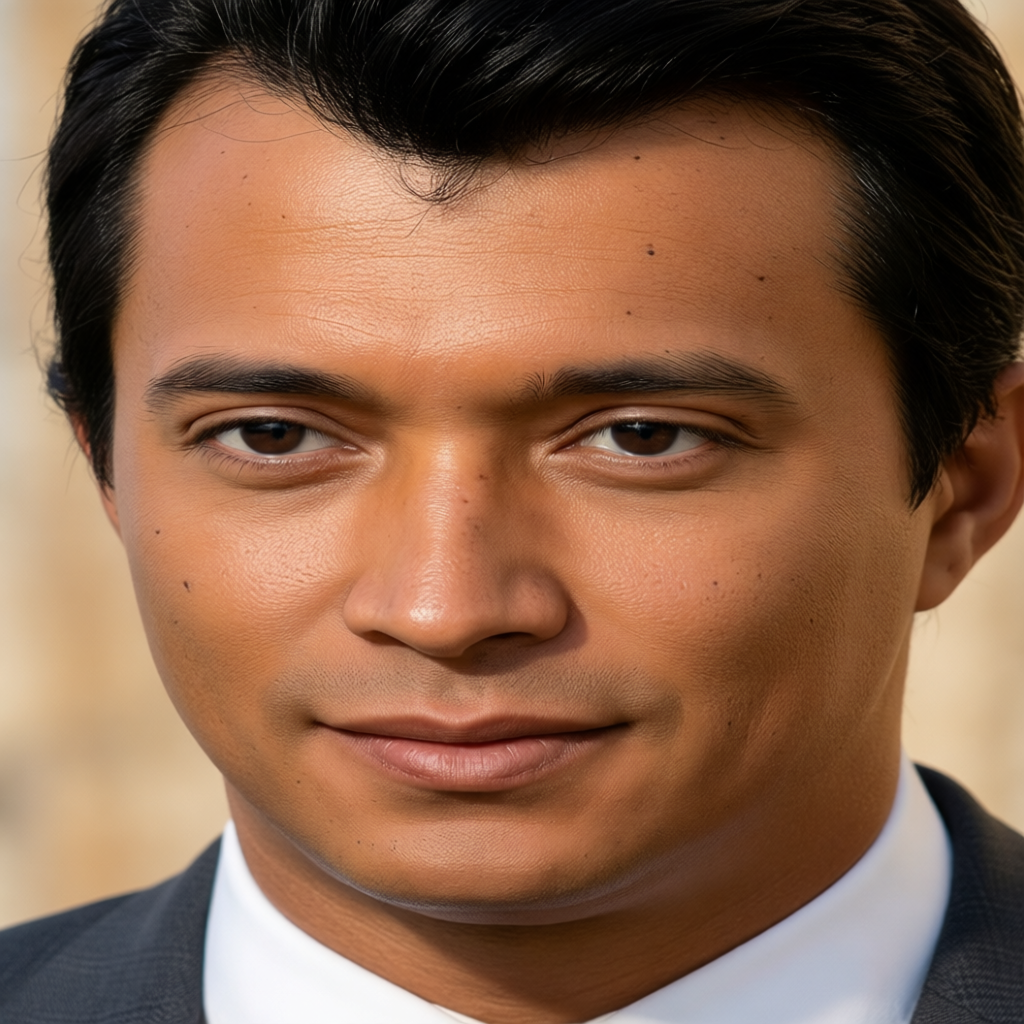} &
\includegraphics[width=0.10\linewidth]{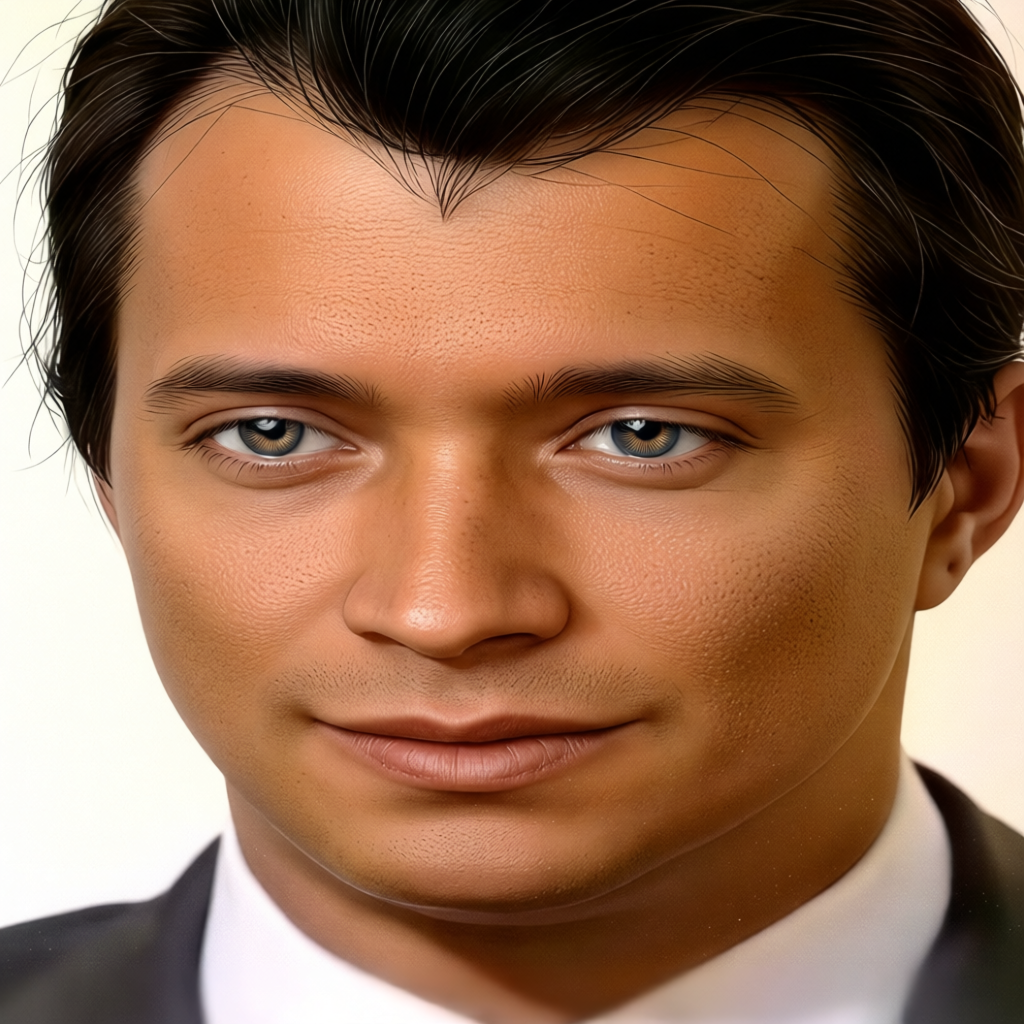} &
\includegraphics[width=0.10\linewidth]{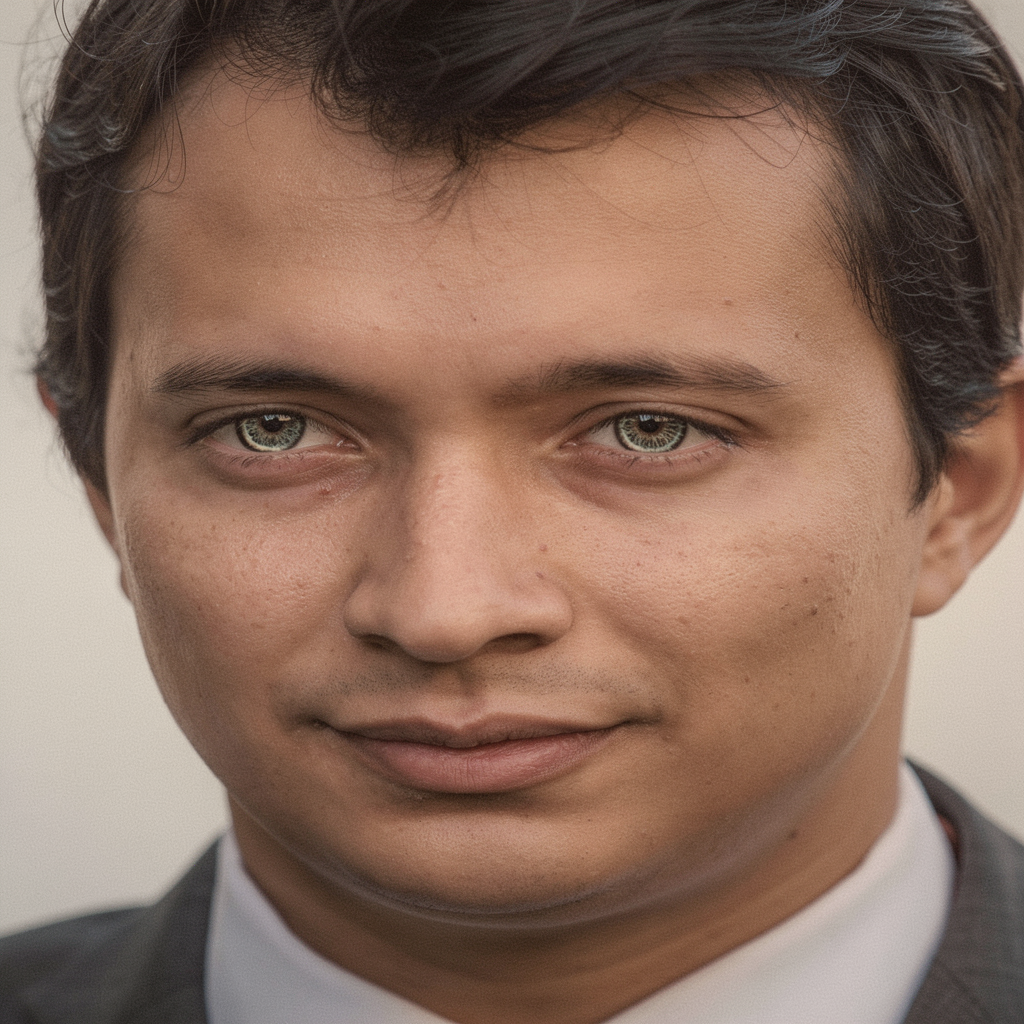} &
\includegraphics[width=0.10\linewidth]{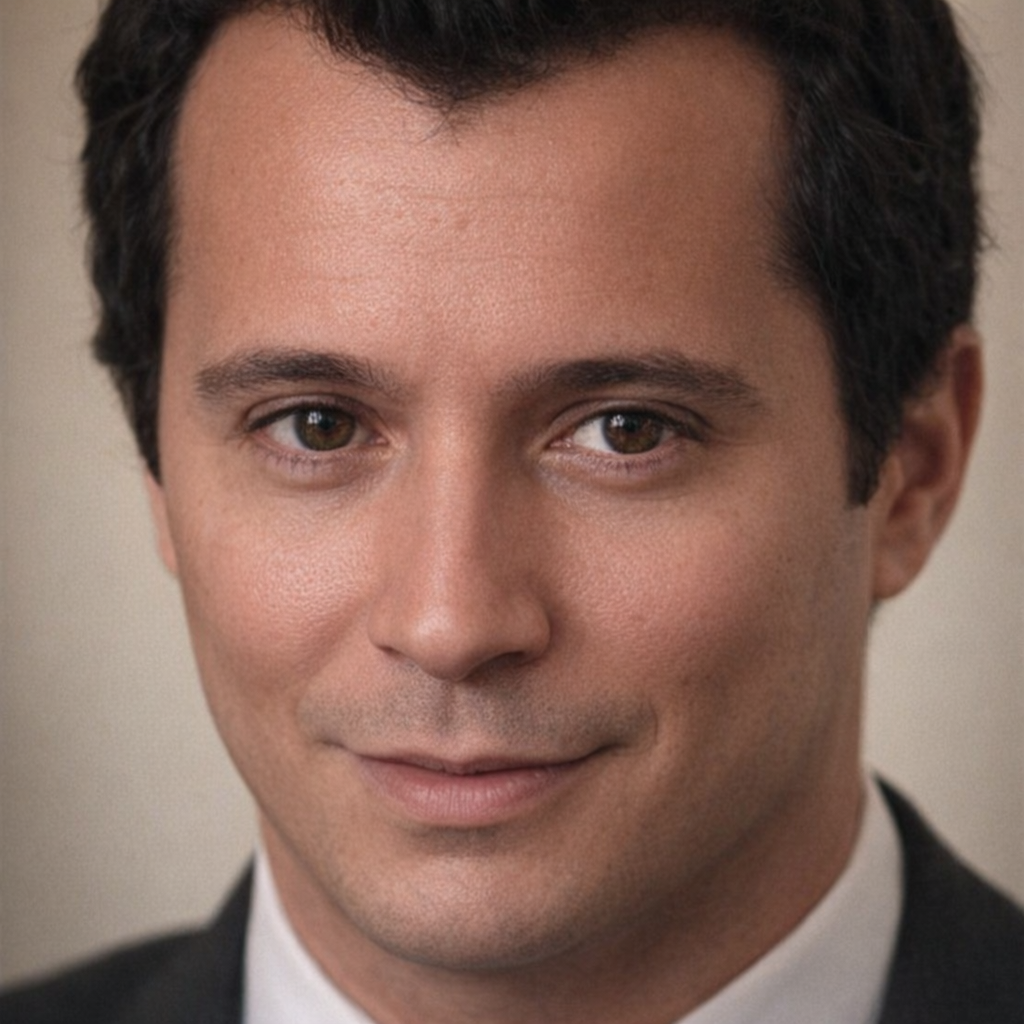} &
\includegraphics[width=0.10\linewidth]{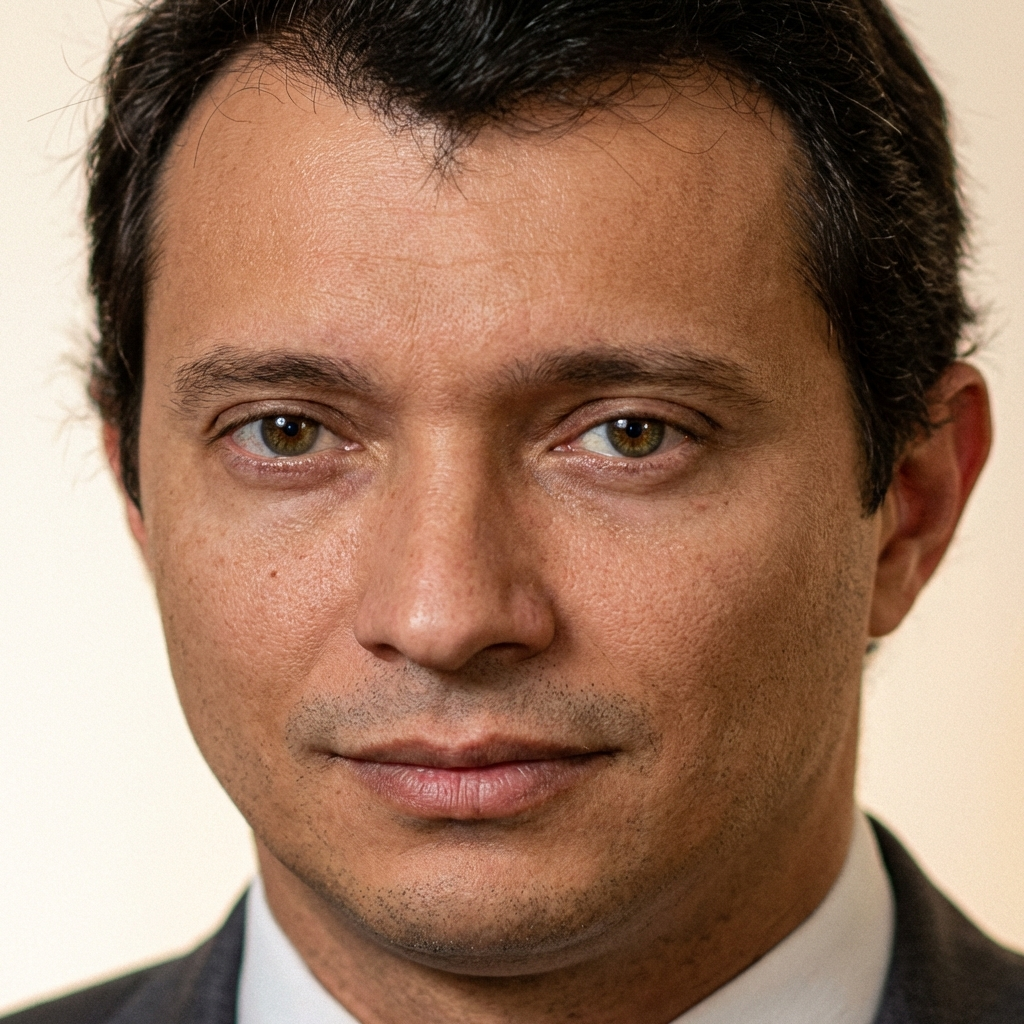} \\

\includegraphics[width=0.10\linewidth]{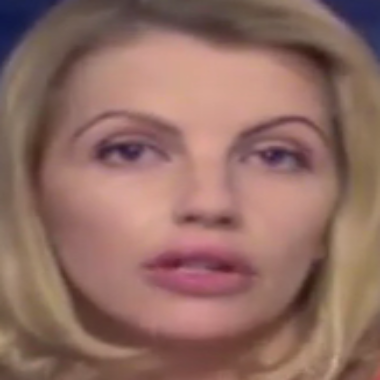} &
\includegraphics[width=0.10\linewidth]{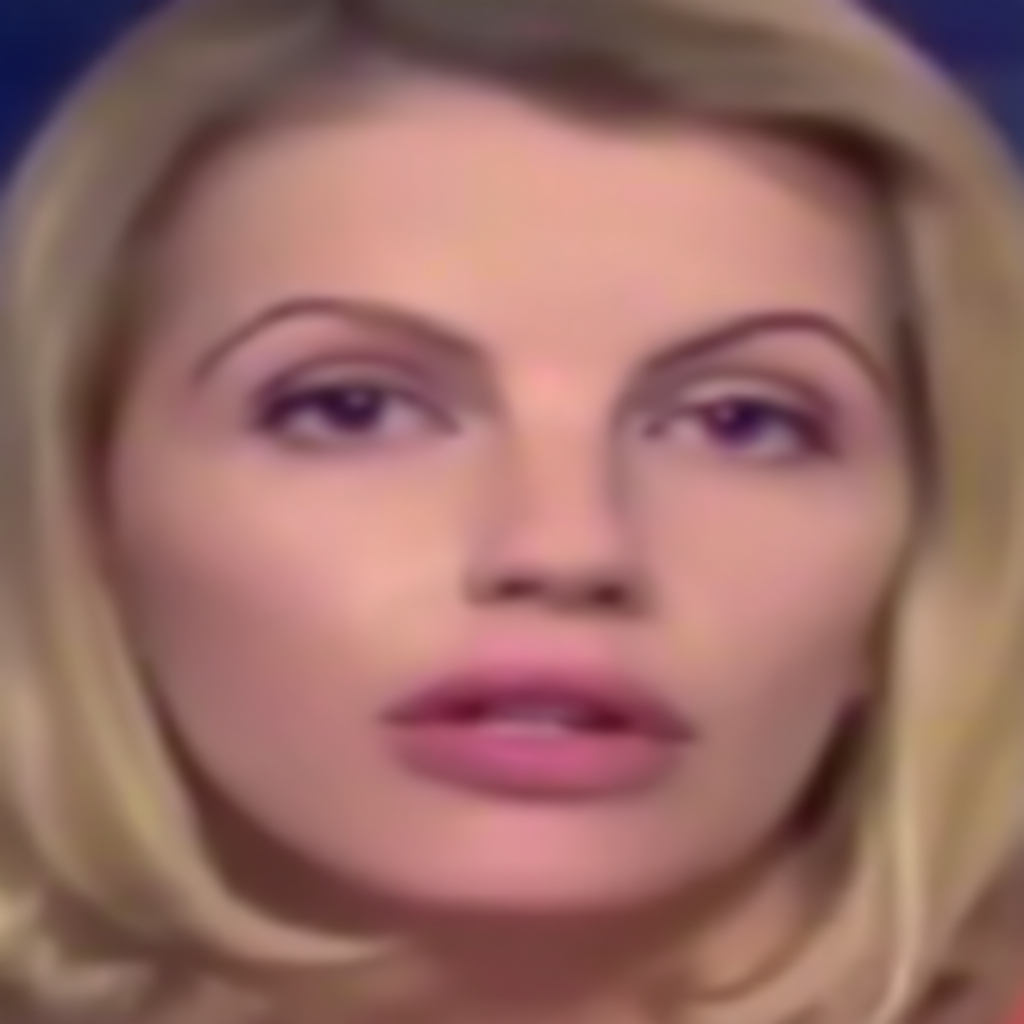} &
\includegraphics[width=0.10\linewidth]{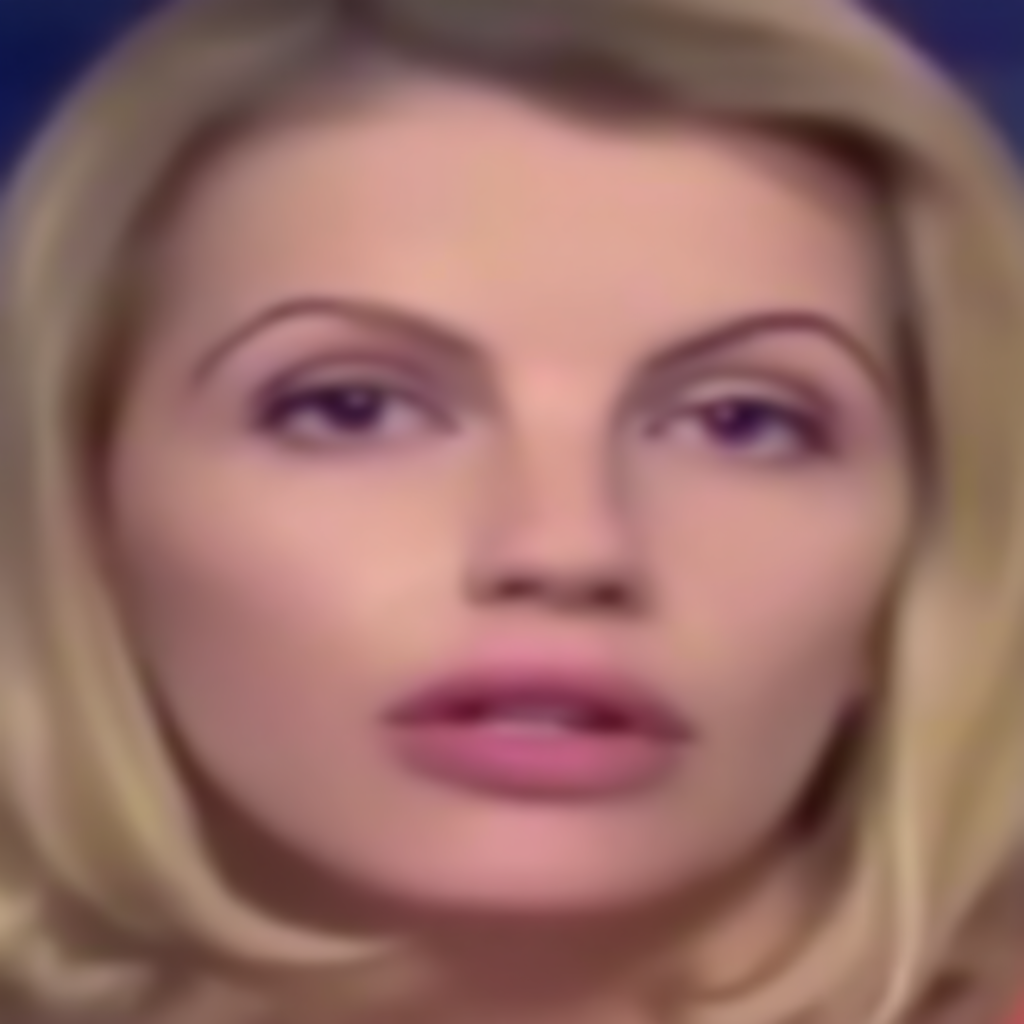} &
\includegraphics[width=0.10\linewidth]{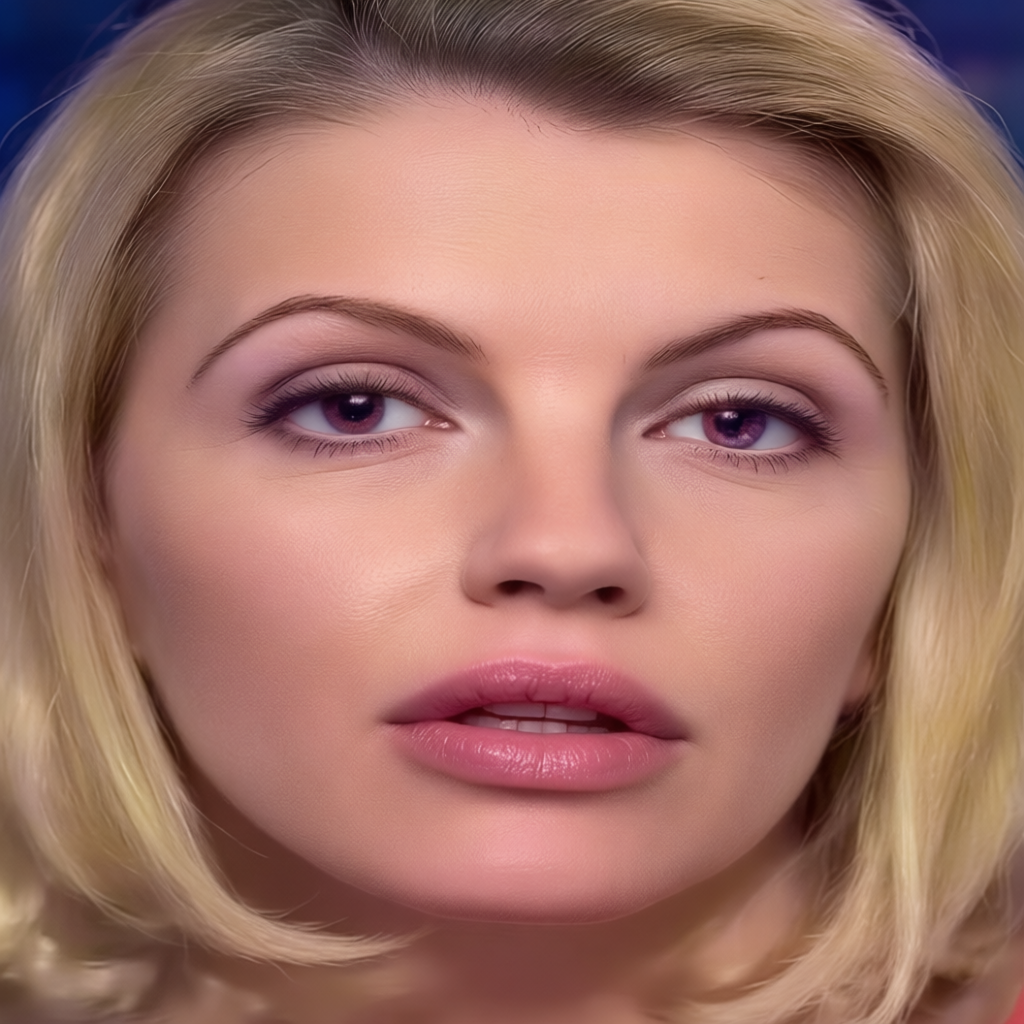} &
\includegraphics[width=0.10\linewidth]{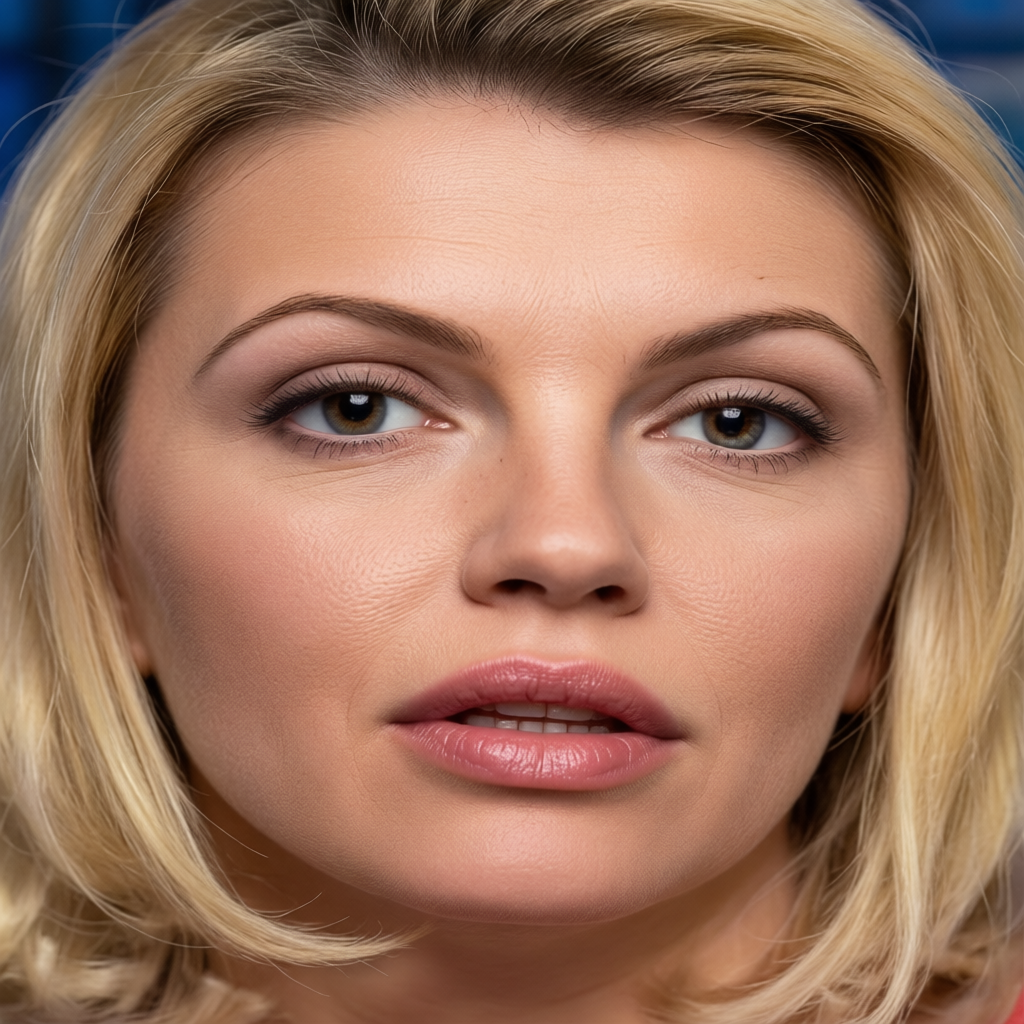} &
\includegraphics[width=0.10\linewidth]{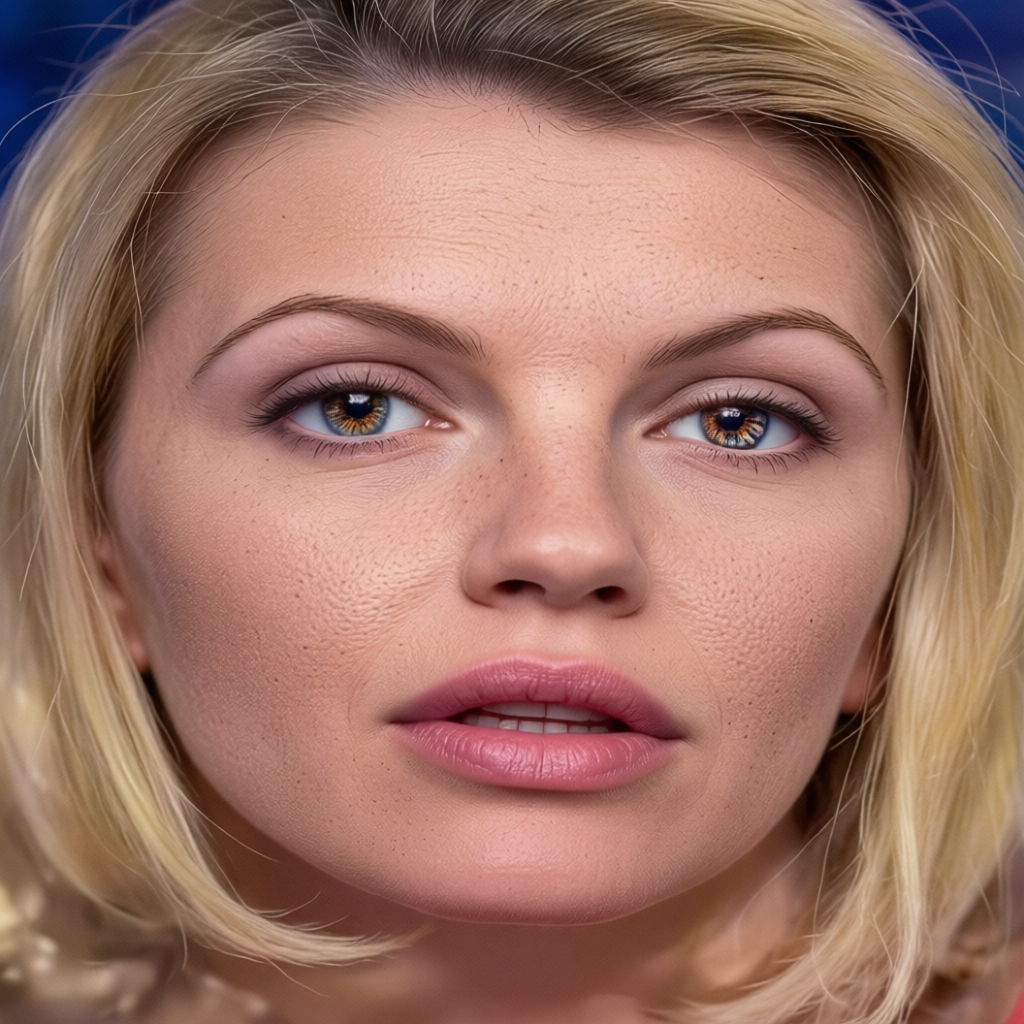} &
\includegraphics[width=0.10\linewidth]{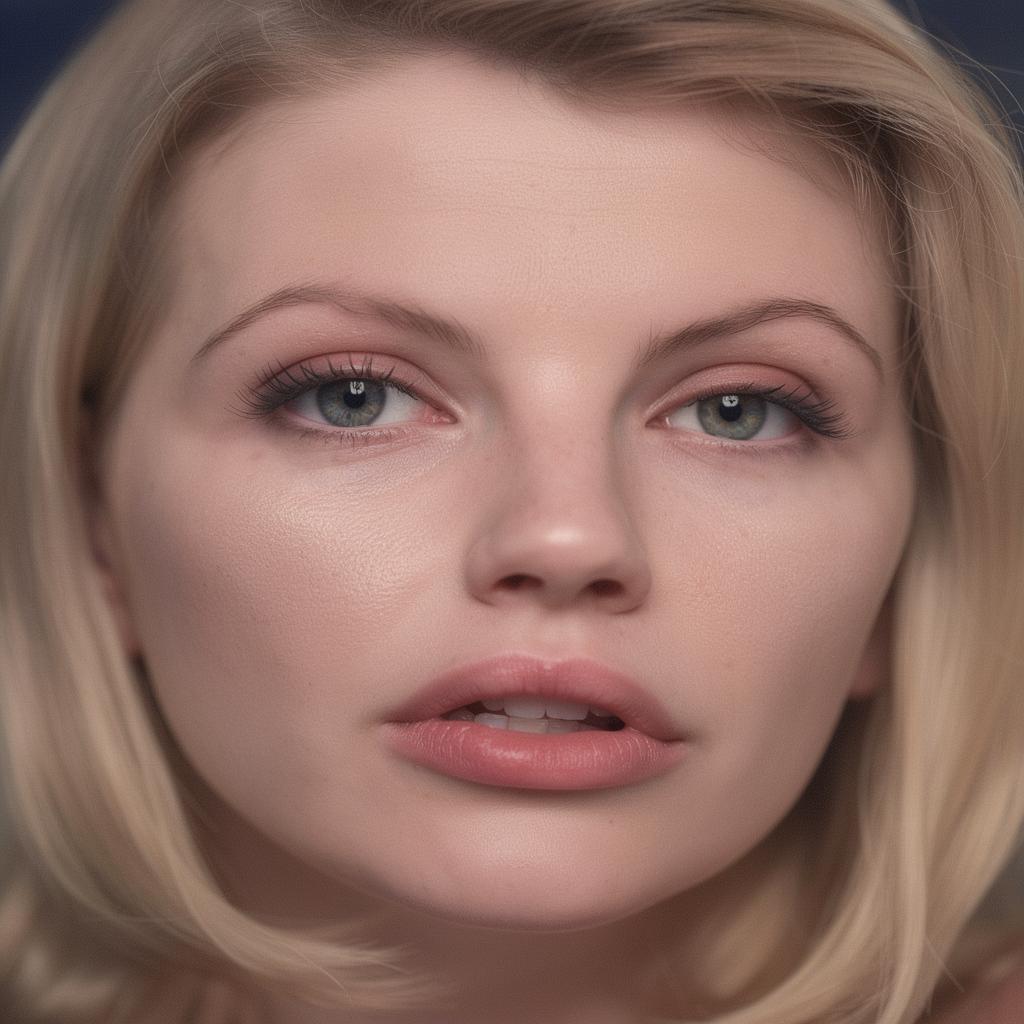} &
\includegraphics[width=0.10\linewidth]{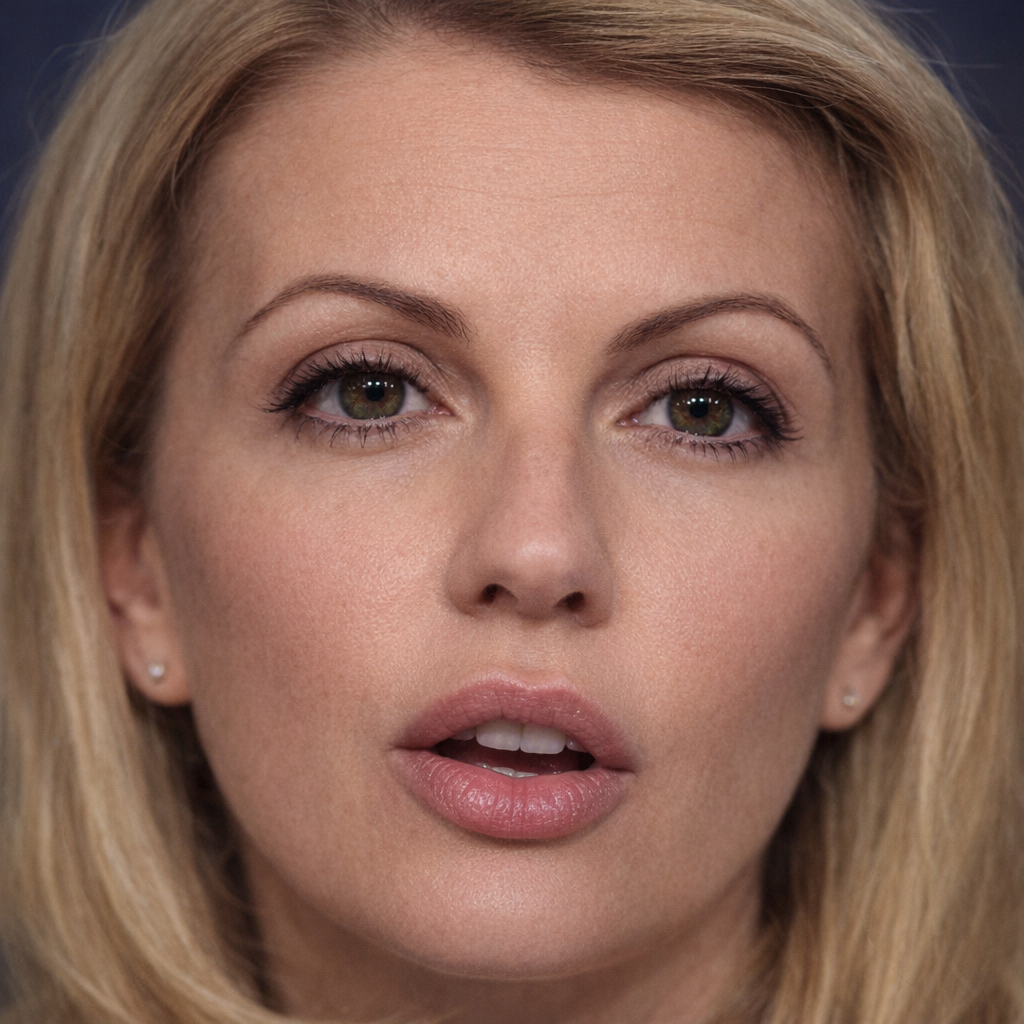} &
\includegraphics[width=0.10\linewidth]{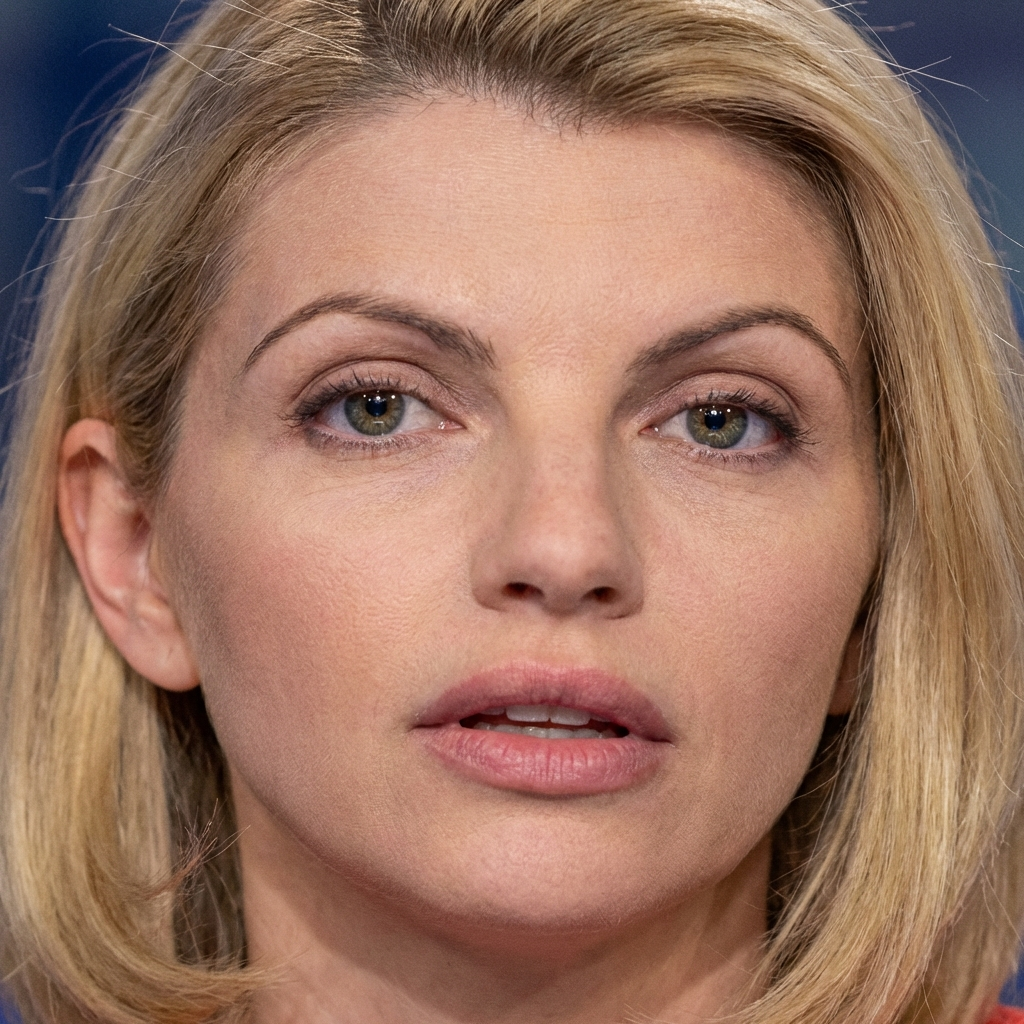} \\
\bottomrule
\end{tabular}
\end{figure*}

We evaluate semantic consistency using the IPR, defined in Section~\ref{sec:metrics}, measured by AWS CompareFaces~\cite{AWS}. 
Figure~\ref{fig:semantic_preservation} and \ref{fig:qual_3x9} reveal three key behaviors.
First, Figure~\ref{fig:semantic_preservation} highlights a substantial improvement from Qwen-v1 to Qwen-v2. 
Qwen-v2 achieves IPR comparable to commercial systems, indicating increased sensitivity to the ``preserve identity'' instruction embedded in our prompts. 
This improvement aligns with the model’s ability to perform detailed texture refinement while maintaining identity.
Second, while Gemini and Flux AI maintain consistently high IPRs, ChatGPT exhibits a sharper decline.
As corroborated by Figure~\ref{fig:qual_3x9}, ChatGPT’s outputs are closer to re-generation than editing, introducing noticeable facial geometry changes despite high perceptual quality.
Finally, Figure~\ref{fig:semantic_preservation} shows that the generic IAP preserves identity slightly better than the adaptive ISP. 
We attribute this effect to the more targeted nature of ISP. 
By explicitly emphasizing biometric landmarks, ISP encourages modifications to regions that are also salient for face recognition. 
In contrast, IAP operates as a holistic refinement, improving realism while largely preserving the underlying facial structure.

\subsubsection{Extension to AI-Generated Image Refinement}

Table~\ref{tab:ai_results} extends our analysis beyond faces to general AI-generated images.
All original images are synthetic samples from Tiny-GenImage~\cite{tinygenimage_kaggle}, and are further re-generated using P4 and P5 by lightly adapting face-specific wording to cover generic objects and scenes.
Across all settings, refinement consistently reduces DRs compared to the original images.
The degradation is particularly pronounced for Hive-AI, where ChatGPT refinement under P4 reduces the DR from 83\% to 14\% at $\tau_{99}$.
These results indicate that our \emph{semantic-preserving image refinement} can substantially weaken AI-generated image detectors even outside the facial domain.

\begin{table}[t]
\centering
\caption{DRs (\%) of AI-generated image detectors on original images and images refined by commercial GAI systems.}
\label{tab:ai_results}
\setlength{\tabcolsep}{4.5pt}
\footnotesize
\begin{tabular}{l cc | cc cc | cc cc}
\toprule
\multirow{3}{*}{\textbf{Detector}} &
\multicolumn{2}{c|}{\cellcolor{gray!10}\textbf{Original}} &
\multicolumn{4}{c|}{\textbf{ChatGPT}} &
\multicolumn{4}{c}{\textbf{Gemini}} \\

\cmidrule(lr){2-3}
\cmidrule(lr){4-7}
\cmidrule(lr){8-11}

& \multicolumn{2}{c|}{\cellcolor{gray!10}\textbf{N/A}} &
\multicolumn{2}{c}{\textbf{P4}} &
\multicolumn{2}{c|}{\textbf{P5}} &
\multicolumn{2}{c}{\textbf{P4}} &
\multicolumn{2}{c}{\textbf{P5}} \\

\cmidrule(lr){2-3}
\cmidrule(lr){4-5}
\cmidrule(lr){6-7}
\cmidrule(lr){8-9}
\cmidrule(lr){10-11}

& \cellcolor{gray!10}$\tau_{99}$ & \cellcolor{gray!10}$\tau_{90}$ &
$\tau_{99}$ & $\tau_{90}$ &
$\tau_{99}$ & $\tau_{90}$ &
$\tau_{99}$ & $\tau_{90}$ &
$\tau_{99}$ & $\tau_{90}$ \\

\midrule
D$^3$ &
\cellcolor{gray!10}50 & \cellcolor{gray!10}96 &
18 & 67 &
10 & 59 &
36 & 72 &
39 & 81 \\

\midrule
UnivFD &
\cellcolor{gray!10}38 & \cellcolor{gray!10}50 &
8 & 20 &
12 & 27 &
23 & 39 &
33 & 46 
 \\

\midrule
Hive-AI &
\cellcolor{gray!10}83 & \cellcolor{gray!10}94 &
14 & 90 &
28 & 86 &
38 & 86 &
41 & 83 \\

\bottomrule
\end{tabular}
\end{table}

\section{Discussion}
\subsection{Why Detectors Fail?}
Our findings suggest that the weaknesses observed in contemporary deepfake detectors are not solely attributable to isolated implementation flaws, but are also influenced by broader assumptions regarding how authenticity cues are defined, surfaced to users, and subsequently used in practice.
In particular, many detectors implicitly assume a relatively stable set of forensic signals, while recent GAI-based image editing pipelines enable users to modify facial attributes in ways that can attenuate or alter such signals.
This discrepancy indicates that detector failures arise not merely from adversarial manipulation, but from shifts in user-driven editing practices that were not anticipated during detector design.
Rather than challenging detector correctness in isolation, our analysis highlights the fragility of detection outcomes under realistic usage scenarios involving GAI-guided image modification.

\subsection{Potential Threats}
Our findings suggest that workflows leveraging existing GAI image editing capabilities can enable facial modifications that largely preserve high-level semantic attributes, including identity-related information.
While such outcomes may be beneficial in benign use cases, they may also raise concerns in certain security-sensitive contexts.

In particular, semantically consistent facial editing, as observed in our study, may have implications for downstream scenarios in which generated or edited images are reused as inputs, such as studies that employ image generation models as sub-components in attacks~\cite{kim2024scores, kim2025non} on face recognition systems.
We do not evaluate the above attacks in this work, nor do we claim the existence of a concrete attack.
Rather, we highlight this possibility to illustrate a direction that may warrant further investigation by the research community.

\subsection{Blind Spots in Existing AI Safety}
\label{subsec:blind}
Contemporary AI safety paradigms and institutional usage policies~\cite{Inan2023LlamaGL, openai_usage_policy, eu2024aiact, nist_ai_rmf_1_0} incorporate multiple safeguards to regulate generative systems, most notably through content-based filtering and prompt-level intent assessment.
In some deployed systems, additional heuristic mechanisms are used to identify known unsafe or misuse-related prompts.
However, these measures are not formalized as comprehensive interaction-level policies and are typically applied at the level of individual inputs or outputs.
As summarized in Table~\ref{tab:policy-dimensions}, risk assessment is typically grounded in whether an output violates prohibited content categories or whether a single instruction explicitly signals malicious intent.

Our findings highlight risks that fall outside both dimensions.
This gap is substantiated by our automated safety filtering analysis (Appendix~\ref{app:safety}),
which shows that both text- and image-based deepfake-related inputs are overwhelmingly classified as benign.
These results suggest that such interactions are weakly aligned with the categories operationalized by current automated safety systems.
They do not involve prohibited content, nor do they rely on explicitly malicious intent at any individual interaction.
Instead, the observed risks arise from how otherwise permitted generative capabilities are combined across usage contexts, becoming visible only when considering how outputs from logically or operationally independent interactions are composed and reused.
For example, a model may provide authenticity critiques, localized artifact descriptions, and benign refinement suggestions across separate interaction contexts, none of which are individually disallowed under existing policies.
By making this gap explicit, our work draws attention to a class of emergent risks that are difficult to capture within current content- or intent-based safety frameworks.

\subsection{Usability vs. Vulnerability}

A key tension revealed by our observations lies in the trade-off between usability and vulnerability.
In principle, certain vulnerabilities identified in this work could be mitigated by introducing additional restrictions, such as limiting image editing prompts, filtering input images through deepfake detection modules, or enforcing explicit indicators that generated images are synthetic.

However, each of these approaches introduces significant challenges.
Relying on deepfake detectors as gatekeepers requires strong assumptions about detector reliability.
Moreover, restricting input images or editing operations can impose substantial constraints on user workflows, undermining the flexibility and expressiveness that motivate the adoption of general-purpose GAI systems.
Similarly, watermarking strategies face inherent trade-offs: invisible watermarks may fail to persist in offline or downstream reuse, while visible markers can degrade user experience and reduce adoption.

These considerations highlight a fundamental difficulty.
The risks we observe are challenging to mitigate not because existing safety mechanisms are absent, but because they arise from benign inputs, naturalistic outputs, and interaction patterns that are central to system usability.
As a result, such risks can remain difficult to detect even within the GAI system itself, while still posing meaningful threats in downstream application contexts.
Rather than prescribing a specific mitigation strategy, our work aims to surface this tension and provide an empirical basis for future research on how to balance usability and security in open-ended, interaction-driven generative systems.

\subsection{Limitations of Our Work}
This study has several limitations that should be considered when interpreting our findings.
First, the extraction of assessment criteria and the execution of image editing rely on natural-language prompts, which do not always admit the correspondence with the intended operational semantics.
As a result, the behaviors we observe should be interpreted as outcomes of prompt-level interactions, rather than as precisely controlled manipulations.

Second, we do not examine the internal mechanisms of image editing models or deepfake detectors.
Accordingly, our analysis remains at the level of observable system behavior and does not attribute the observed effects to specific architectural or training choices.

These limitations delineate the scope of our study and suggest promising directions for future work, including efforts to bridge prompt-level behaviors with model-level analyses and to better characterize how semantic editing practices interact with downstream detection mechanisms.

\section{Conclusion}
In this work, we demonstrate that the naïve exposure of reasoning and refinement capabilities in general-purpose Generative AI (GAI) systems undermines the reliability of modern deepfake detection frameworks. We introduce a realistic threat model showing that adversaries can exploit the interaction between authenticity assessment and semantic-preserving refinement, repurposing system feedback to remove detection cues while preserving identity and expression.

Our results reveal a structural mismatch between current detection paradigms and the capabilities of deployed GAI systems. In particular, commercial services pose greater security risks than open-source alternatives due to their advanced reasoning abilities, alignment-driven assistance, and accessible interfaces. We further show that existing AI safety strategies fail to capture this risk, as they primarily block explicit adversarial intents but overlook reasoning-guided refinement framed as benign image editing.

Overall, our findings suggest that treating deepfake detection as a static classification problem is increasingly insufficient. Future defenses must address the interaction-driven risks introduced by general-purpose GAI systems.

\cleardoublepage
\appendix
\section*{Acknowledgments}
We thank anonymous reviewers for their valuable comments.
\section*{Ethical Considerations}

This work studies the behavior of contemporary deepfake detection systems under realistic, user-facing interactions with general-purpose generative AI systems, including but not limited to image editing workflows.
Our goal is to advance understanding of how detection assumptions interact with flexible and open-ended GAI usage patterns, and to inform the design and evaluation of more robust media authentication systems.
Given the dual-use nature of research on detection limitations, we discuss relevant ethical considerations below.

\paragraph{Research Context and Purpose.}
The purpose of this study is to systematically examine how deepfake detectors behave when exposed to diverse forms of user-driven GAI utilization, where image generation, modification, and assessment are freely combined.
By analyzing detector responses under controlled yet open-ended interaction settings, we aim to clarify the boundary conditions under which current evaluation assumptions hold or break down.
Our findings are intended to support the research community and system designers in developing more comprehensive threat models and evaluation methodologies, rather than to undermine or discredit existing detection approaches.

\paragraph{Potential for Misuse.}
As our work provides a detailed analysis of detector behavior under specific GAI-enabled interaction patterns, there exists a possibility that parts of our findings could be misused or overinterpreted in adversarial contexts.
At the same time, we believe that credible and reproducible security research requires a sufficient level of technical specificity.
Accordingly, we describe our experimental setups and observations in a concrete and transparent manner, while avoiding the presentation of automated attack pipelines, optimization procedures, or step-by-step evasion strategies.
We further note that the scope and granularity of publicly released implementation details may be adjusted based on reviewer feedback and community guidance, in order to balance reproducibility with responsible disclosure.

\paragraph{Scope and Access Assumptions.}
All experiments in this study are conducted under black-box access assumptions that reflect typical user-level interactions with deployed generative AI systems and deepfake detectors.
We do not access or infer internal model parameters, training data, or proprietary detection logic.
As such, our analysis focuses on observable system behavior arising from open-ended GAI access, rather than on internal vulnerabilities tied to specific architectures or implementations.

\paragraph{Human Subjects and Data Sources.}
This study does not involve human subjects or user studies.
All experiments are conducted using publicly available deepfake datasets, AI-generated image datasets, and facial image datasets commonly used in prior research.
No experiments target or profile specific individuals, and no personally identifiable information is collected or introduced beyond what is already present in these public research datasets.

\paragraph{Responsible Reporting.}
We take care to present our findings with appropriate scope and qualification, avoiding claims about concrete or universally effective attacks.
Instead, we emphasize how our observations inform the limitations of current evaluation practices under flexible, user-facing GAI usage conditions.
We believe that presenting these results transparently, while remaining attentive to their broader implications, contributes positively to the responsible development and deployment of deepfake detection technologies.

Overall, we believe that the scientific value of examining deepfake detector behavior under open-ended GAI access outweighs the potential risks, and that this work aligns with the ethical expectations and standards of the USENIX Security community.

\section*{Open Science}
\label{sec:openscience}

All experiments reported in this paper are conducted using publicly available datasets and accessible generative AI systems.
No proprietary datasets, private user data, or internally disclosed system information are used at any stage of the study.
The purpose of this section is to support the credibility of the reported findings.

All datasets used for experimental analysis are obtained from public benchmarks that have been widely adopted in prior research.
These datasets are distributed under licenses that permit academic use, and we adhere to the usage conditions specified by the dataset providers.
Information regarding dataset sources, access links is provided in the accompanying open science repository (\url{https://anonymous.4open.science/r/Naive-GAI-Exposure-Deepfake-9517/}) and in the reference list.

The generative AI systems evaluated in this study are treated strictly as black-box systems, as no internal model details, training data, or implementation specifics are available to the authors.
Accordingly, we avoid making claims about internal mechanisms or design choices of these systems and focus solely on their observable behaviors under controlled and explicitly documented interaction protocols.

To facilitate verification, we provide code necessary to reproduce the experimental procedures and analyses reported in this paper through the same open science repository. 
The scope of the released materials is intentionally limited to what is necessary for assessing the reliability of the findings presented in this paper.
The extent of disclosure may be adjusted in response to reviewer feedback or ethical considerations, as unrestricted release is not always aligned with responsible or meaningful reproducibility.

This study does not involve human subjects or user studies.
All experiments are conducted in accordance with the publicly stated usage policies of the evaluated systems.

\bibliographystyle{plainurl}
\bibliography{ref}
\cleardoublepage
\section{Detailed Prompt Specifications}
\label{app:prompts}

This appendix provides a comprehensive description of all prompts used in the experimental pipeline.
While Section~\ref{subsec:refinement_regimes} introduces the overall prompt design at a high level, the goal of this appendix is to document the exact formulations and structural variations of the prompts for transparency and reproducibility.
The prompts are organized according to the functional stages of interaction: (i) criteria elicitation, (ii) image-level forensic analysis, and (iii) image refinement.

\subsection{Criteria Elicitation}
\label{app:prompt-criteria}

The first stage aims to elicit the model's internal authenticity assessment criteria in a task-agnostic manner.
Rather than providing any image, the model is prompted to articulate the general principles it would rely on when judging whether a face image is AI-generated, manipulated, or real.
This interaction serves as the foundation for subsequent stages, as the articulated criteria are later reused both explicitly and implicitly in downstream prompts.

\begin{quote}
\ttfamily \footnotesize \textbf{User:} You are an expert AI-generated (Deepfake detection) face image detector.\\
What is your key criteria for judgement?
\end{quote}

\subsection{Image Analysis}
\label{app:prompt-analysis}

In the second stage, the model is asked to perform a detailed forensic analysis of a given portrait image.
The prompt explicitly structures the analysis into classification, attribute-level visual inspection, and a concise final summary.
This structured format encourages the model to externalize fine-grained visual cues and produce a rationale that can be directly reused in later refinement stages.

\begin{quote}
\ttfamily \footnotesize
\textbf{User:} You are an expert AI-generated face detector and forensic image analyst.
Analyze the attached portrait by following these internal steps and provide a detailed report:

\textbf{Classification:} Determine whether the image is:
\begin{itemize}
  \item ``AI-generated'' (purely synthetic)
  \item ``Deepfake'' (identity swapped or digitally manipulated)
  \item ``Real'' (a genuine, unedited camera photograph)
\end{itemize}

\textbf{Detailed Forensic Analysis:} Describe the concrete visual cues for each of the following aspects:
\begin{itemize}
  \item \textbf{Skin:} Texture, pore consistency, and light diffusion.
  \item \textbf{Eyes:} Iris patterns, symmetry, and the logic of catchlights (specular reflections).
  \item \textbf{Hair:} Strands definition, interaction with the background, and stylistic consistency.
  \item \textbf{Facial Edges:} Contours of the jawline, nostrils, lips, and eyelids.
  \item \textbf{Background:} Depth-of-field logic, bokeh quality, and any digital artifacts or warping.
\end{itemize}

\textbf{Final Summary:} Conclude with 1--3 sentences highlighting the most decisive cues that led to your final verdict, including a confidence score for your assessment.
\end{quote}

\subsection{Image Refinement}
\label{app:prompt-refinement}

The final stage involves refining a given portrait image to improve perceptual realism while preserving the subject's identity.
We consider two refinement settings: instance-agnostic prompts, which apply a uniform instruction across all images, and instance-specific prompts, which incorporate image-dependent forensic rationales.

\subsubsection{Instance-Agnostic Prompts}
\label{app:prompt-refine-agnostic}

We design a sequence of instance-agnostic prompts (P1--P5) with progressively increasing specificity.
The intent of this progression is to examine how increasingly explicit realism constraints affect the refinement outcome.
P1 represents the most conservative setting, aiming to preserve the original image as closely as possible, while P2--P4 gradually introduce stronger notions of naturalness and photographic realism.
The most detailed prompt, P5, is presented and discussed in Section~4.2 of the main text and is therefore not repeated here.

\begin{quote}
\ttfamily \footnotesize \textbf{P1:} same photo, identical composition
\end{quote}

\begin{quote}
\ttfamily \footnotesize \textbf{P2:} natural
\end{quote}

\begin{quote}
\ttfamily \footnotesize \textbf{P3:} natural camera photograph
\end{quote}

\begin{quote}
\ttfamily \footnotesize \textbf{P4:} Refine this portrait to look like a natural camera photograph.
\end{quote}

\subsubsection{Instance-Specific Prompt Construction}
\label{app:prompt-refine-specific}

In the instance-specific setting, the refinement prompt is dynamically constructed by augmenting a shared base instruction with image-dependent forensic rationales obtained from the analysis stage.
This design allows the model to directly reuse its own articulated authenticity criteria as refinement objectives, while explicitly constraining identity preservation.

The base instruction is fixed across all instances:
\begin{quote}
\ttfamily \footnotesize
\textbf{Base prompt:} Refine this portrait to look like a natural camera photograph.
\end{quote}

For each image, the issues detected during forensic analysis are appended as a bullet-point list.
Conceptually, the final prompt takes the following form:
\begin{quote}
\ttfamily \footnotesize
The following issues were detected in the image. Please correct them while preserving the subject's identity:
\begin{itemize}
  \item Issue \#1 extracted from the forensic rationale
  \item Issue \#2 extracted from the forensic rationale
  \item $\dots$
\end{itemize}
\end{quote}

\noindent
Together, these prompts operationalize a refinement pipeline in which authenticity assessment criteria are first externalized, then instantiated at the image level, and finally reused as explicit optimization targets under semantic-preserving constraints.

\section{Safety Filtering Analysis}
\label{app:safety}

\begin{table}[h]
\centering
\footnotesize
\renewcommand{\arraystretch}{1.15}
\setlength{\tabcolsep}{6pt}
\caption{Risk dimensions in existing AI safety policies.}
\begin{tabularx}{\linewidth}{@{}>{\RaggedRight\arraybackslash}p{0.36\linewidth}
                             >{\RaggedRight\arraybackslash}X@{}}
\toprule
\textbf{Evaluation Dimension} &
\textbf{Operational Focus in Policies} \\
\midrule
Synthesized Output &
Categorical filtering of prohibited content, such as sexual, violent, non-consensual imagery; hate speech; or illegal activities. \\
\midrule
Prompt-level Intent &
Detection of explicit malicious or illicit intent from surface-level wording in individual user instructions. \\
\bottomrule
\end{tabularx}
\label{tab:policy-dimensions}
\end{table}

This appendix analyzes how the interactions and artifacts examined in the main text are treated by contemporary automated safety filtering systems.
To contextualize this analysis, Table~\ref{tab:policy-dimensions} summarizes the primary evaluation dimensions commonly operationalized in current AI safety policies.
These frameworks typically regulate generative systems by filtering outputs that fall into predefined prohibited content categories or by detecting explicit malicious intent expressed within individual user prompts.
While such dimensions capture many well-established misuse scenarios, they are largely defined at the level of isolated inputs or outputs rather than multi-step interaction workflows.

Building on this policy framing, the remainder of this appendix empirically examines whether text-based interactions (e.g., criteria elicitation, image analysis, and refinement prompts) and image-based artifacts (original deepfake images and their refined variants) activate content- or intent-based safety signals under deployed filtering infrastructures.

Notably, all experiments reported in the main text were conducted without interruption or rejection by built-in safety mechanisms, suggesting that these interactions were generally treated as benign by internal filtering layers.
Nevertheless, to avoid relying solely on this implicit signal, we perform an explicit and systematic analysis of automated safety filtering behavior.
The goal of this analysis is not to evaluate policy compliance or normative correctness, but to empirically characterize how deepfake-related workflows are classified by existing safety mechanisms.
By doing so, this appendix provides supporting evidence for the observations discussed in Section~\ref{subsec:blind} regarding the limited alignment between current safety taxonomies and risks that emerge through the composition and reuse of otherwise permitted generative capabilities.

\subsection{Safety Filtering Infrastructures}
\label{app:safety-infra}

We evaluate two widely used automated safety filtering services: the OpenAI Moderation API~\cite{openai_moderation} and Azure AI Content Safety~\cite{azure_safety}.
Although both systems target overlapping categories of harmful or restricted content, they differ substantially in architectural design, modality support, and output representation.
To avoid conflating heterogeneous taxonomies and scoring schemes, results are reported separately for each service throughout this appendix.

\begin{table*}[t]
\centering
\caption{Automated safety filtering services used in our experimental setup.}
\label{tab:safety-infra}
\begin{tabular}{l l l}
\toprule
\textbf{Service} & \textbf{Provider} & \textbf{Primary Capability} \\
\midrule
\mylogo{logo/openai.png} Moderation API & OpenAI & Text-based safety filtering with optional multimodal extension \\
\mylogo{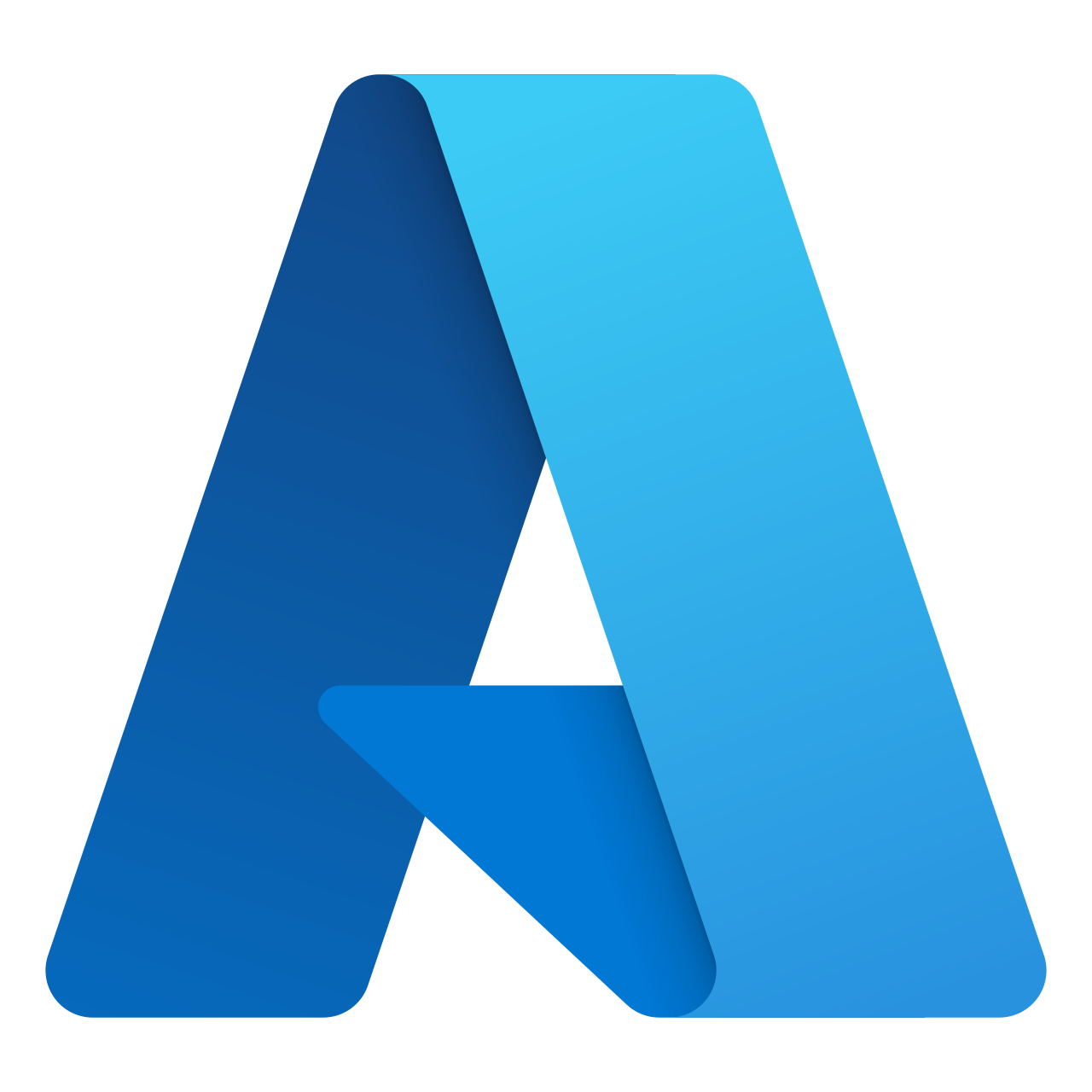} AI Content Safety & Microsoft Azure & Native text and image safety filtering \\
\bottomrule
\end{tabular}
\end{table*}

\subsection{Safety Evaluation Configuration}
\label{app:safety-config}

All safety filtering checks are performed using the default configurations provided by each service.
For OpenAI-based evaluation, we use the \texttt{omni-moderation-latest} endpoint, which evaluates textual inputs and returns both binary flagging decisions and continuous category scores.
All inputs are submitted as standalone text, without custom thresholding, category weighting, or post-processing beyond the API defaults.

While OpenAI safety filtering can be extended to visual content by first generating textual descriptions of images and subsequently moderating those descriptions, we do not adopt this approach in our experiments.
This design choice avoids introducing an additional interpretation layer that may confound image-level safety assessment.

For Azure-based evaluation, we use Azure AI Content Safety with its standard \textit{Analyze Text} and \textit{Analyze Image} endpoints.
Azure directly evaluates textual and visual inputs and reports ordinal severity levels on a discrete scale from 0 to 7 for each category, where a severity of 0 indicates no detectable risk.
All evaluations use the default service configuration, without tuning severity thresholds or category definitions.

Based on these considerations, text-based interactions are evaluated using both services, whereas image-based safety filtering is conducted using Azure AI Content Safety.

\subsection{Text-Based Safety Filtering}
\label{app:safety-text}

This subsection reports safety filtering outcomes for text-only interactions used throughout our study.
We consider four representative prompt types corresponding to different stages of the GAI-based deepfake analysis pipeline:
(i) a criteria elicitation prompt,
(ii) an image analysis request prompt,
(iii) an image refinement prompt formulated independently of any specific image, and
(iv) an image refinement prompt targeting a specific image.
All prompts are evaluated in isolation, without accompanying visual inputs.

Table~\ref{tab:openai-text-all} summarizes results obtained using the OpenAI Moderation API, while Table~\ref{tab:azure-text-all} reports corresponding outcomes from Azure AI Content Safety.

\begin{table}[t]
\centering
\caption{OpenAI safety filtering outcomes for text-based prompts using \texttt{omni-moderation-latest}.(No prompt was flagged by the moderation system.)}
\label{tab:openai-text-all}
\begin{tabular}{l c}
\toprule
\textbf{Prompt Type} &
\textbf{Max Signal} \\
\midrule
Criteria Elicitation & Illicit (0.0166) \\
Image Analysis & Sexual ($2.6\times10^{-5}$) \\
\makecell[l]{Image Refinement\\(Instance-Agnostic)} & Violence ($5.2\times10^{-4}$) \\
\makecell[l]{Image Refinement\\(Instance-Specific)} & Violence ($5.2\times10^{-4}$) \\
\bottomrule
\end{tabular}
\end{table}

\begin{table}[t]
\centering
\caption{Azure AI Content Safety outcomes for text-based prompts.}
\label{tab:azure-text-all}
\begin{tabular}{l c}
\toprule
\textbf{Prompt Type} &
\textbf{Max Severity} \\
\midrule
Criteria Elicitation &
0 \\
Image Analysis&
0 \\
Image Refinement (Instance-Agnostic) &
0 \\
Image Refinement (Instance-Specific) &
0 \\
\bottomrule
\end{tabular}
\end{table}

\paragraph{Interpretation.}
Across all evaluated text-based prompts, both safety filtering services consistently classify the interactions as benign.
Azure AI Content Safety assigns a severity level of zero to all prompt types, while the OpenAI Moderation API does not flag any input and returns uniformly low category scores.
These findings indicate that text-only prompts related to deepfake assessment, structured forensic reasoning, and photorealistic image refinement fall well outside the scope of automated safety filtering, despite explicitly referencing faces, identity, and image manipulation.

\subsection{Image-Based Safety Filtering}
\label{app:safety-image}

This subsection reports image-level safety filtering outcomes obtained using Azure AI Content Safety.
We evaluate the same set of facial deepfake images used in the main experiments, together with their refined variants produced by different image improvement systems.
All images are analyzed directly as visual inputs, without accompanying textual context. Table~\ref{tab:image-safety} reports outcomes from Azure AI Content Safety(Image Analysis).

\paragraph{Evaluated images.}
The evaluation set consists of 100 facial deepfake images used throughout the main text.
In addition to the original deepfake images, we analyze refined versions generated by multiple image improvement systems, with each refined set containing one-to-one counterparts of the original inputs.
All images are evaluated independently using the \textit{Analyze Image} endpoint under the default Azure AI Content Safety configuration. 

\begin{table}[t]
\centering
\caption{Image-based safety filtering outcomes using Azure AI Content Safety. For each image set, we report the proportion of images exhibiting any non-zero severity and the maximum observed severity across all categories.}
\begin{tabular}{l c c}
\toprule
\textbf{Image Set} &
\textbf{\# of Severity $>0$} &
\textbf{Max Severity} \\
\midrule
Original & 0 (0.00\%) & 0 \\
Flux AI & 0 (0.00\%) & 0 \\
ChatGPT & 1 (1.00\%) & 2 \\
Gemini & 0 (0.00\%) & 0 \\
Qwen-v1 & 2 (2.00\%) & 2 \\
Qwen-v2 & 1 (1.00\%) & 2 \\
\bottomrule
\end{tabular}
\label{tab:image-safety}
\end{table}

\paragraph{Interpretation.}
Across all evaluated image sets, the vast majority of images receive a severity level of zero across all safety categories.
For refined images generated by certain systems, a small number of images exhibit non-zero severity scores; however, these cases are limited to low-level signals (severity 2) in the \emph{Violence} category.
No image triggers higher-severity classifications, and no moderation failures are observed during evaluation.

Overall, the image-level results indicate that facial deepfake images and their refined variants are rarely identified as safety-relevant content by current image-based filtering mechanisms.
The observed non-zero signals appear sporadic and do not systematically correspond to identity manipulation or facial synthesis artifacts.

\section{Semantic Preservation with Alternative Face Recognition APIs}
\label{app:semantic_tencent}

To assess the robustness of the semantic preservation results reported in Section~\ref{subsubsec:semantic}, 
we replicate the identity preservation analysis using an alternative commercial face recognition API, 
Tencent CompareFace~\cite{Tencent}.
This experiment follows the same protocol as the main evaluation, differing only in the backend used to compute identity similarity scores.

\begin{figure}[t]
    \centering
    \includegraphics[width=\columnwidth]{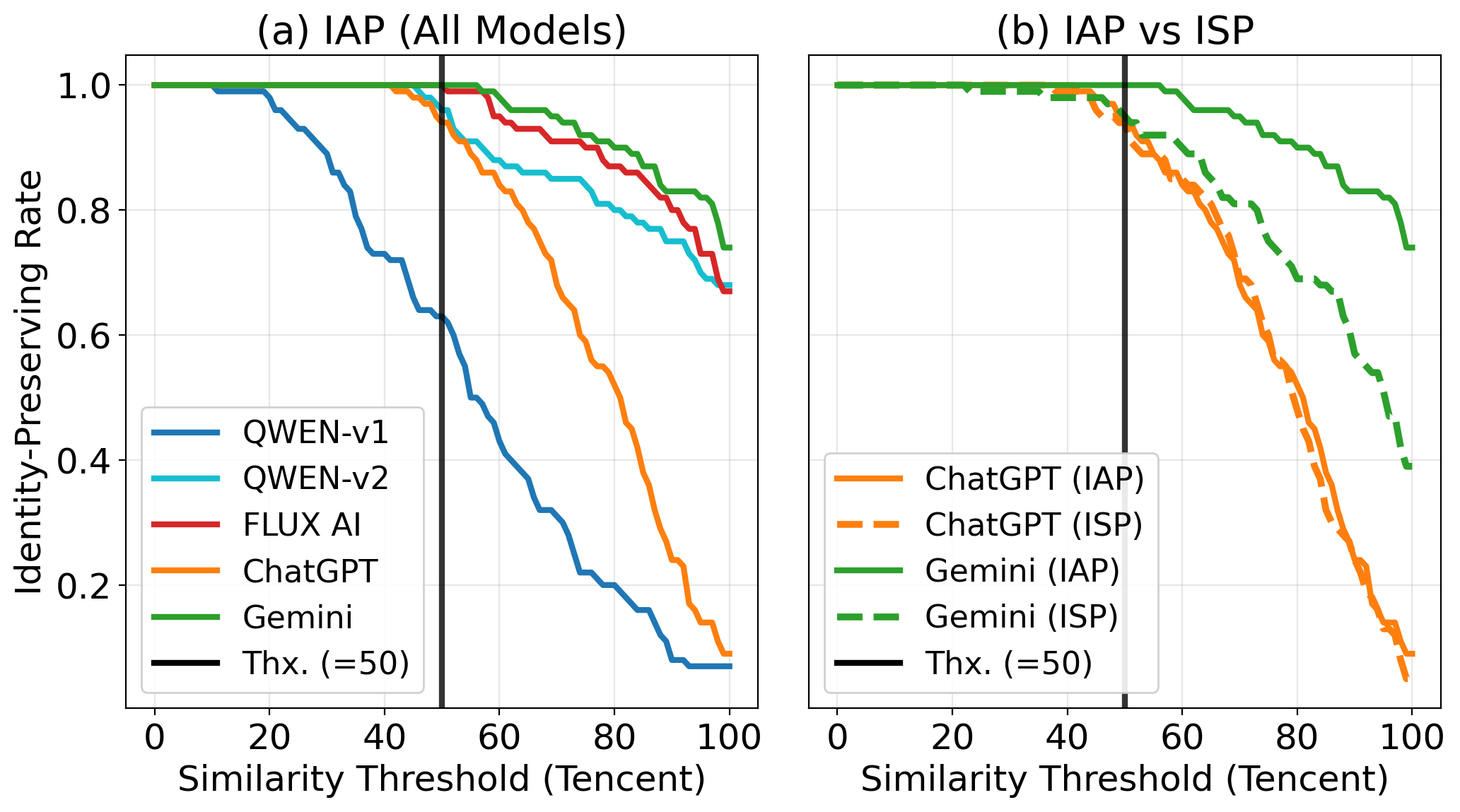}
    \vspace{-6mm}
    \caption{Semantic preservation with IAP and ISP on Tencent CompareFace. (Note that the default thresholds in Tencent CompareFace is $50$)}
    \label{fig:semantic_preservation_tencent}
    \vspace{-3mm}
\end{figure}

Figure~\ref{fig:semantic_preservation_tencent} reports the identity preservation rate (IPR) measured with Tencent CompareFace. The results exhibit trends that are highly consistent with those reported in the main analysis.
Across all models and refinement strategies, the relative ordering of identity preservation is preserved, and no qualitative deviations are observed.
In particular, models that demonstrate strong semantic preservation in the main experiment continue to do so, while models that exhibit larger identity drift show similar behavior in this additional evaluation.

Likewise, the comparative behavior between the two refinement strategies remains unchanged.
These observations indicate that the semantic preservation characteristics analyzed are stable and reproducible, rather than artifacts of a specific evaluation configuration.
Overall, this additional analysis serves to further reinforce the conclusions drawn from the main semantic preservation experiment.

\section{Illustrative Examples of Criterion Articulation and Application}
\label{app:criteria_examples}

This appendix provides illustrative examples supporting the findings discussed in the main text by visualizing two sequential stages of GAI-based facial deepfake assessment:
(i) the articulation of assessment criteria without image input, and
(ii) the application of those criteria to a facial image.
The purpose of these examples is to demonstrate how GAI systems express and operationalize deepfake detection criteria through natural-language interaction, rather than to evaluate detection accuracy.
Importantly, the examples highlight how the criteria articulated in abstract terms are selectively instantiated when producing image-level judgments and explanations.

\subsection{Criterion Articulation Without Image Input}
\label{app:criteria_articulation}

\begin{figure*}[!]
    \centering
    \includegraphics[width=\linewidth]{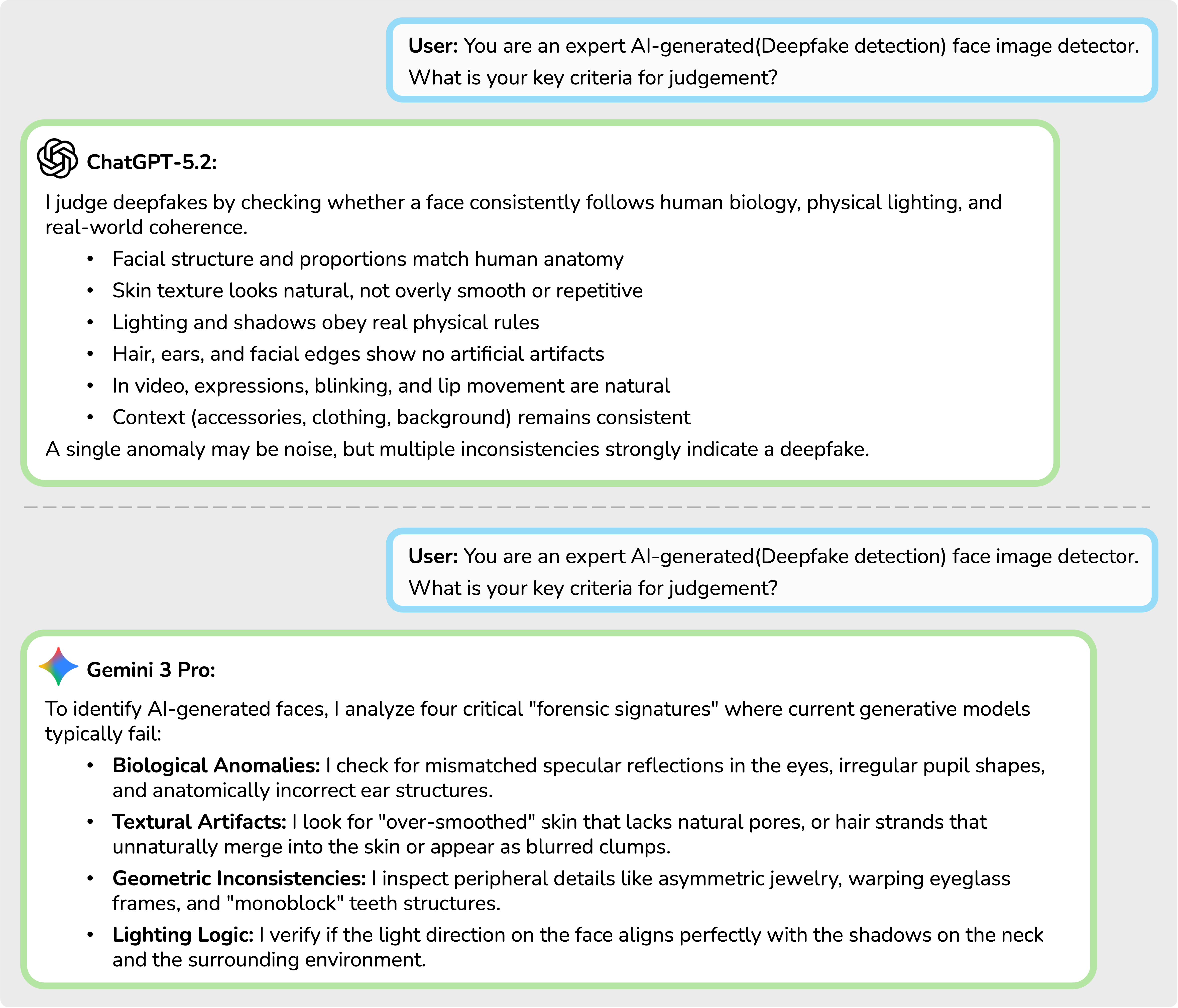}
    \caption{\textbf{Criterion articulation without image input.}
    GAI systems describe their key criteria for facial deepfake detection in the absence of any image input.}
    \label{fig:criteria_articulation}
\end{figure*}

Figure~\ref{fig:criteria_articulation} illustrates how GAI systems articulate high-level assessment principles prior to observing any image.
The criteria are framed as general-purpose forensic heuristics, establishing what the system later treats as relevant evidence for deepfake detection.

\subsection{Criterion-Based Image Analysis}
\label{app:criteria_application}

\begin{figure*}[!]
    \centering
    \includegraphics[width=\linewidth]{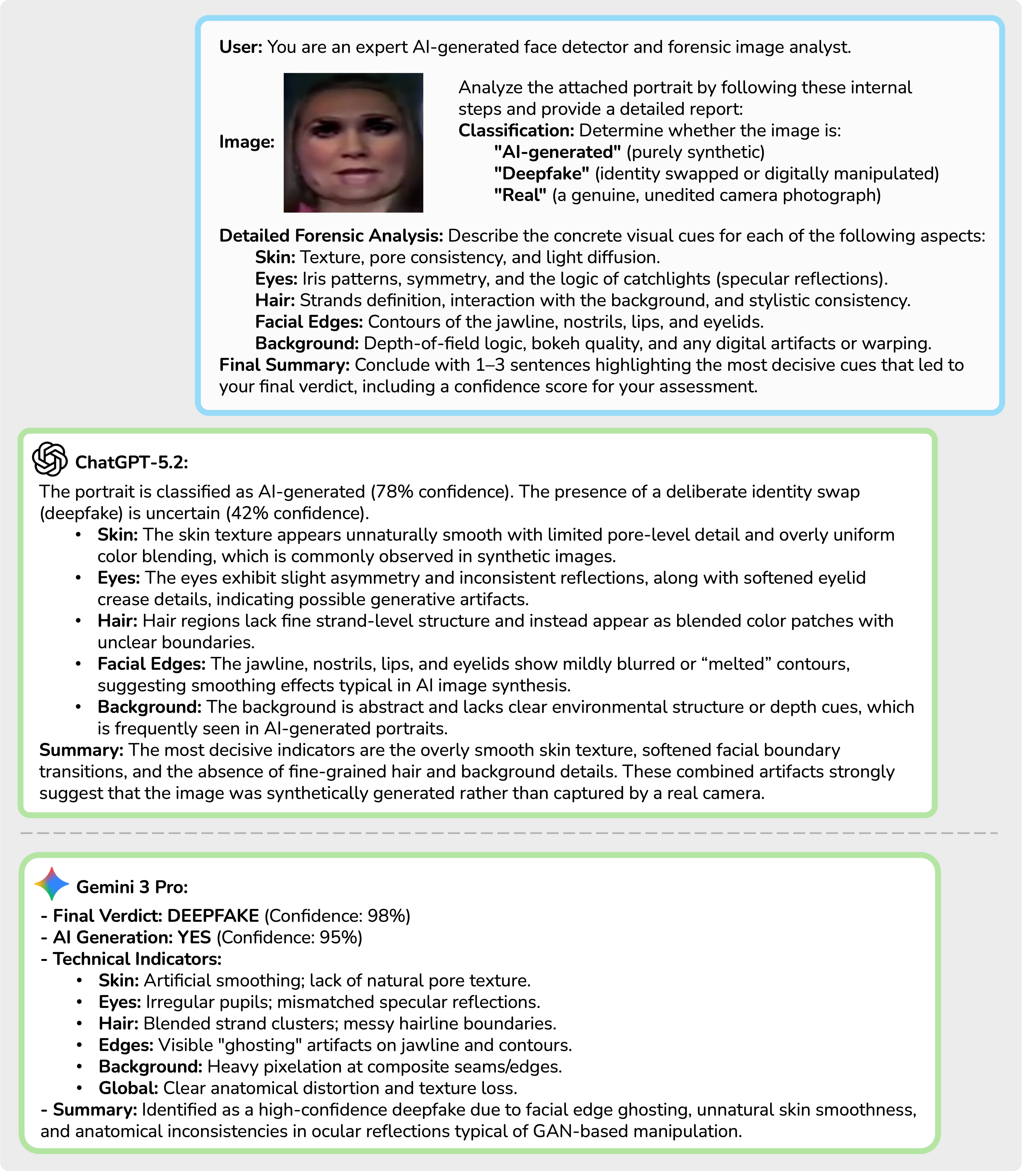}
    \caption{\textbf{Criterion-based image analysis.}
    A GAI system applies the articulated criteria to analyze a facial image that is a known deepfake, producing a structured forensic explanation and final judgment.}
    \label{fig:criteria_application}
\end{figure*}

Figure~\ref{fig:criteria_application} shows how the previously articulated criteria are operationalized during image analysis.
The system selectively maps abstract principles onto concrete visual cues, using them to justify its final judgment, even when such cues are ambiguously related to deepfake-specific artifacts.
\\
\\
Together, Figures~\ref{fig:criteria_articulation} and~\ref{fig:criteria_application} illustrate the two-stage reasoning process analyzed in Sec~\ref{sec:auth}:
GAI systems first define what constitutes evidence for deepfakes and subsequently apply those definitions in a flexible and interpretive manner when producing image-level authenticity judgments.

\end{document}